\documentclass[12pt, twoside]{article}
\usepackage{amsmath,amsthm,amssymb}
\usepackage{times}
\usepackage{enumerate}

\theoremstyle{definition}

\numberwithin{equation}{section}

\usepackage[utf8]{inputenc}
\usepackage{amsmath}
\usepackage{amsfonts}
\usepackage{dsfont}
\usepackage{tikz}
\usepackage{hyperref}
\usepackage{graphicx}
\usepackage{caption}
\usepackage{subcaption}
\usepackage{bm}
\usepackage{braket}
\usepackage{xcolor}
\usepackage{ amssymb }
\usepackage{framed}

\usepackage{appendix}

\usepackage{latexsym}
\usepackage{amsopn}
\usepackage{amsthm,amsmath,amssymb} 
\usepackage{verbatim}
\usepackage{color}
\usepackage{epstopdf}
\usepackage{algorithm}
\usepackage[noend]{algpseudocode}
\makeatletter
\def\BState{\State\hskip-\ALG@thistlm}
\makeatother

\def\bsymbol#1{%
\mbox{\boldmath$\displaystyle#1$\unboldmath}}

\renewcommand{\epsilon}{\varepsilon}
\newcommand{\bmnu}{{\bsymbol{\nu}}}
\newcommand{\sbmnu}{{{\small\nu}}}
\newcommand{\bmomega}{{\bsymbol{\omega}}}
\newcommand{\nv}{V}
\newcommand{\up}{u}
\newcommand{\bmx}{x}
\newcommand{\bmcx}{\mathbf{X}}
\newcommand{\bmnuu}{\tilde{\bmnu}}
\newcommand{\kapp}{\tilde{\kappa}}
\newcommand{\gamm}{\gamma}

\newcommand*\diff{\mathop{}\!\mathrm{d}}

\newcommand{\bOmega}{D}
\newcommand{\ghn}{{\Gamma_h^n}}
\newcommand{\ohn}{{\Omega_h^n}}

\newtheorem{problem}{Problem}[section]
\makeatletter
\newcommand{\setproblemtag}[1]{
	\let\oldtheproblem\theproblem
	\renewcommand{\theproblem}{#1}
	\g@addto@macro\endproblem{
		\addtocounter{problem}{-1}
		\global\let\theproblem\oldtheproblem}
	}
\makeatother

\title{A tractable mathematical model for tissue growth}

\author{Joe Eyles\footnote{Department of Mathematics,
University of Sussex, Brighton BN1 9HQ, UK},~ 
        John R. King\footnote{ 
        Centre for Mathematical Medicine, School of Mathematical Sciences,
University of Nottingham, Nottingham NG7 2RD, UK}
\footnote{
Synthetic Biology Research Centre, University of Nottingham, University Park, Nottingham,
NG7 2RD, UK}~~~and~ 
        Vanessa Styles$^*$}

\date{}

\begin{document}
\maketitle

\begin{abstract}
Using formal asymptotic methods we derive a free boundary problem representing one of the simplest mathematical descriptions of the growth and death of a tumour or other biological tissue.
The mathematical model takes the form of a closed interface evolving via forced mean curvature flow (together with a `kinetic under--cooling' regularisation) where the forcing depends on the solution of a PDE that holds in the domain enclosed by the interface. We perform linear stability analysis and derive a diffuse--interface approximation of the model. 
Finite--element discretisations of two closely related models are presented, together with computational results comparing the approximate solutions. 
\end{abstract}

\section{Introduction}
The aim of this paper is to derive and simulate numerically one of the simplest mathematical descriptions of the growth and death of a tumour (or, more generally, other tissue in a multicellular organism or a bacterial biofilm) 
comprising a nutrient--deprived necrotic (dead) core surrounded by a thin rim of nutrient--rich dividing cells. 
We emphasise we are not seeking biological realism here, but rather to clarify the properties of simple formulations, onto which any level of additional complexity can be built; we also seek to provide proof of concept for the numerical approaches adopted.
Despite the simplicity of the model, the numerical simulations in Section \ref{s:nr} show that the model 
produces many features that are seen in significantly more complicated ones. 
 Applying formal asymptotic methods, the thin rim collapses onto the interface that separates the tumour from its exterior; this 
generates two non--standard terms in the moving--boundary conditions for the tumour (the model otherwise corresponding to a Hele--Shaw reverse squeeze film with surface tension). We assume that the nutrient density is constant on the tumour surface and Darcy flow is used, this being the simplest constitutive assumption appropriate to modelling in a porous tissue--engineering scaffold, and having historically been widely adopted (at least since \cite{greenspan1, greenspan2}) in the modelling of tumour growth. This results (see below, where we clarify the role of nutrient in this formulation) in the following free boundary problem for the tissue pressure $u$,
\begin{eqnarray}
	\Delta \up = 1 &\mbox{ in }& \Omega(t), \label{eq_f2_uOnOmega_intro}\\
	\nabla \up \cdot \bmnu + \frac{\up}{\alpha} + \beta \kappa = Q &\mbox{ on }& \Gamma(t) , \label{eq_f2_uOnGamma_intro} \\
	\nv = \frac{\up}{\alpha} + \beta  \kappa &\mbox{ on }& \Gamma(t). \label{eq_f2_vOnGamma_intro} 
\end{eqnarray}

Here $\Omega(t)$ denotes the tumour, $\Gamma(t)=\partial\Omega(t)$, and $V$, $\kappa$ and $\bmnu$ are respectively the normal velocity, mean curvature and outer unit normal of $\Gamma(t)$, and $\alpha,\beta \in \mathbb{R}$ are positive constants. 
The key ingredients of the model are as follows: the right--hand side of $\Delta \up = 1$ is a volumetric sink that reflects loss of necrotic material and destabilises the free boundary; $Q$ is a surface source, capturing the viable rim around the tumour edge; and $\alpha$ and $\beta$ are regularising parameters, the unregularised case being associated with the limit $\alpha \rightarrow 0$ with $\beta$ fixed - the exact implication 
\[
	\nabla u \cdot \bmnu + V = Q ,
\]
of (\ref{eq_f2_uOnOmega_intro}) -- (\ref{eq_f2_vOnGamma_intro}) is instructive for this limit case.

The novelty of this free boundary problem is that it is derived via a thin--rim limit, resulting in a formulation that does not require additional equations to be solved for the nutrient density in the interior of the tumour. For a brief discussion of the framework of the model see \cite{jkow}, pp. 305.

The model is both simpler and more tractable than many found in the literature, compare for example \cite{cristini, chdTumourModel} and the references therein, and see \cite{byrne_chaplain_SI} and \cite{king_franks_04} for additional background to moving--boundary approaches to tumour growth related to that which is analysed here; indeed, it is perhaps the simplest such model that is able to capture the transition from the linear growth phase to growth saturation: this can be exemplified by the one--dimensional case: imposing symmetry about $x = 0$ and writing $\Gamma(t)$ as $x = s(t)$ gives
\[
	u = \alpha(Q - s(t)) + \frac{1}{2} (x^2 - s^2(t)) ,
\]
with 
\[
	\frac{\diff s}{\diff t} = Q - s, \quad s(t) = Q - (Q - s(0))e^{-t} .
\]
The case $s(0) \ll Q$ is then instructive, giving 
\begin{align*}
	\mbox{(linear growth)}  \quad \ ~ s(t) \sim s(0) + Qt  & \quad \mbox{ for } t = O\left( \frac{s(0)}{Q} \right) , \\
	\mbox{(transition to saturation)}  \quad s(t) \sim Q (1 - e^{-t})  & \quad \mbox{ for } t = O(1) .
\end{align*}

Two--component mixture models of tumour and healthy cells are considered in \cite{cristini, chdTumourModel}, these models taking the form of Cahn--Hilliard type diffuse--interface models in which the interfacial region between the tumour and the healthy cells has a non--zero thickness. 
In \cite{king_franks_04,king_franks_paper} both Darcy and Stokes constitutive assumptions are considered, allowing the tumour to be modelled as a highly viscous fluid moving either through a porous medium or unconstrained; this is also the case in \cite{cristini}, where movement of cells along chemical gradients (chemotaxis) is introduced via a nutrient diffusion equation. 
In \cite{chdTumourModel} a Cahn--Hilliard--Darcy model is considered that models chemotaxis and active transport, and formal matched asymptotic expansions are performed to yield several sharp--interface models. 
In \cite{greenspan1} the authors consider a cylindrical or spherical tumour and focus on the diffusion of nutrients and waste products in order to model tumour growth patterns. 
The geometrically constrained model in \cite{greenspan1} was extended in \cite{greenspan2} whereby, among other things, the geometric constraints are dropped: here the surface normal stress is taken to be proportional to the mean curvature, and Darcy's law is adopted. 
In \cite{byrne_chaplain_SI} the authors consider a single--nutrient sharp--interface model with two growth inhibitors, one external to the tumour, e.g. an anti--cancer drug or immune response, and one internal to the tumour, e.g. a by--product of the degradation of necrotic cells. The diffusion of the nutrient is considered, and its description on the boundary depends upon the curvature.
Although we do not consider anti--cancer drugs or immune responses here, a simplified version of these could potentially be modelled by reducing the size of our surface source.

In Section \ref{s:derivation} we derive the model using formal--asymptotic methods then, in Section \ref{s:stab}, we present calculations pertaining to the linear stability of radially symmetric solutions. 
A diffuse--interface formulation is derived in Section \ref{ss:DI_model} and in Section \ref{s:fea} we present finite--element approximations of the model and its diffuse--interface approximation. Numerical simulations are presented and discussed in Section \ref{s:nr}. In the appendix we perform analysis on the thin--film limit of the model.

\section{Derivation of the model}
\label{s:derivation}
\setcounter{equation}{0}

The model corresponds to a distinguished limit of the following dimensionless formulation (here we elaborate, with some minor differences, on analysis briefly outlined in \cite{king_franks_paper}). We have
\begin{equation}\label{der_L1}
\left. \begin{array}{c}
	\frac{\partial n}{\partial t} + \nabla \cdot (\bm{v} n) = (k_b(c) - k_d(c))n , \quad
	\frac{\partial m}{\partial t} + \nabla \cdot (\bm{v} m) = k_d(c) n - \epsilon \lambda m ,  \\
	n+m = 1 , \\
	\epsilon^2 \nabla \cdot (D(n) \nabla c) = K(c) n , \quad
	\bm{v} = - \nabla p / \mu(n; \epsilon) , 
	\end{array}
	\right\}
\end{equation}
wherein $n$ and $m$ are the volume fractions of live and dead cells, $c$ is the nutrient concentration, $k_b$ and $k_d$ are the cellular birth and death rates (the former being an increasing function of $c$ and the latter a decreasing one), $\epsilon \lambda$ specifies the (slow) degradation rate of the necrotic material, $K(c) / \epsilon^2$ expresses the nutrient consumption rate, which will be taken to be large, $D(n)$ is the nutrient diffusivity (nutrient transport can be treated as quasi--steady), $\bm{v}$ is the velocity field (the two phases being treated as a single continuum, i.e. the model can be characterised as being one--and--a--half phase rather than fully two phase), $p$ is the pressure (i.e. Darcy's law for a Newtonian fluid is adopted as the constitutive assumption, in keeping with many existing models) and $\mu(n; \epsilon)$ is proportional to the tissue viscosity; $\epsilon$ is a small parameter that appears in a number of places, associted with a key distinguished limit. 

The system (\ref{der_L1}) is taken to hold in a finite domain $\Omega(t)$, with boundary $\Gamma(t)$ and dimensionless boundary conditions (the sign convention being such that $\kappa < 0$ for a sphere)
\begin{equation}
\label{der_L2}
	c = 1, \quad
	p = - \epsilon^2 \gamma (n) \kappa, \quad
	q_{\sbmnu} = \bm{v} \cdot \bmnu, \quad 
	\mbox{ on }
	\Gamma(t) ,
\end{equation}
where $\epsilon^2 \gamma(n)$ expresses cell--cell adhesion, $\kappa$ is the mean curvature of $\Gamma(t)$, $\bmnu$ is its unit outward normal, and $q_{\sbmnu}$ its normal velocity. Since we shall be concerned with what is in effect the large--time behaviour, initial conditions are not important, though
\[
	n(\bmx, 0) = 1 \quad \mbox{ in } \Omega(0) ,
\]
would represent a plausible assumption. 

In addition to the powers of $\epsilon$ already introduced in (\ref{der_L1}), (\ref{der_L2}), the final condition that leads to the distinguished limit in question involves 
\begin{equation}
\label{der_L3}
	\mu(n; \epsilon) \sim \mu_0(n) \quad \mbox{ for } \quad 
	n = O(1) , \ n>0 , \ \mu(0; \epsilon) \sim \epsilon \mu_1 \quad \mbox{ as } \epsilon \rightarrow 0,
\end{equation}
for constant $\mu_1$, associated with the physically reasonable assumption that the necrotic material is much less viscous than living tissue.

We now derive the asymptotic structure of the problem in the limit $\epsilon \rightarrow 0$, which comprises two regions, namely a boundary--layer (thin--rim) around the tissue edge in which the living cells are concentrated, resulting from nutrient consumption therein, and a necrotic core. The former (thin--rim) is governed by a one--dimensional travelling--wave balance, described next; the latter generates, on matching to the former, the moving--boundary problem that is the subject of the rest of the paper.

The boundary--layer scalings are as follows
\[
	\nu = \epsilon \zeta, \quad 
	\bm{v} \cdot \bmnu  = \epsilon V_{\sbmnu} , \quad
	q_{\sbmnu} = \epsilon Q_{\sbmnu} , \quad
	p = \epsilon^2 P , \quad
	t = \epsilon^{-1} T ,
\] 
$\nu$ denoting the outward normal distance from $\Gamma(T)$; the tangential velocity components are of $O(\epsilon^2)$ but are not needed in the sequel.
We note that the relevant timescale is $t=O(\varepsilon^{-1})$ due to growth being confined to the thin rim (i.e. only a small minority of the volume is experiencing cell division) and dictating the velocity and pressure scalings above. Thus to leading order the travelling--wave balance 
\begin{align}
	\frac{\diff }{\diff \zeta} (W_{\sbmnu} n) = (k_b (c) - k_d(c)) n , & \quad
	\frac{\diff }{\diff \zeta} (W_{\sbmnu} m) = k_d (c) n , \nonumber \\
	n + m = 1 ,& \label{der_L4} \\
	\frac{\diff }{\diff \zeta} \left( D(n) \frac{\diff c}{\diff \zeta} \right) = K(c) n, & \quad
	V_{\sbmnu} = - \frac{1}{\mu_0 (n)} \frac{\partial P}{\partial \zeta} , \nonumber
\end{align}
holds in $\zeta < 0$ with
\begin{equation}
\label{der_L5}
	c = 1, \quad
	P = - \gamma(n) \kappa , \quad
	W_{\sbmnu} = 0 , \quad \mbox{ on } \zeta = 0 ,
\end{equation}
where $W_{\sbmnu} = V_{\sbmnu} - Q_{\sbmnu}$; we note that the leading--order solution depends only on $\zeta$, so that ordinary derivative notation is indeed appropriate, this being a deduction from the boundary value problem (\ref{der_L4}) -- (\ref{der_L6}) rather than an assumption; in particular $W_\nu$ depends only on $\zeta$ while $Q_\nu$ depends on $T$ and the tangential coordinates, and $V_\nu$, and hence $P$, depends on all the variables, this property of $W_\nu$ being a consequence of the Galilean invariance of the problem when $p$ is disregarded. Indeed, $P$ (and hence the mechanics) decouples, the remaining system being independent of the Darcy constitutive assumption and depending on $V_{\sbmnu}$ and $Q_{\sbmnu}$ only through $W_{\sbmnu}$. It follows from (\ref{der_L5}) that (\ref{der_L4}) implies
\[
	\frac{\diff W_{\sbmnu}}{\diff \zeta} = k_b(1) - k_d(1), \quad
	m \frac{\diff W_{\sbmnu}}{\diff \zeta} = k_d(1)n  \quad
	\mbox{ at } \zeta = 0,
\]
so that
\[
	n = 1 - \frac{k_d(1)}{k_b(1)}, \quad
	m = \frac{k_d(1)}{k_b(1)}  \quad
	\mbox{ at } \zeta = 0,
\]
and (\ref{der_L5}) in consequence involves the constant
\[
	\gamma_0 = \gamma \left( 1 - \frac{k_d(1)}{k_b(1)} \right) ,
\]
which serves as a surface--tension parameter in what follows.

As $\zeta \rightarrow - \infty$ we have
\begin{equation}
\label{der_L6}
	n \rightarrow 0, \quad
	m \rightarrow 1, \quad
	c \rightarrow c_\infty , \quad
	W_{\sbmnu} \rightarrow - Q ,
\end{equation}
where the positive constants $c_\infty$ and $Q$ are to be determined as part of the solution to (\ref{der_L4}) -- (\ref{der_L6}); $Q$ plays a crucial role in what follows, providing a surface source of material that constrasts with the volumetric sink $\lambda$. To complete the matching we must turn to the pressure $P$; the assumption (\ref{der_L3}) implies a non--uniformity here, but it is a consequence of $P$ decoupling that
\[
	V_{\sbmnu} \sim - \frac{1}{\mu(n ; \epsilon)} \frac{\diff P}{\diff \zeta},
\]
holds uniformly in the boundary layer regions (it will be clear that there is an \emph{ad--hoc} flavour to this statement, but it can readily be fully justified in the sense of formal asymptotics). In consequence
\[
	P \sim
	\epsilon \mu_1(Q - Q_{\sbmnu}) \zeta 
	+ Q_{\sbmnu} \int_\zeta^0 \left( \mu(n(\zeta'); \epsilon) - \epsilon \mu_1 \right) \diff \zeta '
	+ \int_\zeta^0 \left( \mu(n(\zeta'); \epsilon) W_{\sbmnu}(\zeta') + \epsilon \mu_1 Q \right) \diff \zeta '
	- \gamma_0 \kappa
\] 
applies, with (\ref{der_L4}) implying that $W_{\sbmnu}(\zeta)$ satisfies
\[
	\frac{\diff W_{\sbmnu}}{\diff \zeta} = k_b (c) n .
\]
Hence as $\zeta \rightarrow - \infty$, $\epsilon \rightarrow 0$ with $\zeta = O(1 / \epsilon)$,
\begin{equation}
\label{der_L7}
	P \sim 
	\epsilon \mu_1 (Q - Q_{\sbmnu}) \zeta
	+ a Q_{\sbmnu} - b - \gamma_0 \kappa ,
\end{equation}
for positive constants
\[
	a \equiv \int_{-\infty}^0 \mu_0 (n(\zeta)) \diff \zeta ,
	\quad
	b \equiv - \int_{-\infty}^0 \mu_0(n(\zeta)) W_{\sbmnu} (\zeta) \diff \zeta.
\]

We can now turn to the necrotic core region, wherein $n$ is exponentially small in $\epsilon$, so (\ref{der_L1}) reduces in $\Omega(t)$ at leading order to, on setting $\bm{v} = \epsilon \bm{V}$, 
\begin{equation}
\label{der_L8}
	\nabla \cdot \bm{V} = - \lambda , \quad
	\bm{V} = - \nabla P / \mu_1 ,
\end{equation}
with $m = 1$ and 
\begin{equation}
\label{der_L9}	
	\Delta c = 0.
\end{equation}

The boundary conditions on (\ref{der_L8}) -- (\ref{der_L9}) result from matching to (\ref{der_L6}) -- (\ref{der_L7}) (i.e. by the usual matching arguments from singular--perturbation theory, (\ref{der_L1}) does not apply to the core). Thus (\ref{der_L8}) is subject on $\Gamma(t)$ to 
\begin{equation}
\label{der_L10}
	V_{\sbmnu} = Q_{\sbmnu} - Q, \quad
	P = a Q_{\sbmnu} - b - \gamma_0 \kappa,
\end{equation}
(the creation of extra volume through cell division necessitating the deformation of the surrounding tissue), while the regularising $aQ_\nu$ term (mathematically equivalent to kinetic undercooling, but arising for reasons distinct from those relevant in widely studied phase change problems) is associated with the mechanical interactions between the large low--viscosity core and the high--viscosity rim (i.e. the former cannot readily deform the latter despite their relative sizes, (\ref{der_L3}) reflecting the associated distinguished limit). We stress that the `kinetic--undercooling' regulariastion arises from the thin--rim limit, whereas the `surface--tension' term $\gamma_0\kappa$ is imposed through the original boundary condition (\ref{der_L2}). $b$ can be viewed as a proliferation--driven contribution to the pressure, while (\ref{der_L6}) -- (\ref{der_L9}) simply give $c = c_\infty$; note that the first term on the right--hand side of (\ref{der_L7}) gives the matching condition
\[
	\bmnu \cdot \nabla P = \mu_1 V_{\sbmnu},
\]
which is automatically satisfied given (\ref{der_L8}) and (\ref{der_L10}). Finally we obtain, on translating $P$ by $b$, the moving--boundary problem
\begin{align}
\label{der_L11}
	\Delta P = \lambda \mu_1 , \quad \mbox{ in } \Omega(t), \\
	Q_{\sbmnu} = Q - \frac{1}{\mu_1} \frac{\partial P}{\partial \nu} ,
	\quad
	P = a Q_{\sbmnu} - \gamma_0 \kappa , 
	\quad \mbox{ on } \Gamma(t) \nonumber ,
\end{align}
where $\frac{\partial }{\partial \nu}$ denotes the outward--normal derivative. For $a = \gamma_0 = Q = 0$, this is the classical Hele--Shaw reverse (negative) squeeze--film problem (see, e.g., \cite{ok_ho_la_03}), which is ill--posed; the $a$ term corresponds to a kinetic--undercooling regularisation and $\gamma_0$ to a surface--energy regularisation (cf. \cite{ho_92}, for example, for a discussion of such regularisations in the Hele--Shaw context) - both are stabilising, but for small enough values (\ref{der_L11}) will be susceptible to fingering instabilities. The surface source term $Q$ is novel, making (\ref{der_L11}) distinct from previous Hele--Shaw formulations; since the instabilities are associated with shrinking fluid domains (due to $\lambda \mu_1 > 0$), it might be expected to play a stabilising role, though the linear--stability analysis in Section \ref{s:stab} quantifies the extent to which this intuition is misleading  (indeed, the growth associated with $Q$ has the potential to stabilise by filling inward--facing fingers but also allows more room for the destabilising mechanism to manifest itself).
Setting $P=u$, $a=\alpha$, $Q_{\sbmnu}=V$, $\beta=\gamma_0/a$ and scaling such that $\mu_1 = \lambda = 1$ yields (\ref{eq_f2_uOnOmega_intro}) -- (\ref{eq_f2_vOnGamma_intro}), with $t$ reinstated as the time variable, in place of (and equivalent to) $T$ above.

We regard the rigorous justification of the above `thin--rim' limiting process for (\ref{eq_f2_uOnOmega_intro}) -- (\ref{eq_f2_vOnGamma_intro}), from (\ref{der_L1}) as a worthwhile open problem. The limit has features in common with both sharp--interface and thin--film limits, without being exactly of either type (for example an interface is present both before and after the limit is taken, but the conditions that apply thereon are modified by the limit process).

\section{Model analysis}\label{s:stab}
\setcounter{equation}{0}
\subsection{Radially symmetric solutions and their stability}
\subsubsection{Circular solutions}\label{rad}

We limit ourselves here, for brevity, to the two--dimensional case, though the three--dimensional one is of course also tractable. Taking the cylindrically symmetric solution to be $u = u_c(r, t)$ in $\Omega$, with $r = s_c(t)$ corresponding to $\Gamma$,  (\ref{eq_f2_uOnOmega_intro}) -- (\ref{eq_f2_vOnGamma_intro}), reduces to 
\begin{equation}
	\dot{s}_c = Q - \frac{1}{2} s_c, \quad  u_c = \frac{1}{4} r^2 - \frac{1}{4} s_c^2 - \frac{\alpha \beta}{s_c}  + \alpha \dot{s}_c,
	\label{radusol}
\end{equation}
so that
\begin{equation}
\label{radrad}
	s_c = 2Q + (s_0 - 2Q) e^{-\frac{1}{2} t} ,
\end{equation}
for initial data $s_c(0) = s_0$.

\subsubsection{Linear stability}\label{sstab}

Setting
\[
	u \sim u_c(r , t) + \delta U(r, \theta, t), 
	\quad s \sim s_c (t) + \delta S(\theta, t),
\]
where $r = s(\theta, t)$ parametrises $\Gamma$, and retaining only those terms that are linear in $\delta$ implies
\[
	\Delta U = 0, \quad 0 \leq r \leq s_c (t) ,
\]
\[
	\frac{\partial S}{\partial t} = - \frac{1}{2} S - \frac{\partial U}{\partial r} , 
	\quad \frac{\partial S}{\partial t} = \frac{1}{\alpha} \left( \frac{1}{2} s_c S + U \right) + \frac{\beta}{s_c^2} \left( S + \frac{\partial^2 S}{\partial \theta^2} \right) \quad \mbox{ on } r = s_c(t) ,
\]
on projecting $\Gamma$ onto $r = s_c(t)$ in the usual way. Fourier decomposing in the form
\[
	U = A_n(t) \left( \frac{r}{s_c} \right) ^n 
	\begin{cases}
		\cos(n \theta) \\
		\sin(n \theta)
	\end{cases},
	\quad S = B_n (t) 
	\begin{cases}
		\cos(n \theta) \\
		\sin(n \theta)
	\end{cases} 
\]
for non--negative integer $n$ then implies that
\[
	\dot{B}_n = -\frac{1}{2} B_n - \frac{n}{s_c} A_n, 
	\quad
	\dot{B}_n = \frac{1}{\alpha} \left( \frac{1}{2} s_c B_n + A_n \right) - \frac{\beta}{s_c^2} (n^2 - 1) B_n ,
\]
so that
\begin{equation}
\label{stab:N2}
	(s_c + n \alpha) \dot{B}_n = \frac{n-1}{2 s_c^2} \left( s_c^3 - 2 \alpha \beta n (n + 1) \right) B_n 
	.
\end{equation}
Hence 
\[
	\dot{B}_0 = - \frac{1}{2} B_0, \quad \dot{B}_1 = 0 ,
\]
the former corresponding to perturbing $s_0$ in (\ref{radrad}), and the latter arising because the $n = 1$ modes simply correspond to small shifts in the coordinate origin. $\dot{B}_n$ is negative for $n \geq 2$, and the circular solution is hence linearly stable, if 
\begin{equation}
\label{stab:N3}
	12 \alpha \beta > s_c^3,
\end{equation}
i.e. the steady state $s_c = 2 Q$ is linearly stable if
\begin{equation}
\label{stab:N4}
	3 \alpha \beta > 2 Q^3.
\end{equation}

In view of (\ref{stab:N2}), it is instructive to determine the faster growing mode, i.e. value of $n$ that maximises 
\begin{equation}
\label{stab:N4_again}
	\frac{n-1}{2 s_c^2 (s_c + n Q)} \left( s_c^3 - 2 \alpha \beta n (n + 1) \right),
\end{equation}
whenever the latter is positive, and this may change as $s_c$ evolves; for $s_0$ sufficiently small (and less than $2 Q$), all modes will initially decay, but when (\ref{stab:N4}) is violated the $n = 2$ mode will grow for sufficiently large $t$, and other modes may subsequently become unstable. For a wide range of parameters the $n = 2$ mode will be the most unstable one for all $t$, but an instructive regime in which this is not the case arises for $\beta$ small. Thus for $\beta = 0$ all modes with $n \geq 2$ are unstable, with the limit $n \rightarrow \infty$ giving the fastest growth rate, namely
\[
	 \frac{s_c}{2 \alpha}
\]
(this expression clarifies the role of $\alpha$ as a regularising parameter). For $\beta \ll 1$ and $n \gg 1$, (\ref{stab:N4_again})  is given asymptotically by 
\begin{equation}
\label{stab:N5}
	\frac{s_c}{2 \alpha} - \left( 1 + \frac{s_c}{\alpha} \right) \frac{s_c}{2 \alpha n} - \frac{\beta}{s_c^2} n^2 ,
\end{equation} 
for $n = O(\beta^{-\frac{1}{3}})$, so the fastest growing mode at a given $t$ has
\[
	n \sim s_c \left( (1 + \frac{s_c}{\alpha}) \frac{1}{4 \alpha \beta}\right) ^{\frac{1}{3}}.
\]
Because this quantity changes with $t$, the dominant wavelength observed  at large time will presumably be dependent on how the relative magnitudes of the modes in the initial data feed through into the ultimate nonlinear evolution, and it is important to note that the leading term in (\ref{stab:N5}) is independent of $n$. While this instability mechanism is closely related to the well known Saffman--Taylor instability in Hele--Shaw and Darcy flows, its relevance to tissue growth applications is less widely recognised, a key point being that (in contrast to the Hele--Shaw case) instabilities can arise for growing domains (due to the surface source $Q$) as well as shrinking ones; see \cite{king_franks_04, king_franks_book} for a detailed classification of instability mechanisms for such tissue--growth moving--boundary problems.

\subsection{Other properties}
\label{s:other_matters}

We briefly note some other properties here (additionally the appendix records a thin--film limit). The result
\begin{align}
	\frac{\diff }{\diff t} \mbox{Vol}(\Omega(t)) &= \int_{\Gamma(t)} V \diff S \nonumber \\
	& = Q \mbox{Area}(\Gamma(t)) - \int_{\Gamma(t)} \nabla u \cdot \bmnu \diff S \nonumber \\
	& = Q \mbox{Area}(\Gamma (t)) - \mbox{Vol} (\Omega(t)) ,
\label{eq_mass_conserv_other_prop}
\end{align}
which readily follows on applying the divergence theorem, expresses the overall mass balance arising from the surface source and the volumetric sink. Thus tissue can enhance its growth by maximising its surface area for a given volume, making spheres the most sub--optimal shapes in this sense, and providing an interpretation of interface fingering (this does not characterise a fingering mechanism but does clarify why such instabilities may be favoured).

Secondly, steady state solutions satisfy
\[
	\Delta u = 1,
\]
\[
	u = -\alpha \beta \kappa, \quad  \nabla u \cdot \bmnu = Q ,
\]
and this free--boundary problem is itself of interest.
The circular case is addressed in the linear--stability section, Section \ref{sstab}, but there are presumably also non--trivial (but probably unstable) steady configurations. Applying the linear stability analysis may, by identifying bifurcation points, be instructive in classifying these solutions and in facilitating weakly nonlinear analyses but, given that they seem likely to be unstable, we shall not pursue their analysis here. Nevertheless, we highlight their potential interest from the free--boundary problem perspective.

Finally, we record the interfacial dynamics limit that arises for small domains (though of course nevertheless much larger than the rim thickness) with $\alpha \gg 1$, whereby the leading--order problem involves the evolution of the free boundary only. Setting $\alpha = 1 / \epsilon$, $\beta = \epsilon \gamma$, and rescaling $\bm{x} \rightarrow \epsilon \bm{x}$, $V \rightarrow \epsilon V$, $Q \rightarrow \epsilon Q$ to give
\[
	\Delta u = \epsilon^2
\]
\[
	\nabla u \cdot \bm{\nu} + \epsilon ^2 V = \epsilon^2 Q , 
	\quad V = u + \gamma \kappa,
\]
so that
\[
	u \sim U(t), \quad V \sim U + \gamma \kappa ,
\]
and the divergence theorem (or, equivalently, the solvability condition at $O(\epsilon^2)$) gives (as above)
\begin{align*}
	\mbox{Vol}(\Omega(t)) 
	&= Q \mbox{Area}(\Gamma(t)) - \int_{\Gamma(t)} V \diff S \\
	&= (Q - U) \mbox{Area}(\Gamma(t)) - \gamma \int_{\Gamma(t)} \kappa \diff S ,
\end{align*} 
implying the novel non--local interfacial dynamics law of forced mean--curvature--flow type
\begin{equation}
\label{eq_non-local_dynamics}
	V = Q - \frac{\mbox{Vol}(\Omega(t))}{\mbox{Area}(\Gamma(t))} 
	+ \gamma \left( \kappa - \frac{\int_{\Gamma(t)} \kappa \diff S}{\mbox{Area}(\Gamma(t))} \right)
\end{equation}
that involves the total mean curvature; this becomes more explicit in two dimensions, whereby 
\[
	V = Q - \frac{\mbox{Area}(\Omega(t))}{\mbox{Length}(\Gamma(t))} 
	+ \gamma \left( \kappa - \frac{2 \pi}{\mbox{Length}(\Gamma(t))} \right) .
\]
We note that the exact result (\ref{eq_mass_conserv_other_prop}) is readily recovered from the limit case (\ref{eq_non-local_dynamics}).

\section{A diffuse-interface model}
\label{ss:DI_model}
\setcounter{equation}{0}
In this section we derive a diffuse--interface formulation of (\ref{eq_f2_uOnOmega_intro}) -- (\ref{eq_f2_vOnGamma_intro}). 
In the diffuse--interface paradigm we approximate the interface $\Gamma(t)$ by an interfacial region $\Gamma_\varepsilon(t)$ with width $C \varepsilon\ll 1$. 
To this end a phase--field parameter $\varphi$ is introduced such that, within the thin interfacial region, $\varphi$ varies smoothly from $\varphi = -1$ to $\varphi = 1$, while outside it $|\varphi| = 1$, thus we can define $\Gamma_\varepsilon$ by 
$$
\Gamma_\varepsilon := \Set{\bmx \in \mathbb{R}^d|   \left|\varphi(\bmx)\right| < 1} .
$$
We set $\varphi=1$ in the region inside $\Gamma_\varepsilon$ and $\varphi=-1$ exterior to $\Gamma_\varepsilon$. 
In addition we let $\bOmega$ denote a bounded domain in $\mathbb{R}^d$, with a polygonal boundary, such that $\bOmega$ is chosen large enough to contain $\Gamma_\varepsilon(t)$ for all $t$. We denote the outward unit normal to $\partial\bOmega$ by $\tilde{\bsymbol{\nu}}$.

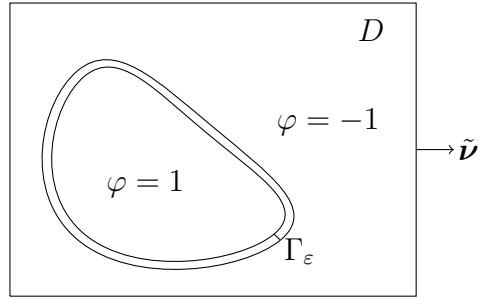
\begin{figure}[H]
    \centering  
\begin{tikzpicture}[scale = 0.5]
   \draw plot coordinates {(0,0) (10.8,0) (10.8,7.8) (0,7.8) (0,0)};
\draw plot [smooth cycle, tension=1] coordinates {(1.8,1.5) (7.2,1.5) (5.4,4.5) (1.8,6)};
\draw plot [smooth cycle, tension=1] coordinates {(1.98,1.65) (7.02,1.65) (5.22,4.35) (1.98,5.85)};
\draw[-] plot (7.178,1.503) -- (7.02,1.65);
\node[draw] [fill=none,draw=none] at (7.7,1.2) {$\Gamma_\varepsilon$};
\node[draw] [fill=none,draw=none] at (9.6,7.075) {$\bOmega$};
\node[draw] [fill=none,draw=none] at (12.2,3.9) {$\tilde{\bsymbol{\nu}}$};
\draw[->] plot (10.8,3.9) -- (11.8,3.9);
\node[draw] [fill=none,draw=none] at (3.6,3) {$\varphi = 1$};
\node[draw] [fill=none,draw=none] at (8.46,4.65) {$\varphi = -1$};
\end{tikzpicture}
\caption{Diffuse--interface configuration.}
    \label{fig:CurveInSpace}
\end{figure}

To derive a diffuse--interface approximation of (\ref{eq_f2_uOnOmega_intro}), (\ref{eq_f2_uOnGamma_intro}) we first introduce the variational form:
\begin{equation}
\label{equation_ParametricWeak_u_f2}
\int_\Omega \nabla u \cdot \nabla \eta \diff \bmx + \frac{1}{\alpha} \int_\Gamma u \eta \diff \bmx = \int_\Gamma (Q - \beta \kappa) \eta \diff \bmx -  \int_\Omega \eta \diff \bmx  ~~ \forall \eta \in H^1(\Omega). 
\end{equation}
We now follow the techniques described in \cite{alam} 
to obtain a diffuse--interface approximation of (\ref{equation_ParametricWeak_u_f2}). Setting $\tilde{u}\in \bOmega$ to be a diffuse--interface approximation to $u\in\Omega$ we have  
\begin{equation}
\label{equation_phase_weighted_zeta_approx_integral}
	\int_{\Omega} u(\bmx) \diff \bmx \approx \int_{\bOmega} \zeta(\varphi) \tilde{u}(\bmx) \diff \bmx ~~~\mbox{and}~~~
	\int_{\Gamma} u(\bmx) \diff \bmx \approx \int_{\bOmega} \delta(\varphi) \tilde{u}(\bmx) \diff \bmx 
\end{equation}
where
\[
	 \zeta(\varphi) := \frac{1+\varphi(\bmx)}{2}~~~\mbox{and}~~~ \delta(\varphi)  := \frac{2}{\pi \epsilon} (1 - (\varphi (\bmx))^2) .
\]
Using (\ref{equation_phase_weighted_zeta_approx_integral}) we arrive at the following diffuse--interface approximation of (\ref{equation_ParametricWeak_u_f2}):
\begin{equation}
\label{equation_phase_weak_u7}
	\int_{\bOmega} \zeta(\varphi) \nabla \tilde{u} \cdot \nabla \eta \diff \bmx + \frac{1}{\alpha} \int_{\bOmega} \delta(\varphi) \tilde{u} \eta \diff \bmx = \int_{\bOmega}  \delta(\varphi)(Q - \beta \kapp) \eta \diff \bmx - \int_{\bOmega} \zeta(\varphi) \eta \diff \bmx , 
	\quad \forall \eta \in H^1(\bOmega).
\end{equation}
Here $\kapp$ approximates an extension of $\kappa$ to $\Gamma_\varepsilon$ such that 
at $t=0$, $\kapp$ is obtained by constantly extending $\kappa$ in the normal direction via the signed distance function to $\Gamma(0)$, while for $t>0$ we use the  Allen--Cahn approximation of forced--mean--curvature flow, presented below in (\ref{equation_phase_allencahn}), to set 
\begin{equation}
\kapp = \frac{1}{\beta}\left(\varepsilon\varphi_t- \frac{\pi \tilde{u}}{4  \alpha}\right)\qquad |\varphi|<1.
\label{curv_ext}
\end{equation}
 
For $\beta=0$, the strong convergence of the solution $\tilde{u}$ of (\ref{equation_phase_weak_u7}) to the solution $u$ of (\ref{equation_ParametricWeak_u_f2}), as the approximation parameter $\varepsilon$ tends to zero, is shown in \cite{alam}, while in \cite{Burger2014} 
a convergence result of the form
\[
	\int_{\bOmega} \zeta(\varphi) \left( |\nabla (\tilde{u}-u)|^2 + |\tilde{u}-u|^2 \right) \diff \bmx \leq C\varepsilon^p
\]
for a constant $C > 0$ independent of $\varepsilon$, and some $p > 0$ is shown for a slightly different diffuse interface approximation in which $\delta(\varphi)=|\nabla\varphi|$.

We now present a diffuse--interface approximation or, as it is more commonly known, a phase--field approximation, of (\ref{eq_f2_vOnGamma_intro}). Since (\ref{eq_f2_vOnGamma_intro}) describes forced--mean--curvature flow, we approximate it using an Allen--Cahn equation with forcing,
\begin{equation}
\label{equation_phase_allencahn}
	\varepsilon \varphi_t + \beta ( \frac{1}{\varepsilon} W' (\varphi) - \varepsilon \Delta \varphi ) =  \frac{\pi\tilde{u}}{4\alpha}, \quad |\varphi|<1,
\end{equation}
with boundary data 
\begin{equation}
\label{eef}
	\nabla \varphi \cdot \bmnuu= 0 , \quad \mbox{ on } \partial \bOmega. 
\end{equation}
Here $\tilde{u}$ is the solution to (\ref{equation_phase_weak_u7}), 
$W(\varphi)$ is the double--obstacle potential, see \cite{Blowey_Elliott_double_obs}, 
$$
W(\varphi)=\frac12(1-\varphi^2)+I_{[-1,1]}(r)
$$
where $I_{[-1,1]}(u)$ is the indicator function 
$$
	I_{[-1, 1]} (\varphi) := 
	\begin{cases}
	+ \infty & \mbox{ for } |\varphi| > 1\\
	0 & \mbox{ for } |\varphi| \leq 1
	\end{cases}
$$
and the factor of $\pi/4$ on the right hand size of (\ref{equation_phase_allencahn}) is chosen
such that as $\varepsilon \rightarrow 0$, we recover the velocity law (\ref{eq_f2_vOnGamma_intro}). 

We define the initial data for the phase--field parameter $\varphi$ by
\begin{equation}
	\varphi_0(\bmx) = 
	\begin{cases}
	1 & \mbox{ if } d(\bmx) \geq \frac{\varepsilon \pi}{2} \\
	\sin (\frac{d(\bmx)}{\varepsilon}) & \mbox{ if } -\frac{\varepsilon \pi}{2} < d(\bmx) < \frac{\varepsilon \pi}{2} \\
	-1 & \mbox{ if } d(\bmx) \leq -\frac{\varepsilon \pi}{2}  
	\end{cases}  \label{phiid}
\end{equation}
where $d(\bmx)$ is the signed distance function to $\Gamma(0)$. 
This initial condition corresponds to a diffuse--interface of width $\varepsilon \pi$, with the level set $\varphi = 0$ approximating $\Gamma(0)$.\\

Combining \eqref{equation_phase_weak_u7} with the weak formulation of (\ref{equation_phase_allencahn}), \eqref{eef} yields the following diffuse--interface approximation of (\ref{eq_f2_uOnOmega_intro}) -- (\ref{eq_f2_vOnGamma_intro}):\\

Given $\Gamma(0)\in \mathbb{R}^d$, find $(\tilde{u},\varphi)$ such that $\varphi$ satisfies the initial data (\ref{phiid}) and for all $t\in (0,T)$, $\varphi(t)\in K$,
\begin{equation}
\label{equation_phase_weak_u}
	\int_{\bOmega} \zeta(\varphi) \nabla \tilde{u} \cdot \nabla \eta \diff \bmx + \frac{1}{\alpha} \int_{\bOmega} \delta(\varphi) \tilde{u} \eta \diff \bmx = \int_{\bOmega}\delta(\varphi) (Q-\beta \kappa)  \eta \diff \bmx - \int_{\bOmega} \zeta(\varphi) \eta \diff \bmx , 
	\quad \forall \eta \in H^1(\bOmega),
\end{equation}
and
\begin{equation}
	   \int_{\bOmega} \left(\varepsilon\varphi_t (\rho - \varphi) 
	+ \varepsilon \beta   \nabla \varphi \cdot \nabla (\rho - \varphi) 
	 -  \frac{\beta}{\varepsilon}    \varphi (\rho - \varphi)\right) \diff \bmx
	 \geq\frac{\pi}{4 \alpha}   \int_{\bOmega}  \tilde{u} (\rho - \varphi) \diff \bmx, ~~\forall \rho \in K\label{equation_phase_weak_v}
\end{equation} 
where
$$
K:=\{\eta\in H^1(\bOmega)|~ |\eta|\leq 1~\mbox{a.e. in~}\bOmega\}.
$$
The well-posedness of the variational problem (\ref{equation_phase_weak_u}), (\ref{equation_phase_weak_v}) is an open problem. 
Setting $\beta=0$ in (\ref{equation_phase_weak_u}) and adding a time regularisation term, $\varepsilon \tilde{u}_t$, to the left hand side of  the resulting equation would result in a diffuse--interface system similar to the one studied in \cite{digm} in which a parabolic variational inequality is coupled to a degenerate diffusion equation. In \cite{digm} the authors 
use a regularisation technique to prove an existence and uniqueness result for the coupled system.

\section{Finite--element approximations}\label{s:fea}
\setcounter{equation}{0}
In this section we derive finite--element approximations to (\ref{eq_f2_uOnOmega_intro}) -- (\ref{eq_f2_vOnGamma_intro}), and to the diffuse--interface model (\ref{equation_phase_weak_u}), (\ref{equation_phase_weak_v}).

\subsection{Parametric approach}\label{ss:parametric}

We partition the time interval $[0,T]$ into $N$ equidistant steps, $0 = t_0 < t_1 < \ldots < t_{N-1} < t_N = T$, so that for $i=1,\ldots,N$, $\Delta t :=t_n-t_{n-1}$. 
We let $\Gamma_h^n$ be a polyhedral surface approximating $\Gamma(t_n)$, with ${\Gamma_h^n = \bigcup_{i=1}^I \overline{\sigma}_i^n}$, where $\{ \sigma_i^n \}^I_{i=1}$ is a family of disjoint open simplices (straight line segments in $\mathbb{R}^2$ and triangles in $\mathbb{R}^3$). 
We let $\mathcal{J}$ be the set of nodes of $\{ \sigma_i^n \}^I_{i=1}$ and let $\{ \bm{q}_j^n \}_{j\in \mathcal{J}}$ be the coordinates of these nodes.

We denote the interior of $\Gamma_h^n$ by $\Omega_h^n$ and we partition $\Omega_h^n$ into a family of disjoint open simplices $\{ \mu_k^n \} ^K_{k=1}$ (triangles in $\mathbb{R}^2$ and tetrahedra in $\mathbb{R}^3$) such that $\Omega_h^n = \bigcup_{k=1}^K \overline{\mu}_k^n$. Here $\{ \mu_k^n \} ^K_{k=1}$ is chosen such that the nodes and the faces (straight line segments in $\mathbb{R}^2$ and triangles in $\mathbb{R}^3$) of the elements of $\{ \mu_k^n \} ^K_{k=1}$ that make up the boundary of $\Omega_h^n$ coincide with nodes and simplices of $\{ \sigma_i^n \}^I_{i=1}$.
We let $\mathcal{L}$ be the set of nodes of $\{ \mu_k^n \}^K_{k=1}$ and let $\{ \bm{p}_l^n \}_{l\in \mathcal{L}}$ be the coordinates of these nodes.

We define $\bmnu^n$ to be the outward unit normal to $\Gamma_h^n$, such that $\bmnu_i^n:=\bmnu^n|_{\sigma_i}$, for $i\in 1,\ldots,I$ and 
for $n\in 0,\ldots,N-1$ we introduce the finite--element spaces
\[
	V_h(\Gamma_h^n) := \set{ \rho_h \in C(\Gamma_h^n) | \rho_h \mbox{ is affine on each } \sigma_i^n \in \Gamma_h^n },
\]
and
\[
	V_h(\Omega_h^n) := \set{ \eta_h \in C(\overline{\Omega_h^n}) | \eta_h \mbox{ is affine on each } \mu_k^n \in \Omega_h^n} 
\]
with $\displaystyle{\{\chi_j^n\}_{j\in\mathcal{J}}}$ 
denoting the standard basis for $V_h(\Gamma_h^n)$.

We describe $\Gamma_h^n$ by the piecewise linear vector function $\bm{X}^n_h$ such that $\bm{X}_h^n\in [V_h(\Gamma_h^{n-1})]^d$.\\

We denote the $L^2$--inner product over $\ohn$ by ${(f,g)_\ohn:=\int_{\ohn}fg dx}$ and the discrete $L^2$--inner product by
$$
(f,g)_\ohn^h:=\int_{\ohn}\Pi^h(fg)dx
$$ 
with $\Pi^h:C(\overline{\ohn})\to S_h$ denoting the interpolation operator, such that $(\pi^h\eta)(\bm{p}_l)=\eta(\bm{p}_l)$ for all $l\in \mathcal{L}$. We adopt similar notation for $(u_h,v_h)_\ghn^h$. \\

We now present a finite element approximation of (\ref{eq_f2_uOnOmega_intro}) -- (\ref{eq_f2_vOnGamma_intro}) in which we used (\ref{equation_ParametricWeak_u_f2}) in order to approximate (\ref{eq_f2_uOnOmega_intro}) -- (\ref{eq_f2_uOnGamma_intro}) and we followed the authors in \cite{Dziuk1990} to approximate (\ref{eq_f2_vOnGamma_intro}). \\

Given $\Gamma^0$ and the identity function $\bm{X}^0_h\in [V_h(\Gamma_h^0)]^d$ on $\Gamma^0$, for $n=0\to N-1$ find $\{u_h^n, \bm{X}_h^{n+1}\} \in V_h(\Omega_h^n)\times [V_h(\Gamma_h^n)]^d$ such that 
\begin{equation}
\label{equation_ParametricFEM_u_f2}
\left( \nabla u_h^n, \nabla \eta_h \right)_\ohn+ \frac{1}{\alpha} \left( u_h^n,\eta_h^n\right)_\ghn^h=\left( Q - \beta \kappa_h^n,\eta_h^n\right)_\ghn^h-\left(1,\eta_h\right)_\ohn~~~\forall \eta_h \in V_h(\Omega_h^n),
\end{equation}
and
\begin{equation}
 \label{equation_ParametricFEM_v_f2} 
\frac1{\Delta t}\left(\bm{X}_{h}^{n+1} - \bm{X}_{h}^{n},\bm{\rho}_{h} \right)_{\Gamma_h^n}^h+\beta\left( \nabla_{\Gamma_h^n} \bm{X}_h^{n+1} ,\nabla_{\Gamma_h^n} \bm{\rho}_h \right)_{\Gamma_h^n}=\frac{1}{\alpha} \left( u_h^n \bmomega^n_{h},\bm{\rho}_{h} \right)_{\Gamma_h^n}^h
	~~~ \forall \bm{\rho}_h \in [V_h(\Gamma_h^n)]^d,
\end{equation}
where $\displaystyle{\bmomega^n_{h}:=\sum_{j\in\mathcal{J}}\bmomega_j^n\chi_j^n(x)\in [V_h(\Gamma_h^n)]^d}$ with 
$$
	\bmomega^n_j = \frac1{|\Lambda_j^n|} \left( \displaystyle{\sum_{\sigma_i\in \mathcal{T}_j^n}\bmnu^n_i|\sigma_i|} \right),~~~~~~\mbox{for}~j\in \mathcal{J},~n=0,\ldots,N,
$$
with $|\sigma_i|$ denoting the measure of $\sigma_i$,  $\mathcal{T}_j^n:=\{\sigma_i^n:\bm{q}_j^n\in \overline{\sigma_i^n}\}$ and $\Lambda_j^n:=\cup_{\sigma_i\in \mathcal{T}_j^n}\overline{\sigma_i^n}$, such that $\bmomega_j^n$ can interpreted as a weighted normal defined at the node $\bm{X}_{h}^{n}(\bm{q}_j^n)=\bm{q}_j^n$ of the surface $\Gamma_h^n$.\\

Noting from (\ref{eq_f2_vOnGamma_intro}) that $\beta \kappa=V-u/\alpha$, 
the approximation, $\displaystyle{ \beta \kappa_h^n= \beta\sum_{j\in\mathcal{J}}\kappa_j^n\chi_j^n(x)\in V_h(\Gamma_h^n)}$, of the curvature is given by
$$
\beta \kappa_j^n= \left(\frac{\bm{X}_{j}^{n} - \bm{X}_{j}^{n-1}}{\Delta t}\right)\cdot \bmomega_j^n - \frac1{\alpha} u_j^n,~~~\mbox{for}~j\in \mathcal{J},~n=1,\ldots,N,
$$
and we set $\kappa_h^0\in V_h(\Gamma_h^0):=\Pi^h(\kappa_0)$, where $\kappa_0$ is the curvature of $\Gamma(0)$.\\

\subsection{Diffuse--interface approach}\label{ss:diffuse}

We partition the time interval $[0,T]$ into $N$ equidistant steps, $0 = t_0 < t_1 < \ldots < t_{N-1} < t_N = T$, so that for $i=1,\ldots,N$, $\Delta t :=t_{n}-t_{n-1}$. 
Let 
 $\{{\cal T}^h\}_{h>0}$ be a family of partitionings of $\bOmega$ into disjoint open simplices $\sigma$ with $h_{\sigma}:={\rm diam}(\sigma)$ and $h:=\max_{\sigma\in {\cal T}^h}h_\sigma$.

Associated with ${\cal T}^h$ is the finite element space 
$$
S_h := \set{ \rho_h \in C(\overline{\bOmega}) | \rho_h|_\sigma \mbox{ is affine } \forall \sigma \in {\cal T}^h }\subset H^1(\bOmega).
$$
We denote the standard basis for $S_h(\bOmega)$ by $\displaystyle{\{\psi_k^n\}_{k=1}^M}$ and we introduce 
$$
K_h := \set{ \rho_h \in S_h | |\rho_h| \leq 1 \mbox{ in }\bOmega }\subset K.
$$

Using standard finite element approximations of (\ref{equation_phase_weak_u}) and (\ref{equation_phase_weak_v}) we obtain the following. \\

Given an approximation $\varphi_h^0\in K_h$ of $\varphi_0\in K$,  for $n=0\to N-1$ find $\{\tilde{u}_h^n,\varphi_h^{n+1}\} \in S_h\times K_h$ such that 
for all $\eta_h \in S_h$,
\begin{equation}
\label{equation_phase_FEM_u}
\big( \zeta(\varphi^{n}_h) \nabla \tilde{u}_h^n, \nabla \eta_h \big)_{\bOmega}^h+\frac{1}{\alpha} \big(\delta(\varphi_h^n)\tilde{u}_h^n,\eta_h\big)_{\bOmega}^h=\big(\delta(\varphi_h^n)(Q - \beta \kapp_h^n),\eta_h\big)_{\bOmega}^h-\big(\zeta(\varphi_h^n),\eta_h\big)_{\bOmega}^h\end{equation}
and
\begin{eqnarray}
\frac\varepsilon{\Delta t}\left(\varphi_h^{n+1} - \varphi_h^{n},\rho_h - \varphi_h^{n+1}\right)_{\bOmega}^h\!\!\!&+&\!\!\!\beta \varepsilon(\nabla \varphi_h^{n+1} , \nabla (\rho_h - \varphi_h^{n+1}))_{\bOmega} -  \frac{\beta}{\varepsilon} 
(\varphi_h^{n+1}, \rho_h  - \varphi_h^{n+1})_{\bOmega}^h\nonumber\\
&&\hspace{10mm}\geq \frac{\pi}{4  \alpha} \left(\tilde{u}_h^n,\rho_h - \varphi_h^{n+1}\right)_{\bOmega}^h ~~ \forall \rho_h \in K_h,\label{equation_phase_FEM_v}
\end{eqnarray}
where $\tilde{u}_h^n(x)$ is set to zero if $\varphi_h^n(x)=-1$, $(\cdot,\cdot)_{\bOmega}^h$ is discrete inner product over $\bOmega$ and noting (\ref{curv_ext}) the approximation $\displaystyle{\beta \kapp_h^n=\beta \sum_{j=1}^M\kapp_j^n\psi_j\in S_h}$, of the curvature is given by
$$
 \kapp_j^n = \frac{1}{\beta}\left(\frac\varepsilon{\Delta t}(\varphi_j^{n} - \varphi_j^{n-1}) - \frac{\pi \tilde{u}_j^n}{4  \alpha}\right)~~~\mbox{for}~j=1\ldots,M,~n=1,\ldots,N,
$$
with 
$\kapp_h^0\in S_h=\Pi^h(\tilde{\kappa}_0)$, where $\kappa_0$ is the curvature of $\Gamma(0)$ and $\tilde{\kappa}_0$ is such that 
$\tilde{\kappa}_0(x):=\kappa_0(\hat{p}(x))$, where $\hat{p}(x)$ denotes the closest point projection of a point $x\in \bOmega$ onto $\Gamma(0)$.\\[0mm]

\section{Numerical Results}\label{s:nr}
\setcounter{equation}{0}

In this section we first briefly describe the implementation of the finite element schemes presented in Section \ref{s:fea}, and then 
we present a number of numerical simulations. The results displayed are visualised in Paraview \cite{paraview_book}.

\subsection{Implementation of the finite element schemes}
\subsubsection{Implementation of the parametric scheme in $\mathbb{R}^2$}

For $\Omega_h^n\in \mathbb{R}^2$ the parametric finite--element scheme (\ref{equation_ParametricFEM_u_f2}), (\ref{equation_ParametricFEM_v_f2}) was implemented with the finite--element toolbox ALBERTA 2.0, see \cite{ALBERTABook}. %
As we are solving on an evolving domain, care has to be taken to ensure good mesh properties for the bulk mesh $\Omega_h^n$. To achieve this we implemented Algorithm 1, as introduced in \cite{Elliotta} in which the authors present a variant of the so-called the DeTurck trick using the the harmonic map heat flow that leads to computational meshes of high quality for both for the bulk mesh $\Omega_h^n$ and its boundary $\Gamma_h^n$.

A simple mesh refinement algorithm is also presented in \cite{Elliotta}, in which we seek to refine element $\mu^n \in \Omega_h^n$ if
$|\mu^n| \geq 1.5\overline{|\mu^0|}$
and coarsen element  $\mu^n \in \Omega_h^n$ if $|\mu^n| \leq 0.5\overline{|\mu^0|}$,
where $\mu^0 \in \Omega_h^0$, $|\mu^n|$ is the area of element $\mu^n$, and $\overline{|\mu^0|}$ is the mean area of all elements at time $t=0$. 

When the mesh reached a point at which the algorithm presented in \cite{Elliotta} was unable to maintain the required mesh quality, we used the re--meshing software GMSH, see \cite{GMSH_paper}, to re--mesh $\Omega_h^n$,  \textcolor{black}{whilst keeping the nodes on $\Gamma_h^n$ fixed}. 
We performed this re--meshing procedure when
\[
	\max_{\mu^n \in \Omega_h^n} \frac{H(\mu^n)}{h(\mu^m)} \geq 0.2 ,
\]
where 
$H(\mu^n)$ is the length of the longest edge of $\mu^n$ and $h(\mu^n)$ is the length of the shortest edge. 

\subsubsection{Implementation of the parametric scheme in $\mathbb{R}^3$}
For $\Omega_h^n\in \mathbb{R}^3$ the parametric finite--element scheme (\ref{equation_ParametricFEM_u_f2}), (\ref{equation_ParametricFEM_v_f2}) was implemented with the finite--element toolbox ALBERTA 3.1; the quality of the mesh was maintained by moving the nodes with a velocity $\bm{u}$, using the harmonic extension method. We compute $\bm{u}$ by solving a standard finite element approximation to
\begin{eqnarray*}
		\Delta \bm{u}= \bm{0} & \mbox{ in }  \Omega,\\ 
		\bm{u} = \bm{v} & \mbox{ on } \Gamma,
\end{eqnarray*}
where $\bm{v}$ is the velocity of $\Gamma$. This extends the velocity on $\Gamma$ to $\Omega$ in a `smooth' manner.

\subsubsection{Implementation of the diffuse--interface scheme}
\begin{figure}[htbp]
\begin{center}
\includegraphics[width=.24\textwidth,angle=0]{{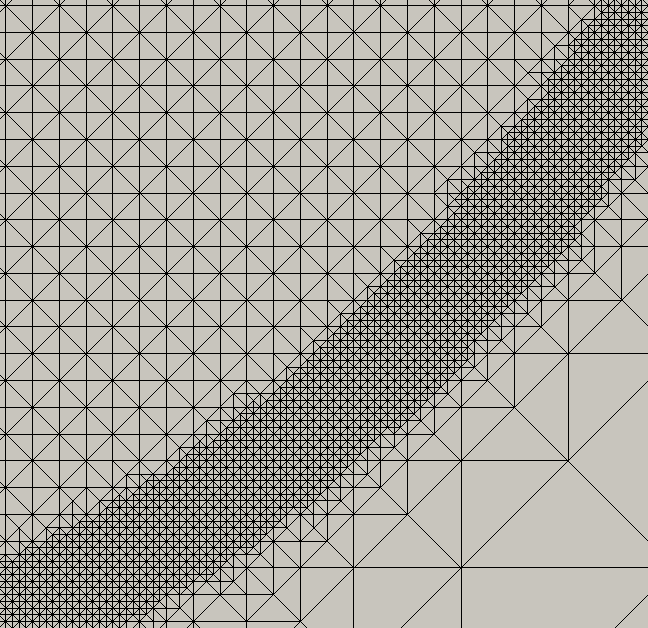}}\hspace{20mm}
\includegraphics[width=.24\textwidth,angle=0]{{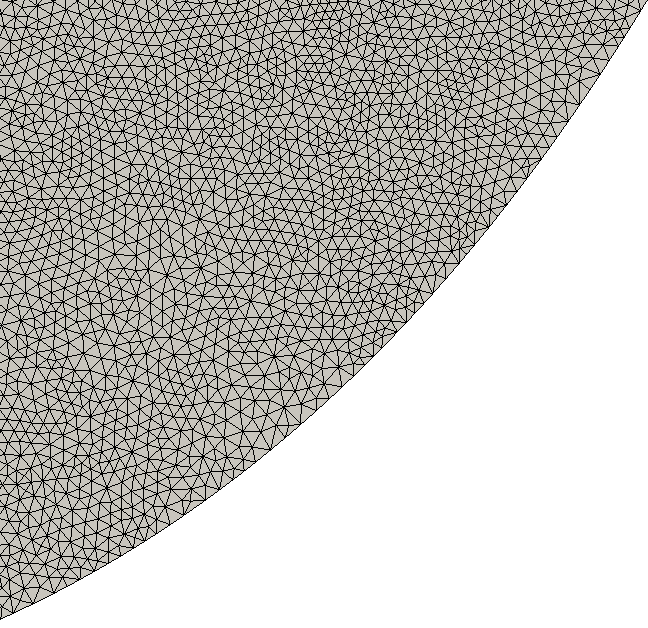}}\hspace{0mm}
\end{center}
\caption{
An enlarged section of the diffuse--interface mesh (left plot) and parametric mesh (right plot) associated with the results in Figure \ref{image_a10b01_old} at $t=45$.}
\label{f:meshes}
\end{figure}

The diffuse--interface finite--element scheme (\ref{equation_phase_FEM_u}), (\ref{equation_phase_FEM_v}) was implemented with the finite--element toolbox ALBERTA 2.0.
Since the interfacial thickness is proportional to $\varepsilon$, in order to resolve the interfacial layer, we need to choose the mesh size $h$ such that $h \ll \varepsilon$,  see \cite{Deckelnick2005} for details. Away from the interface, $h$ can be chosen larger and hence adaptivity in space can be used to speed up the computations. We use the built in algorithms from ALBERTA 2.0 for adaptivity to implement a mesh refinement strategy in which the following meshes are constructed: a fine mesh in the interfacial region where $|\varphi_h^n|<1$, a coarse mesh exterior to the tumour where $\varphi_{h}^{n} = -1$ and a standard sized mesh in the interior of the tumour where $\varphi_{h}^{n} = 1$, see the left hand plot in Figure \ref{f:meshes} in which an enlarged section of the diffuse--interface mesh associated with the results in Figure \ref{image_a10b01_old} at $t=45$ is displayed, in addition an enlarged section of the corresponding parametric mesh is displayed in the right hand plot.  
We denote the maximum diameter of the triangles in the three meshes by $h_{max,f}=\max_{\sigma\in\mathcal T_f^n}h_\sigma, h_{max,m}=\max_{\sigma\in\mathcal T_m^n}h_\sigma$ and $h_{max,c}=\max_{\sigma\in\mathcal T_c^n}h_\sigma$, where
$$
\mathcal T^n_f:= \lbrace \sigma \in \mathcal T_h \, | \, |\varphi_h^n(x)|<1 \mbox{ for some } x\in \sigma \rbrace, \quad\mathcal T^n_m:= \lbrace \sigma \in \mathcal T_h \, | \, \varphi_h^n(x)=1 \mbox{ for some } x \in \sigma\rbrace, 
$$
and
$$
\mathcal T^n_c:= \lbrace \sigma \in \mathcal T_h \, | \, \varphi_h^n(x)=-1 \mbox{ for some } x\in \sigma \rbrace.
$$
For more details on a similar mesh refinement strategy see \cite{Barrett2005}.

\subsection{Generalisation of the model}

\textcolor{black}{In the numerical results presented below we consider the following generalisation of (\ref{eq_f2_uOnOmega_intro}) -- (\ref{eq_f2_vOnGamma_intro})
\begin{eqnarray}
	\Delta \up = 1 &\mbox{ in }& \Omega(t), \label{eq_f2_uOnOmega_alt} \\
	\nabla \up \cdot \bmnu + \frac{\up}{\alpha} + \beta \kappa= Q &\mbox{ on }& \Gamma(t) , \label{eq_f2_uOnGamma_alt} \\
	\nv = \frac{\up}{\alpha} + (\beta+\gamm) \kappa &\mbox{ on }& \Gamma(t). \label{eq_f2_vOnGamma_alt}
\end{eqnarray}
Here $\gamm\in\mathbb{R}$ is a nonnegative regularisation parameter, such that setting $\gamm=0$ recovers the original model (\ref{eq_f2_uOnOmega_intro}) -- (\ref{eq_f2_vOnGamma_intro}). 
We introduce this generalised model for numerical purposes as it enables us to run computations with $\beta=0$, thus simplifying the numerical schemes by removing the curvature term from the Robin boundary conditions for $u$, whilst maintaing a velocity law of forced mean curvature flow.}

\subsection{Parameter values}
Unless otherwise specified, in all simulations in $\mathbb{R}^2$ the following hold:
we set $Q = 1.0$ and the initial geometry $\Gamma(0)$ was set to be an ellipse with length $0.5$ and height $1.0$, 
in the parametric examples we set $\Delta t= 10^{-3}$ and the mesh size at $t=0$ was taken to be $h \approx 0.022$. 
In the diffuse--interface simulations we set $h_{max,c}\approx 1.77$, $h_{max,m}\approx0.02$, and used the values of $\epsilon,~\Delta t$ and $h_{max,f}$ given in Tables \ref{t:param} and \ref{t:param2}. 
\begin{table}[!h]\begin{center}
 \begin{tabular}{ |c||c|c|c| }
 \hline
~ & $\epsilon$ & $\Delta t$ & $h_{max,f}$ \\
 \hline
 \hline
$\gamm=0$, $\alpha=1$, $\beta=0.1$ & $0.075$ & $1.0\times10^{-4}$ &0.0048\\
$\gamm=0$, $\alpha=0.1$, $\beta=0.1$ & $0.01$ & $1.0\times10^{-5}$ & 0.0028\\
$\gamm=0$, $\alpha=1$, $\beta=1$ & $0.1$ & $3.0\times10^{-6}$ & 0.0099\\
$\gamm=0$, $\alpha=0.1$, $\beta=1$ & $0.1$ & $1.0\times10^{-5}$ & 0.0042\\
\hline
\end{tabular}
\end{center}
\caption{Parameters for the diffuse--interface simulations for $\gamm=0$.}
\label{t:param}
\end{table}

\begin{table}[!h]\begin{center}
 \begin{tabular}{ |c||c|c|c| }
 \hline
~ & $\epsilon$ & $\Delta t$ & $h_{max,f}$ \\
 \hline
 \hline
$\beta=0$, $\alpha=1$, $\gamm=0.1$ & $0.04$ & $1.0\times10^{-4}$ &0.0048\\
$\beta=0$, $\alpha=0.1$, $\gamm=0.1$ & $0.01$ & $1.0\times10^{-5}$ & 0.0028\\
$\beta=0$, $\alpha=1$, $\gamm=1$ & $0.1$ & $3.0\times10^{-6}$ & 0.0099\\
$\beta=0$, $\alpha=0.1$, $\gamm=1$ & $0.1$ & $1.0\times10^{-5}$ & 0.0042\\
\hline
\end{tabular}
\end{center}
\caption{Parameters for the diffuse--interface simulations with $\beta=0$.}
\label{t:param2}
\end{table}

\subsection{Simulations for $\gamm=0$}
\begin{figure}[htbp]
\begin{center}
\includegraphics[width=.32\textwidth,angle=0]{{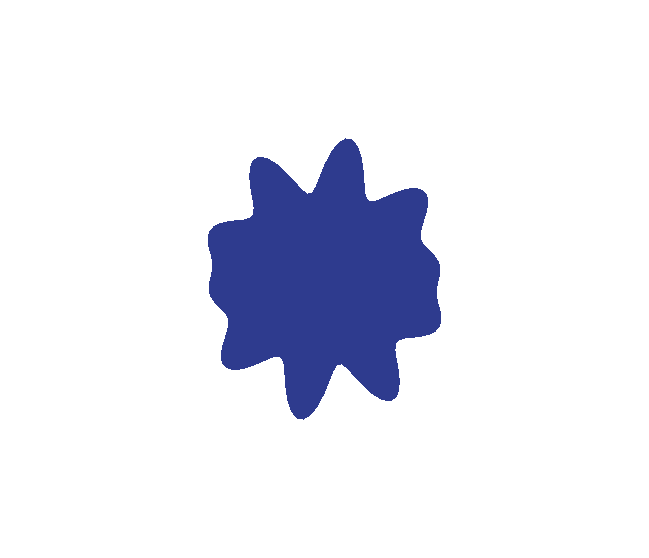}}\hspace{0mm}
\includegraphics[width=.32\textwidth,angle=0]{{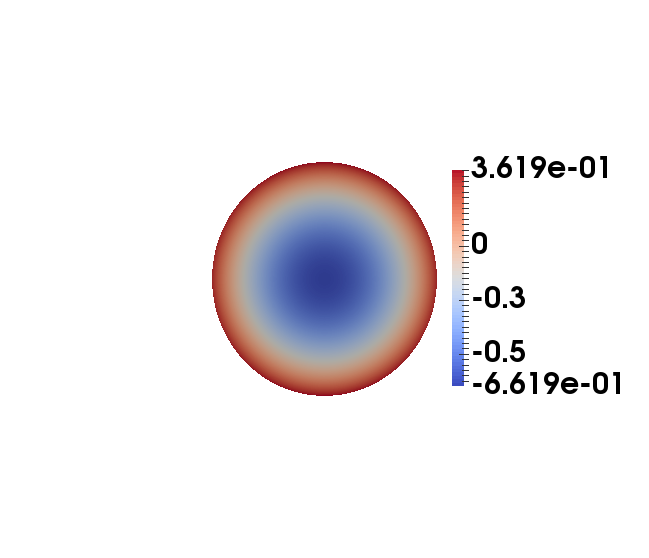}}\hspace{0mm}
\includegraphics[width=.32\textwidth,angle=0]{{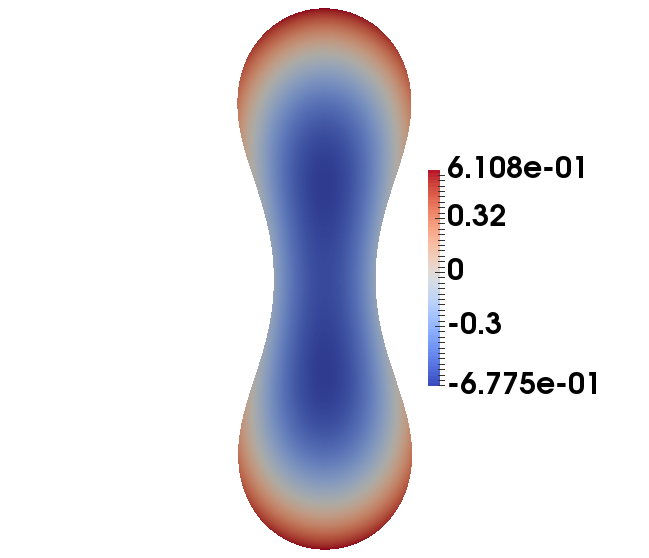}}\hspace{0mm}
\end{center}
\caption{
Linear stability, (\ref{stab:N4}), with $Q=1$, $\gamm=0$ and $\beta=0.1$. The left plot displays the initial geometry $R(\theta,0)=2Q+0.25\sin(\theta)\cos(9\theta)$, the centre plot displays $u_h$ at $t = 700$ with $\alpha = 0.68$ such that $3\alpha \beta>2$ (linearly stable) and the right plot displays $u_h$ at $t = 700$ with $\alpha = 0.65$ such that $3\alpha \beta<2$ (linearly unstable). The instability manifest in the right plot is associated with the higher perimeter--length to area ratio at the tips than in between, with higher pressures at the tips.
}
\label{f:stab}
\end{figure}

\begin{figure}[htbp]
\begin{center}
\includegraphics[width=.24\textwidth,angle=0]{{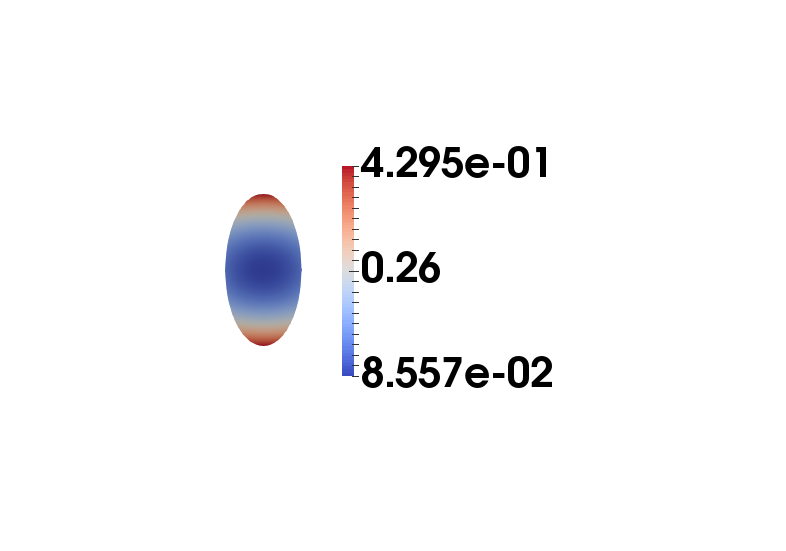}}\hspace{0mm}
\includegraphics[width=.24\textwidth,angle=0]{{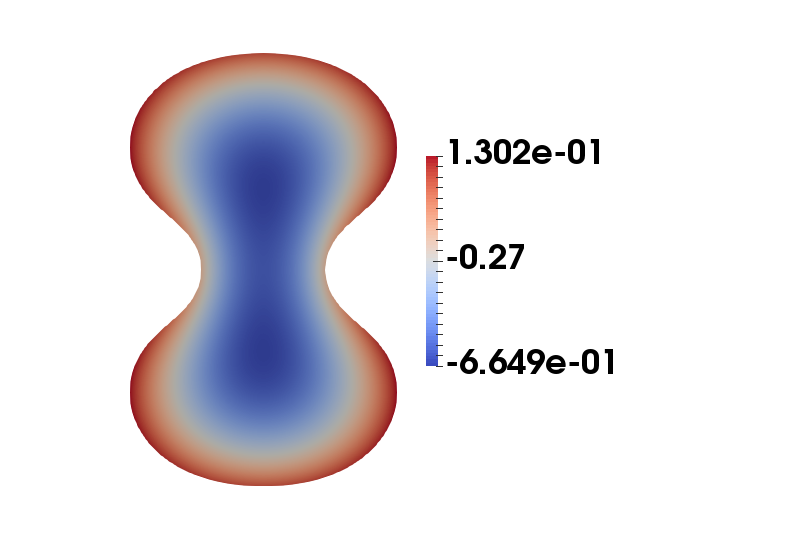}}\hspace{0mm}
\includegraphics[width=.24\textwidth,angle=0]{{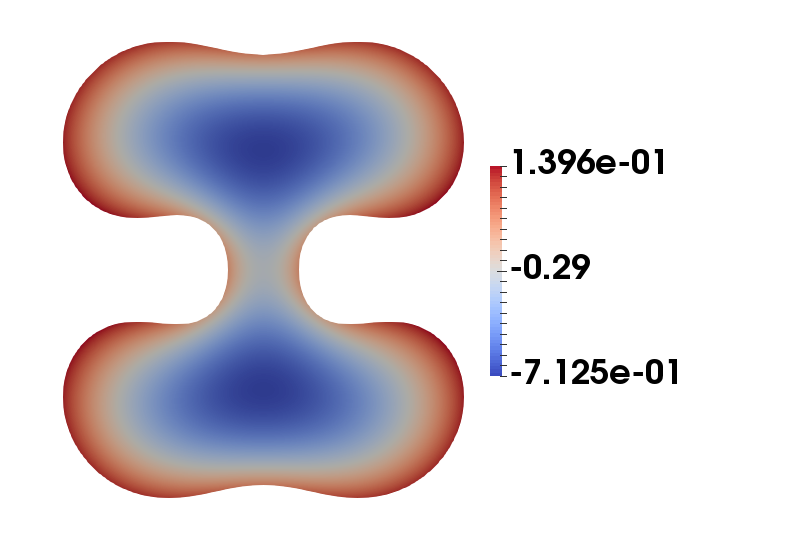}}\hspace{0mm}
\includegraphics[width=.24\textwidth,angle=0]{{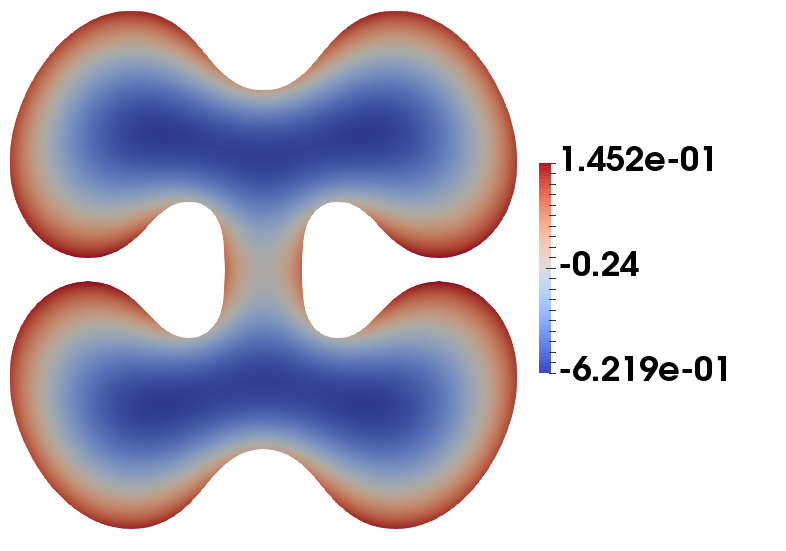}}\hspace{0mm}
\end{center}
\caption{Results with  $\gamm=0$, $\alpha = 0.1$ and $\beta = 1.0$: $u_h$ given by the parametric scheme at $t = 0,7,10,12.6$. This simulation exhibits repeated tip splitting akin to that arising in a variety of moving boundary problems.}
\label{image_a01b10_new}
\end{figure}

In the first set of simulations, Figures \ref{f:radial} -- \ref{image_a01b01_thin_film_new},  we take $\gamm=0$ in the generalised model to obtain results for the original model (\ref{eq_f2_uOnOmega_intro}) -- (\ref{eq_f2_vOnGamma_intro}).

\subsubsection{Radially symmetric solutions}

To test the accuracy of the numerical schemes, we set $\Gamma(t) \subset \mathbb{R}^2$ to be a circle with radius $R(t)$ and we express $u$ in polar coordinates such that $u(r, \theta)=u(r)$. In this setting (\ref{eq_f2_uOnOmega_intro}) -- (\ref{eq_f2_vOnGamma_intro}) reduces to 
\begin{equation}
	u(r) = \frac{1}{4} r^2 + \alpha (Q +\frac{\beta}{R}) - \frac{\alpha}{2} R - \frac{1}{4} R^2,
\label{ueqn}
\end{equation}
\begin{equation}
	R'(t) = - \frac{\beta}{R} + \frac{1}{\alpha} u(r) = Q - \frac{R}{2}, ~~~R(0) := R_0,
\end{equation}
so that
\begin{equation}
	R(t) = 2Q + ( R_0 - 2Q )e^{-t / 2} .
\label{reqn}
\end{equation}

For some choices of parameters, 
typically $\alpha,\, \beta \ll1$ and $Q=O(1)$, instabilities occurred in the parametric scheme that gave rise to non-radially symmetric solutions, these instabilities seem likely to have stemmed from the initial triangulation of $\Omega(0)$. 
In the following results we avoided such parameter combinations.   

\begin{figure}[htbp]
\begin{center}
\includegraphics[width=.24\textwidth,angle=0]{{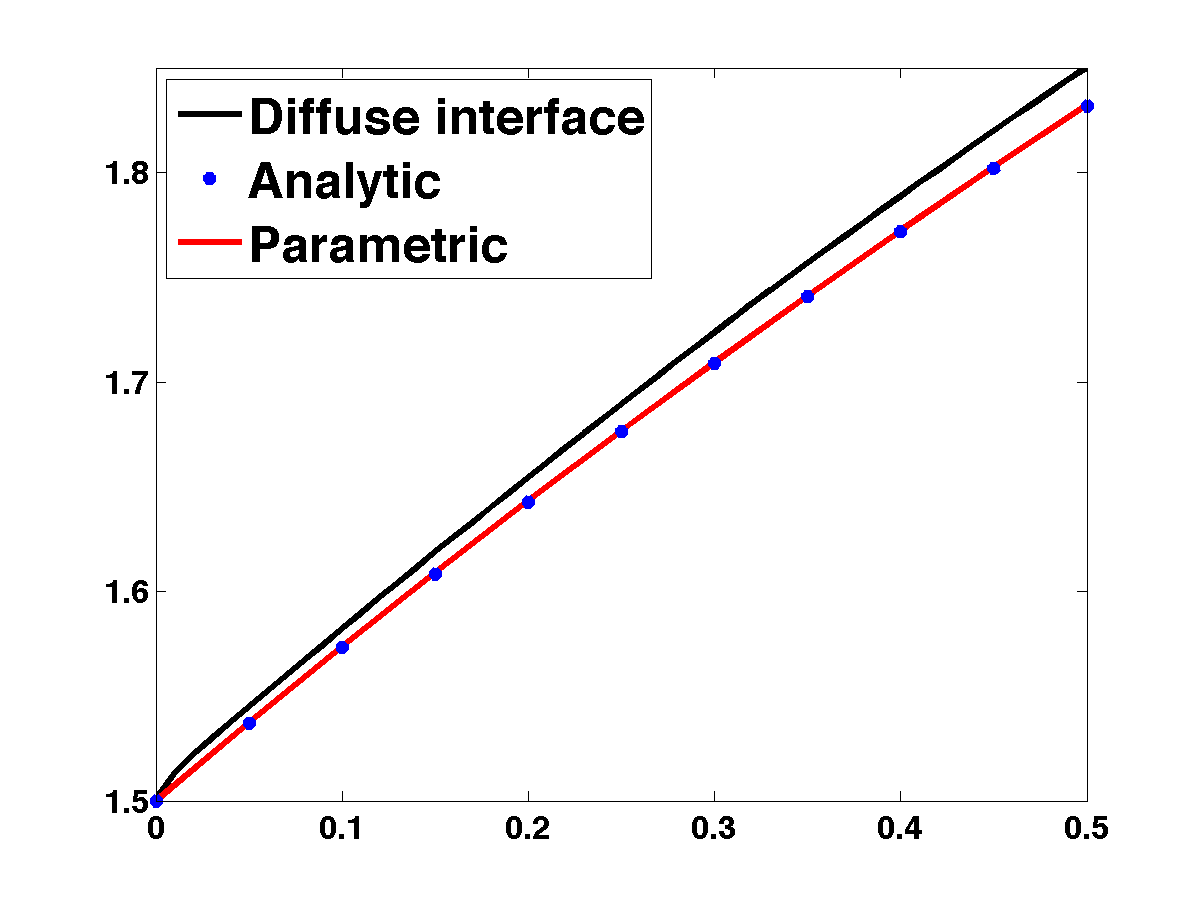}}\hspace{0mm}
\includegraphics[width=.24\textwidth,angle=0]{{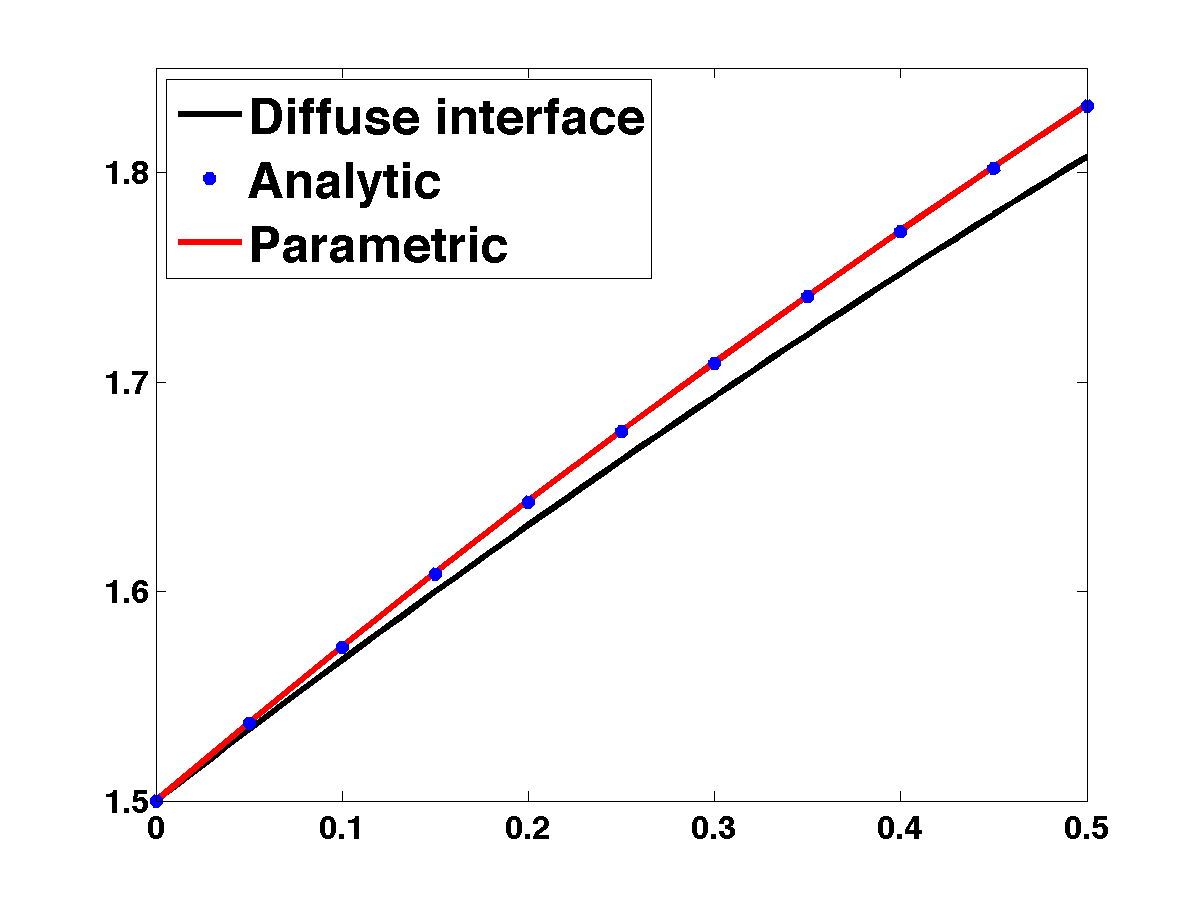}}\hspace{0mm}
\includegraphics[width=.24\textwidth,angle=0]{{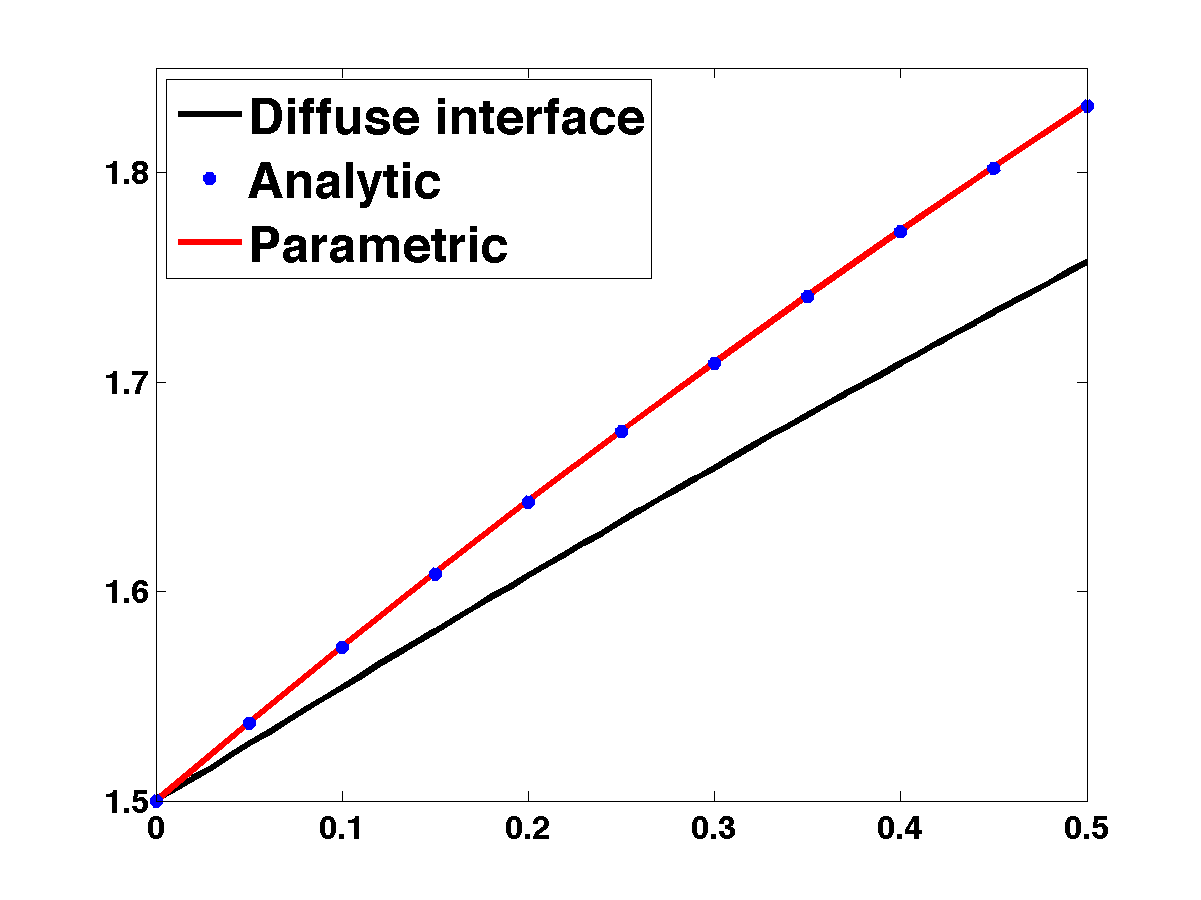}}\hspace{0mm}
\includegraphics[width=.24\textwidth,angle=0]{{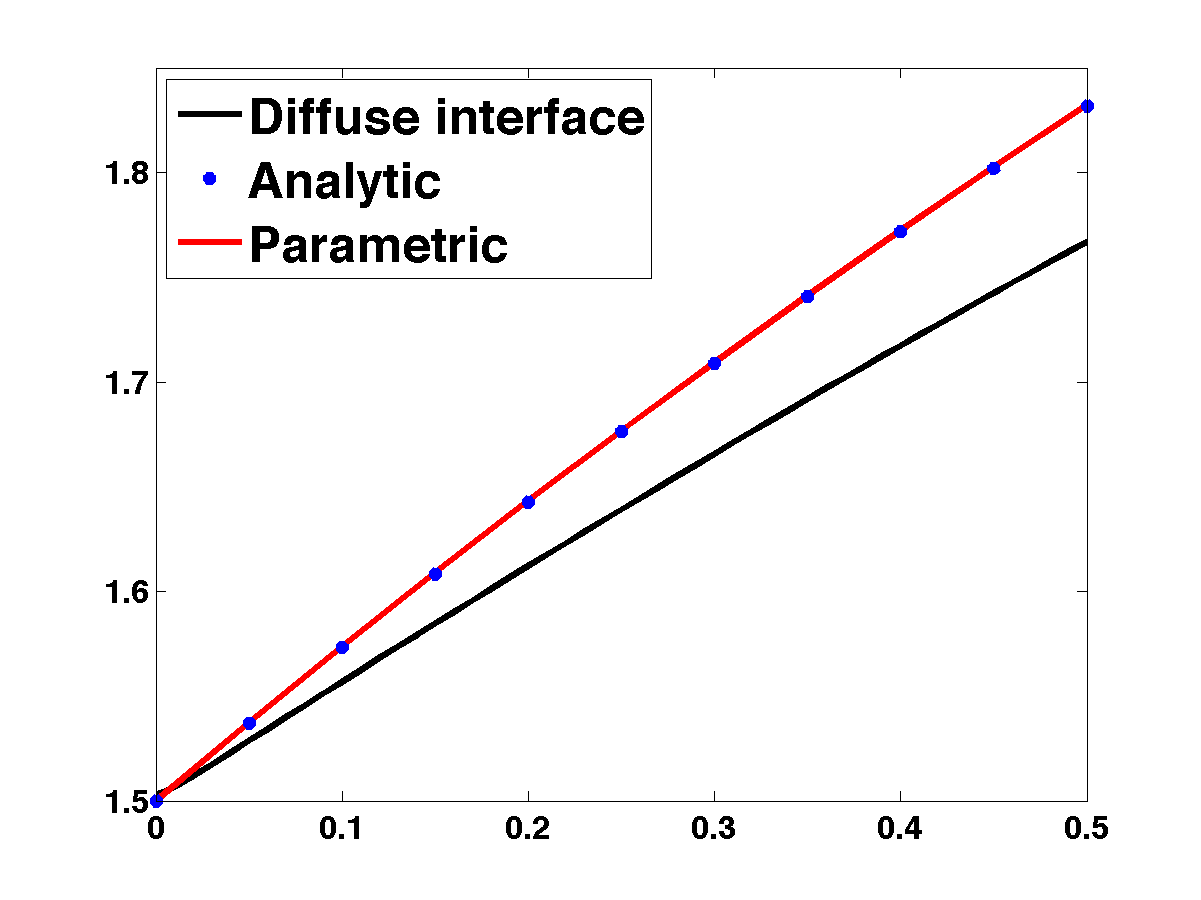}}\hspace{0mm}\\
\includegraphics[width=.24\textwidth,angle=0]{{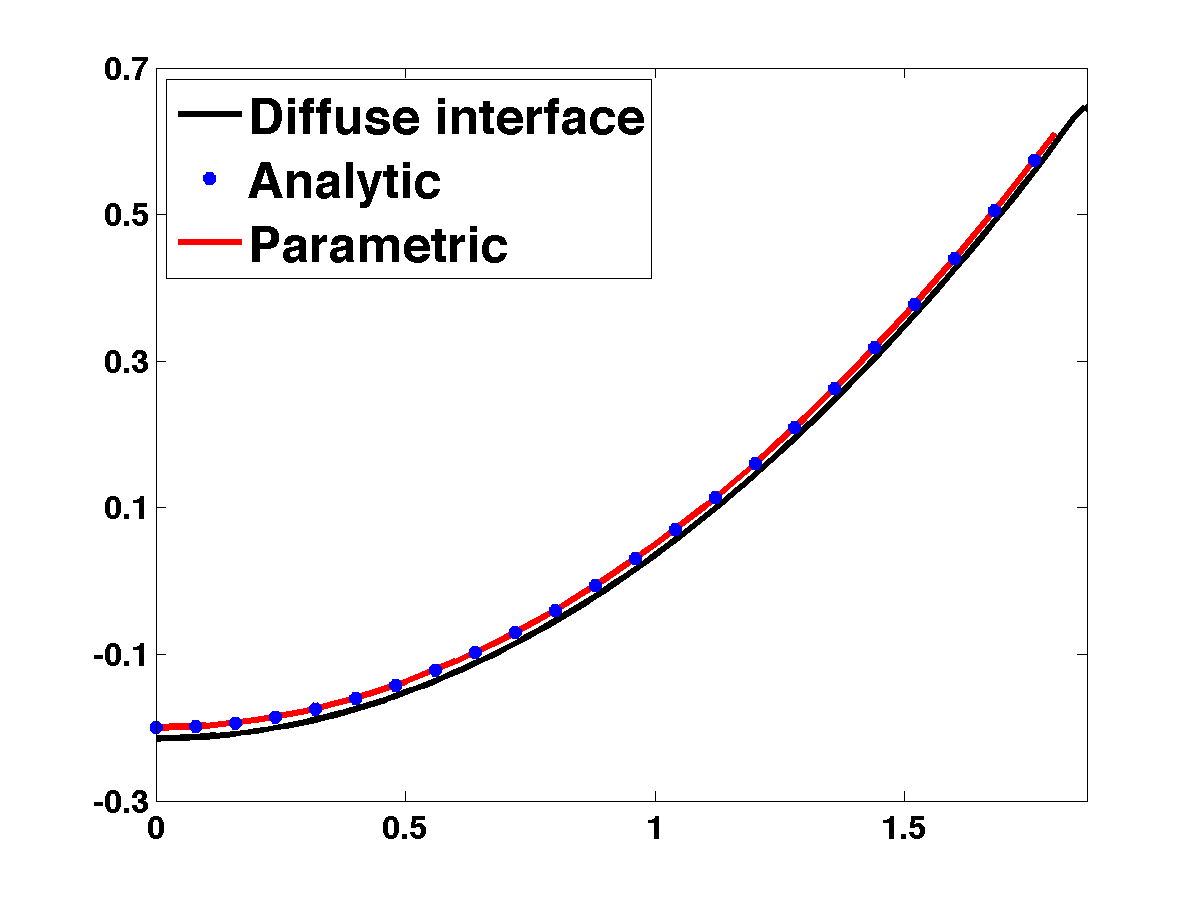}}\hspace{0mm}
\includegraphics[width=.24\textwidth,angle=0]{{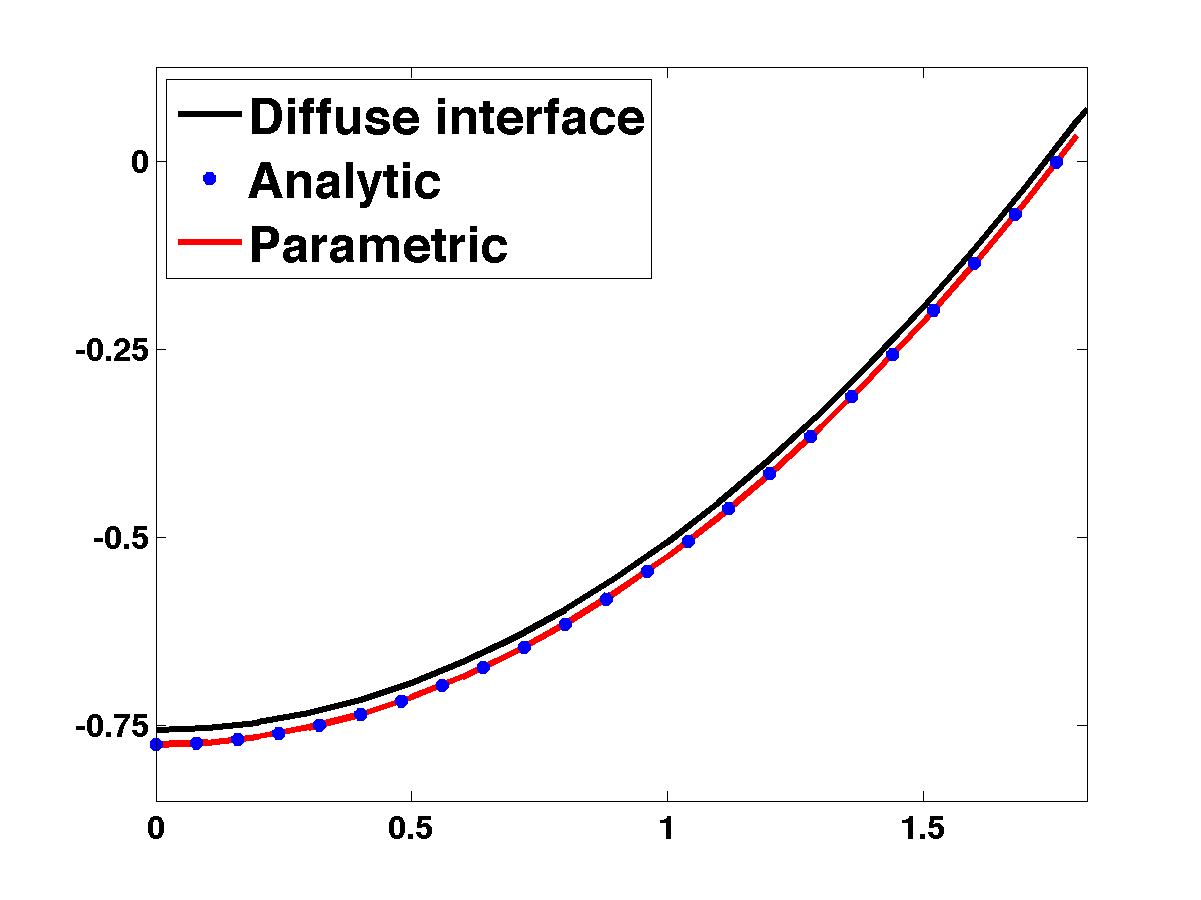}}\hspace{0mm}
\includegraphics[width=.24\textwidth,angle=0]{{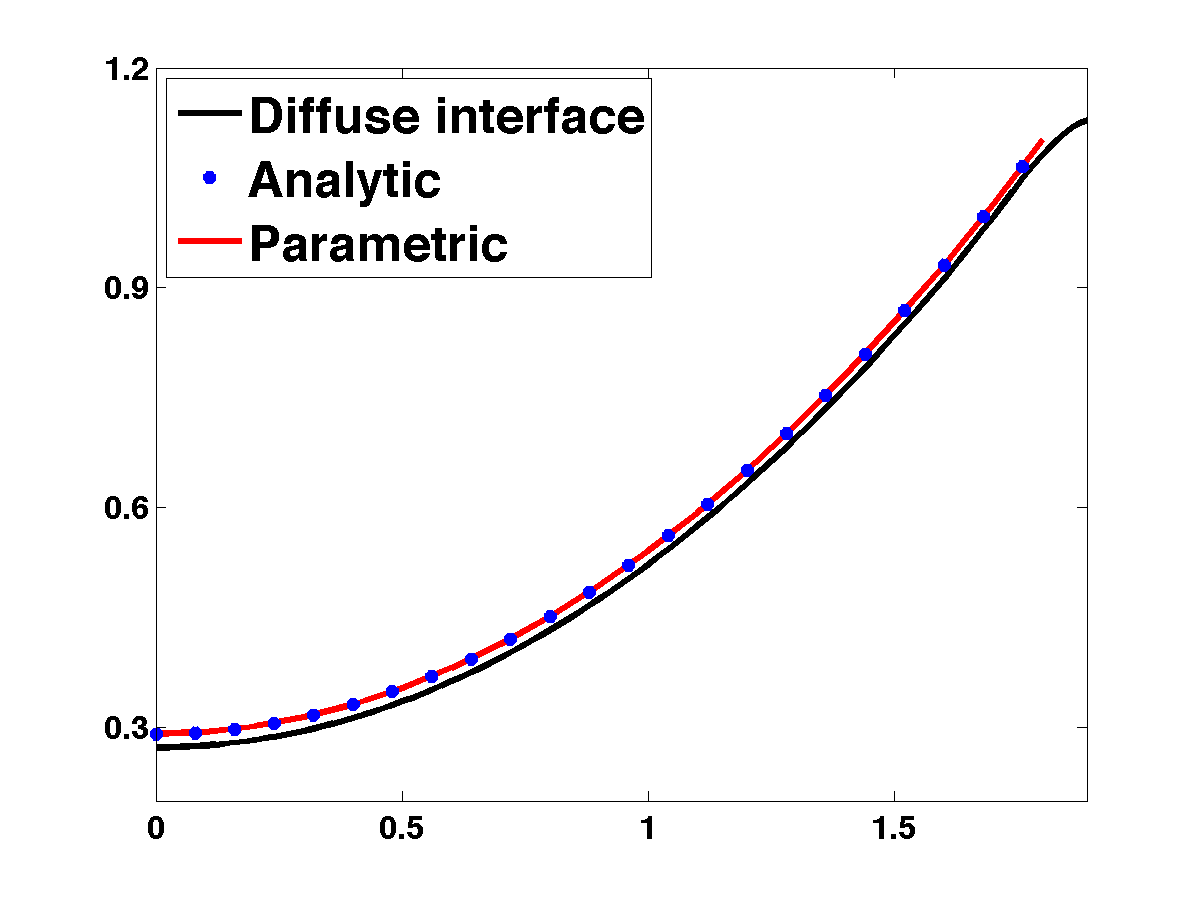}}\hspace{0mm}
\includegraphics[width=.24\textwidth,angle=0]{{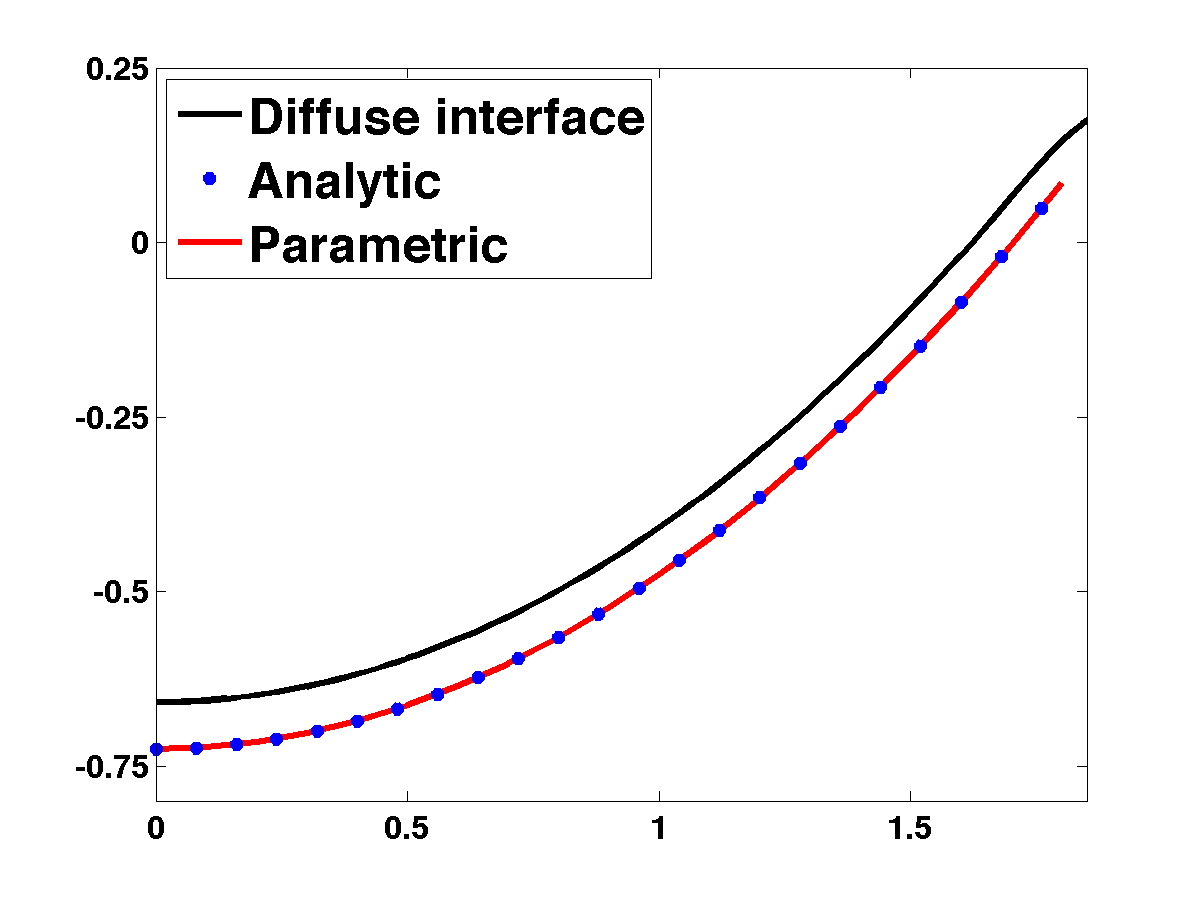}}\hspace{0mm}
\end{center}
\caption{
Comparison of the radius (upper plots) and the pressure $u$ at $t=0.5$ (lower plots) with $\gamma=0$, $\alpha=1$, $\beta=0.1$ (first column), $\gamma=0$, $\alpha=\beta=0.1$, (second column), $\gamma=0$, $\alpha=\beta=1$ (third column) and $\gamma=0$, $\alpha=0.1$, $\beta=1$ (fourth column). The upper plots display $t$ against $R$, on the $x$-- and $y$--axis, respectively. The lower plots display $r$ against $u$, again on the $x$-- and $y$--axis, respectively.
}
\label{f:radial}
\end{figure}

In Figure \ref{f:radial} we compare the radii obtained from the finite element approximations of the parametric and the diffuse--interface formulations for $\gamm=0$ 
to the corresponding  analytical solutions. 
Similarly we compare the pressure solutions obtained from the parametric scheme and the diffuse--interface scheme to the analytical solution. 

We consider four combinations of $\alpha$ and $\beta$ and set $Q = R_0=1.5$. In 
the upper plots we display the radius of the circle for the parametric model (red line), the radius of the circle for the diffuse--interface model, obtained from the zero--level line of $\varphi$, (black line), and the radius of the analytical solution of (\ref{reqn}) (blue dots). 
In the lower plots we display $u_h$, $\tilde{u}_h$, and $u$ at $t=0.5$, where $u$ is the solution of \eqref{ueqn}, 
again, we use a red line for the parametric scheme, a black line for the diffuse--interface scheme, and blue dots for the analytical solution.
We display the results for $\gamm=0$, $\alpha=1$, $\beta=0.1$ in the first column, $\gamm=0$, $\alpha=\beta=0.1$ in the second column, $\gamm=0$, $\alpha=\beta=1$ in the third column and $\gamm=0$, $\alpha=0.1$, $\beta=1$ in the fourth column. 
In the parametric simulations the mesh size at $t=0$ was taken to be  $h\approx 0.1$.

From Figure \ref{f:radial} we see that for all four combinations of $\alpha$ and $\beta$ there is very good agreement between the solutions of the parametric scheme and the analytical solution, while the solutions of the diffuse--interface scheme compare reasonably well to the analytical solution for $\gamm=0$, $\alpha=1$, $\beta=0.1$ and $\gamm=0$, $\alpha=\beta=0.1$ but not well for $\gamm=0$, $\alpha=\beta=1$  and $\gamm=0$, $\alpha=0.1$, $\beta=1$. 
We believe the poor approximation of the diffuse--interface scheme for $\beta=O(1)$ is due to the fact that the scheme is only an approximation to forced mean curvature flow and hence larger errors occur if the curvature term in the evolution becomes more dominant. Based on these results in the subsequent simulations we solved the diffuse--interface model only with $\beta=0.1$.

\subsubsection{Linear stability}\label{sf:ls}
In Figure \ref{f:stab} we display results from the parametric scheme (\ref{equation_ParametricFEM_u_f2}), (\ref{equation_ParametricFEM_v_f2}) relating to the linear stability result, (\ref{stab:N4}). 
We set $Q=1$ and $\beta=0.1$ and set the initial geometry to be a perturbation of a circle with radius $R(0)=2Q=2$ such that $R(\theta,0)=2Q+0.25\sin(\theta)\cos(9\theta)$, (left plot) and 
we set $h \approx 0.1$. 
We consider two values of $\alpha$; $\alpha = 0.68$ (centre plot) such that $3\alpha \beta>2$ and $\alpha = 0.65$ (right plot) such that $3\alpha \beta<2$, in each case we display the solution $u_h$ at $t=700$. 
From the figure we see that when $3\alpha \beta>2$ the stable state in which $\Gamma(t)$ is a circle of radius $R=2Q=2$ is obtained, whereas for $3\alpha \beta<2$ the radial symmetry of the problem is lost. We note that for this parameter set with $\alpha = 0.65$, $n=2$, recall (\ref{stab:N2}), is the only unstable mode, the simulations leading to the same (reflectional) symmetries as those of an ellipse.

\begin{figure}[htbp]
\begin{center}
\hspace{12mm}
\includegraphics[width=.24\textwidth,angle=0]{{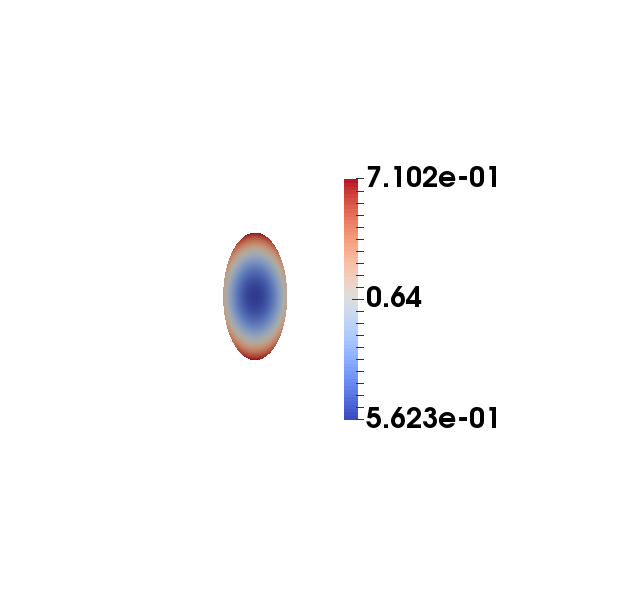}}\hspace{0mm}
\includegraphics[width=.24\textwidth,angle=0]{{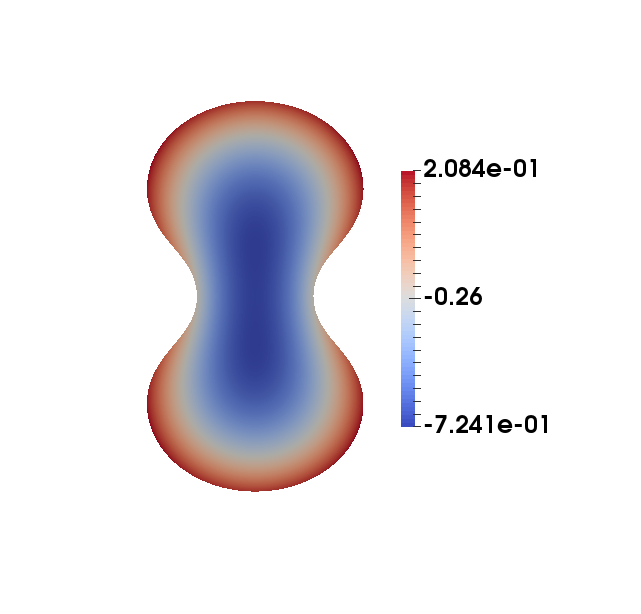}}\hspace{4mm}
\includegraphics[width=.24\textwidth,angle=0]{{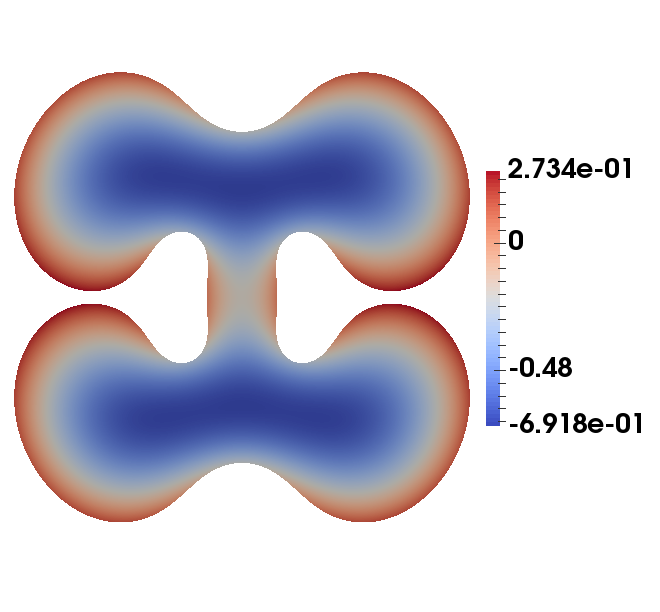}}\hspace{0mm}
\end{center}
\begin{center}
\hspace{12mm}
\includegraphics[width=.24\textwidth,angle=0]{{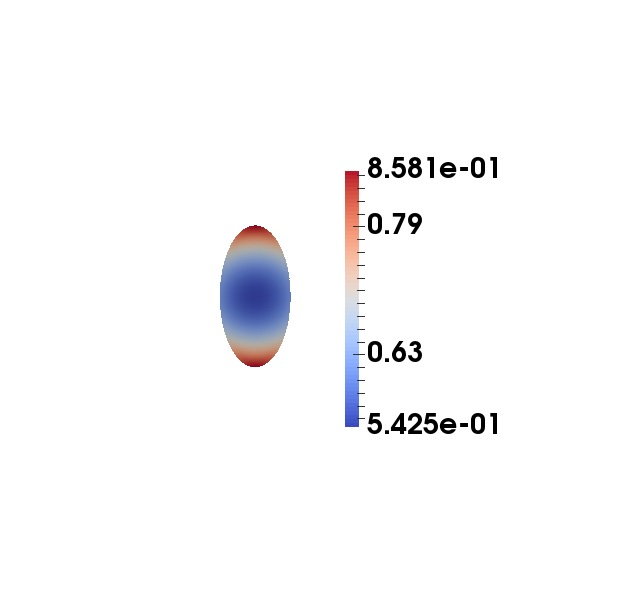}}\hspace{0mm}
\includegraphics[width=.24\textwidth,angle=0]{{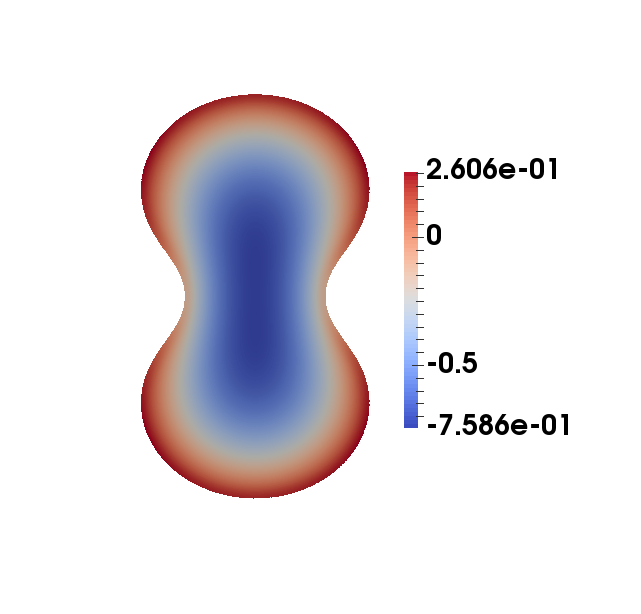}}\hspace{4mm}
\includegraphics[width=.24\textwidth,angle=0]{{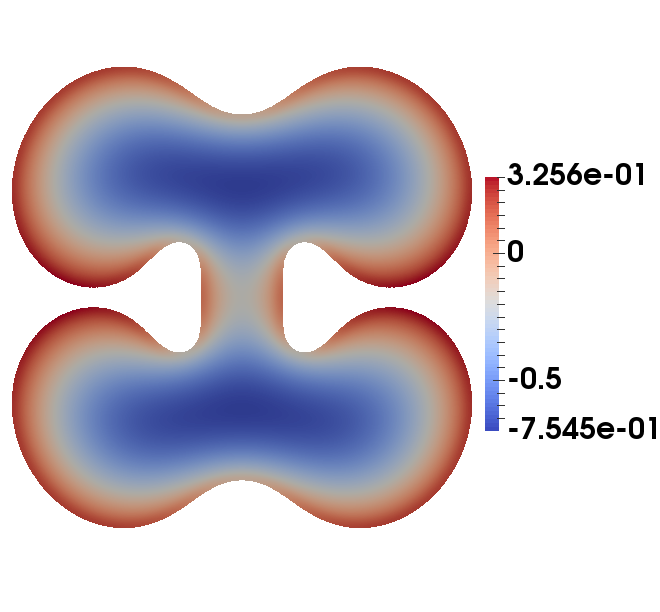}}\hspace{0mm}
\end{center}
\begin{center}
\includegraphics[width=.24\textwidth,angle=0]{{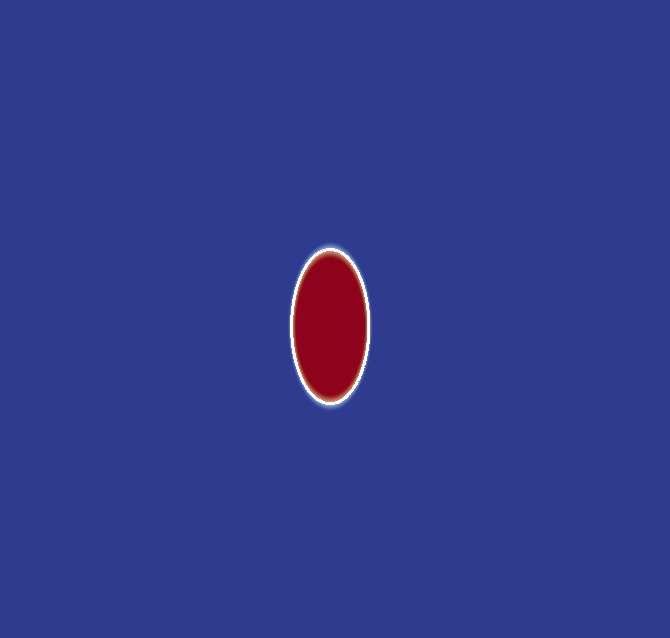}}\hspace{0mm}
\includegraphics[width=.24\textwidth,angle=0]{{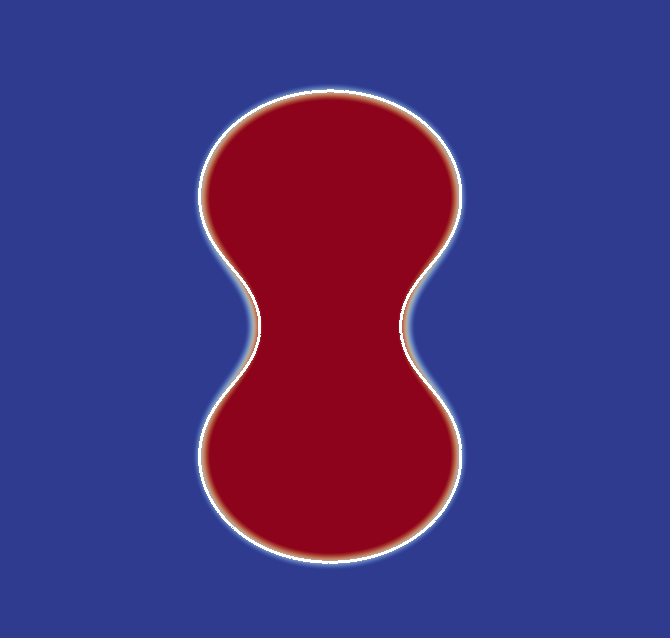}}\hspace{0mm}
\includegraphics[width=.24\textwidth,angle=0]{{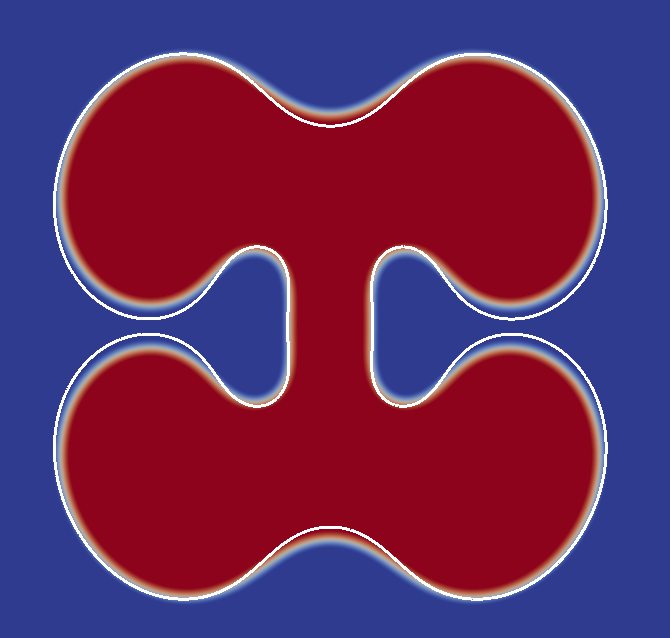}}\hspace{0mm}
\end{center}
\caption{Results with $\gamm=0$, $\alpha = 1.0$ and $\beta = 0.1$ at $t = 0,11,23$ (columns $1$ to $3$): $u_h$ given by the parametric scheme (top row), $\tilde{u}_h$ given by the diffuse--interface scheme (middle row), $\varphi_h$ (in red and blue) from the diffuse--interface scheme and $\bmcx_h$ (in white) from the parametric scheme (bottom row). }
\label{image_a10b01_new}
\end{figure}

\subsubsection{Dependence on $\alpha$ and $\beta$}

We now investigate the morphologies that arise using differing combinations of $\alpha$ and $\beta$. 

We first note that, for the initial data chosen, an ellipse with length $0.5$ and height $1.0$, our simulations show that for $\gamm = 0$, $\alpha = \beta = 1.0$ the geometry tends to a radially symmetric steady state, with the radius of the resulting circle decreasing as $\beta$ increases; this radius can be calculated by setting $R'(t) = 0$ in (\ref{reqn}). 

In Figure \ref{image_a01b10_new} we display the solution $u_h$ at $t = 0, 7, 10, 12.6$, computed using the parametric scheme (\ref{equation_ParametricFEM_u_f2}), (\ref{equation_ParametricFEM_v_f2}) with $\gamm = 0$, $\alpha=0.1$ and $\beta=1$ 

\begin{figure}[htbp]
\begin{center}
\includegraphics[width=.24\textwidth,angle=0]{{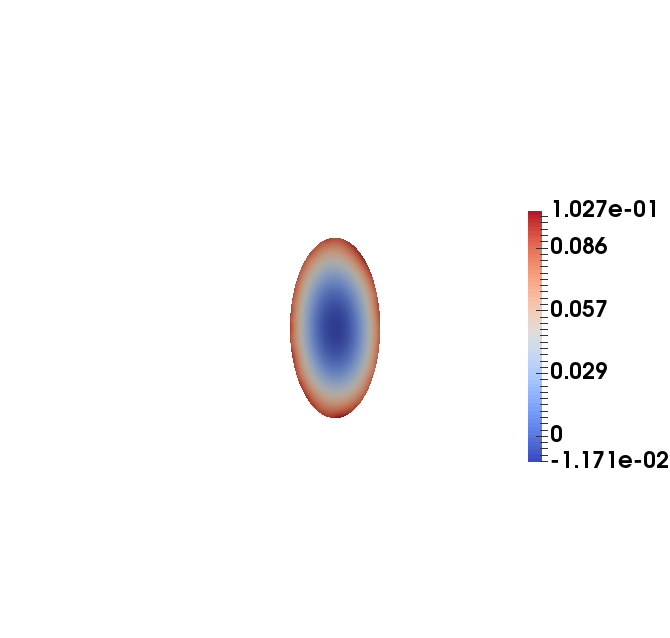}}\hspace{0mm}
\includegraphics[width=.24\textwidth,angle=0]{{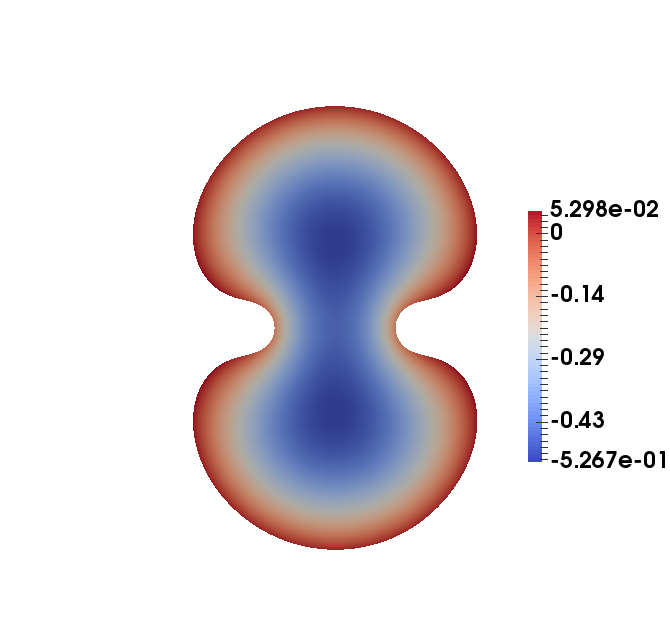}}\hspace{0mm}
\includegraphics[width=.24\textwidth,angle=0]{{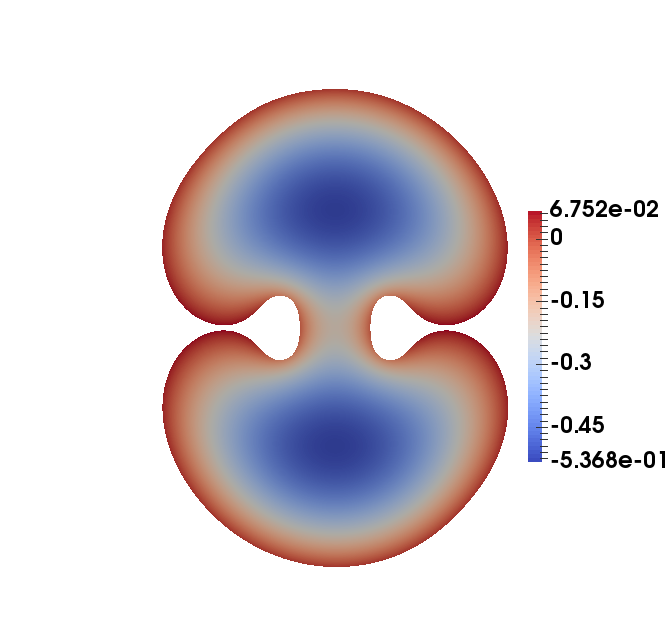}}\hspace{0mm}
\end{center}
\begin{center}
\includegraphics[width=.24\textwidth,angle=0]{{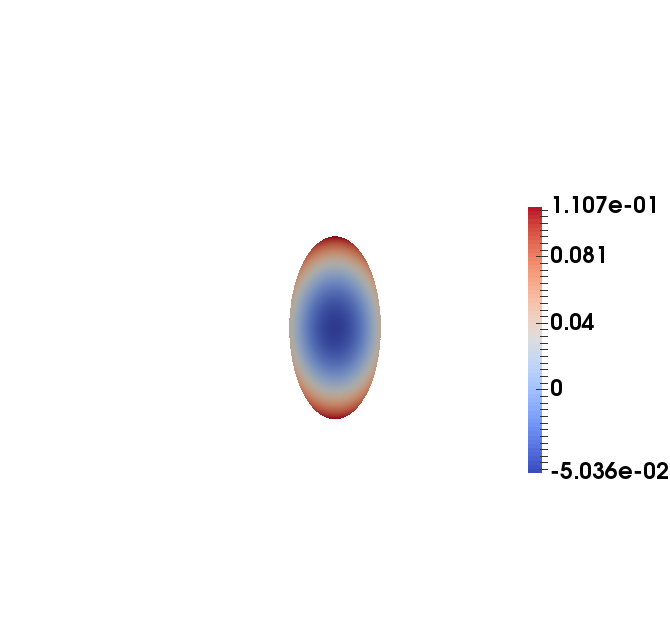}}\hspace{0mm}
\includegraphics[width=.24\textwidth,angle=0]{{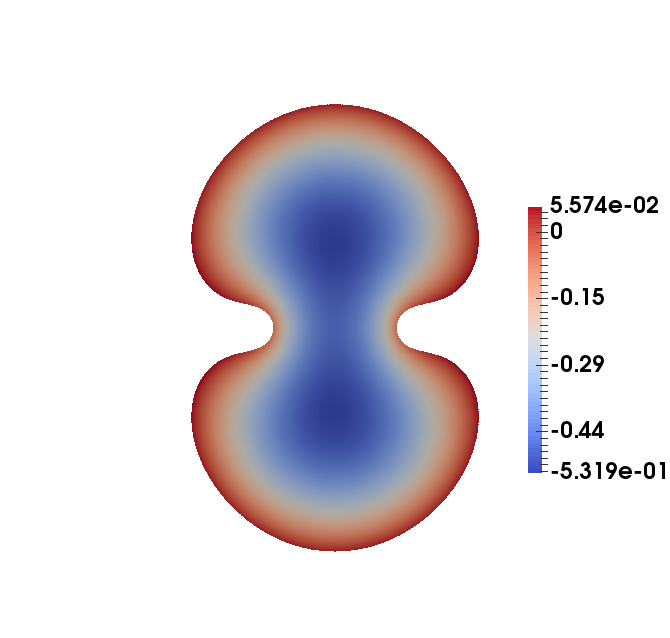}}\hspace{0mm}
\includegraphics[width=.24\textwidth,angle=0]{{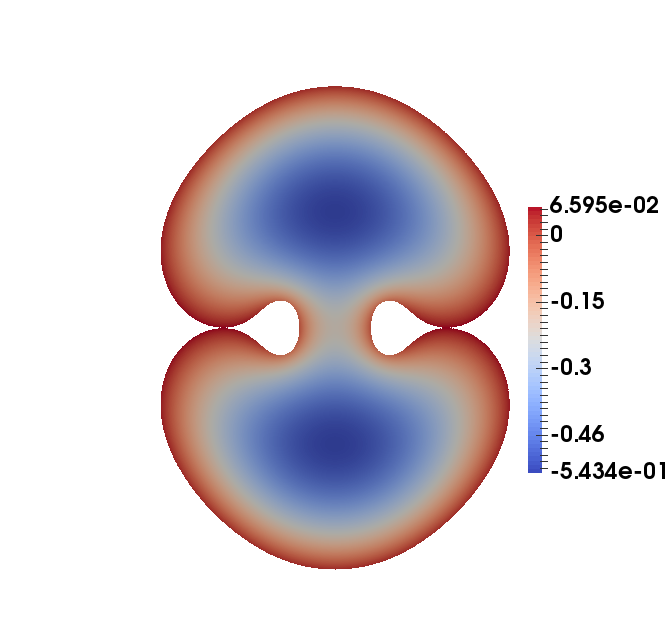}}\hspace{0mm}
\end{center}
\begin{center}
\includegraphics[width=.24\textwidth,angle=0]{{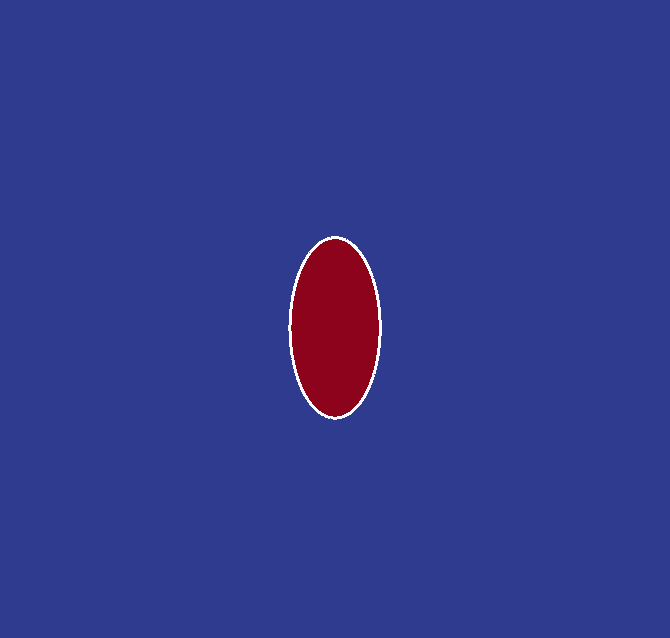}}\hspace{0mm}
\includegraphics[width=.24\textwidth,angle=0]{{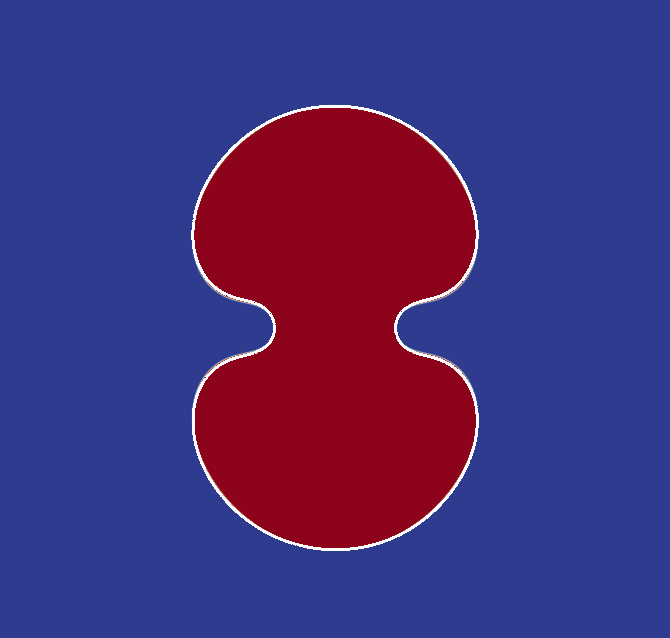}}\hspace{0mm}
\includegraphics[width=.24\textwidth,angle=0]{{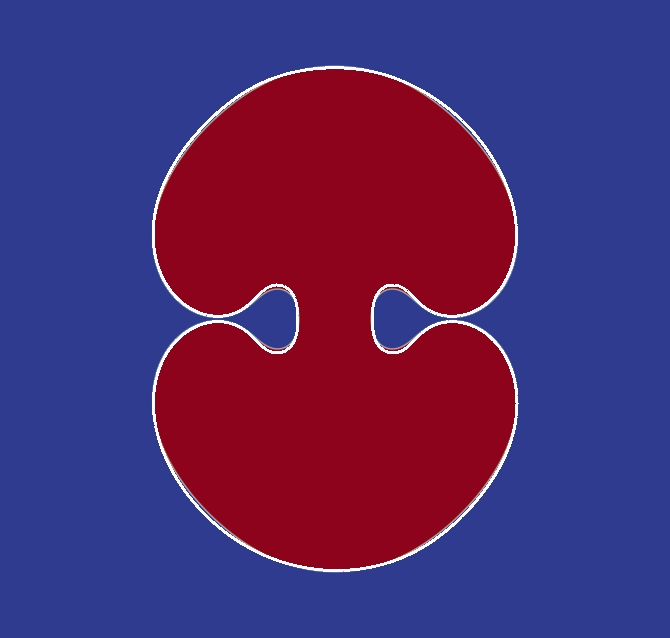}}\hspace{0mm}
\end{center}
\caption{Results with $\gamm=0$, $\alpha = 0.1$ and $\beta = 0.1$: $u_h$ given by the parametric scheme (top row), $\tilde{u}_h$ given by the diffuse--interface scheme (middle row), $\varphi_h$ (in red and blue) from the diffuse--interface scheme and $\bmcx_h$ (in white) from the parametric scheme (bottom row). The diffuse--interface solutions are at $t = 0,4,5$ and the parametric solutions are at $t=0,3.5,4.4$.}
\label{image_a01b01_new}
\end{figure}

In Figure \ref{image_a10b01_new} we compare the parametric and diffuse--interface numerical solutions obtained by setting $\gamm = 0$, $\alpha=1$ and $\beta=0.1$.  
The solution $u_h$ given by the parametric scheme (\ref{equation_ParametricFEM_u_f2}), (\ref{equation_ParametricFEM_v_f2}) is displayed in the top row, while the solution $\tilde{u}_h$ given by the diffuse--interface scheme (\ref{equation_phase_FEM_u}), (\ref{equation_phase_FEM_v}) is displayed in the middle row. 
The bottom row displays the order parameter $\varphi_h$ (in red and blue) given by the diffuse--interface scheme and $\bmcx_h$ (in white) given by the parametric scheme. 
The solutions are displayed at times $t = 0$ (left) $t=11$ (centre) and $t=23$ (right). At $t=11$ there is good agreement between the parametric and diffuse--interface schemes, though for $t=23$ the difference becomes more pronounced. \\
The results obtained by setting $\gamm=0$, $\alpha = 0.1$ and $\beta = 0.1$ are displayed in Figure \ref{image_a01b01_new}. For this choice of parameters the agreement between the parametric and diffuse--interface schemes is quite pronounced; in particular, the evolution of the diffuse--interface solution is slower than that of the parametric solution, however, the geometries of the two solutions compare well if the diffuse--interface solution is displayed at a different time to the parametric solution. To this end in Figure \ref{image_a01b01_new}, the diffuse--interface results are displayed at $t = 0,4,5$ while the parametric results are displayed at $t=0,3.5,4.4$. We believe the disparity between the two sets of solutions is predominantly caused by the diffuse--interface approximation of the pressure that is used in the Allen-Cahn approximation of the velocity law, since 
for $\alpha\ll 1$, a small error in the approximation of $u$ on $\Gamma(t)$ can result in a significant error in the velocity of $\Gamma(t)$. 
When we solved the model with $\beta=0$ with the same set of parameters we saw good agreement between the parametric and diffuse--interface schemes, see Figure \ref{image_a01b01_old} below, and thus we conclude that the diffuse interface approximation of the curvature term in the Robin boundary conditions is the main reason for the differences.

\subsubsection{Thin--film limit (see the appendix)}

In Figure \ref{image_a01b01_thin_film_new} we display results from the parametric scheme (\ref{equation_ParametricFEM_u_f2}), (\ref{equation_ParametricFEM_v_f2}), relating to the thin--film limit from the appendix. Here the initial geometry $\Gamma(0)$ is given by an ellipse of length $1.0$ and height $0.1$. We set   $Q=0.1$, $\gamm=0$ and $\alpha = \beta = 0.1$. We chose $\Delta t = 10^{-4}$ and $h \approx 0.005$ in order to maintain a stable evolution.
We present the solutions at $t = 0, 0.2,0.3$, demonstrating how the large--aspect--ratio of the domain is lost, consistent with the analysis of the appendix.

\begin{figure}[htbp]
\begin{center}
\includegraphics[width=.36\textwidth,angle=0]{{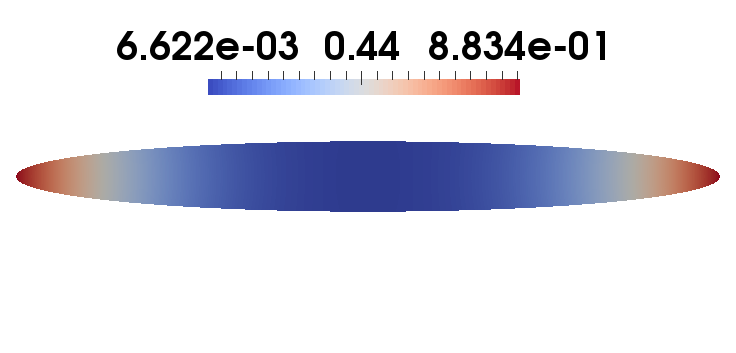}}\hspace{-2mm}
\includegraphics[width=.36\textwidth,angle=0]{{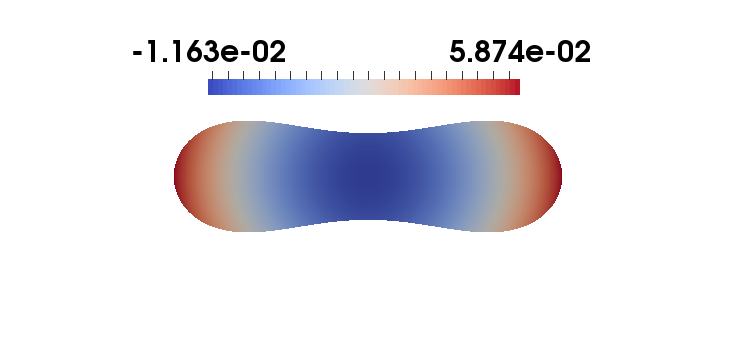}}\hspace{-14mm}
\includegraphics[width=.36\textwidth,angle=0]{{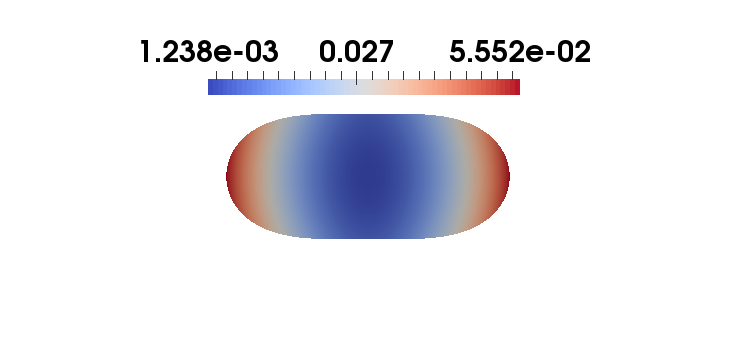}}\hspace{0mm}
\end{center}
\caption{Simulation relating to the thin--film limit using the parametric scheme (\ref{equation_ParametricFEM_v_f2}), (\ref{equation_ParametricFEM_u_f2}), with $Q=0.1$, $\gamm=0$ and $\alpha = \beta = 0.1$. The initial geometry, $\Gamma(0)$, is given by an ellipse of length $1.0$ and height $0.1$. The solution, $u_h$, is displayed at $t = 0, 0.2,0.3$. The middle plot illustrates similar behaviour to that on the right of Figure \ref{f:stab}, though the tendency to fatten for this parameter set subsequently overcomes the tip bulges.}  
\label{image_a01b01_thin_film_new}
\end{figure}

\subsection{Simulations for $\beta=0$ and $\gamm\neq 0$}
We conclude our numerical results with some simulations with $\beta=0$ and $\gamm\neq 0$, such that the curvature term from the Robin boundary conditions for $u$ is removed, whilst a velocity law of forced mean curvature flow is maintained. 

\subsubsection{Radially symmetric solutions in $\mathbb{R}^2$}

\begin{figure}[htbp]
\begin{center}
\includegraphics[width=.24\textwidth,angle=0]{{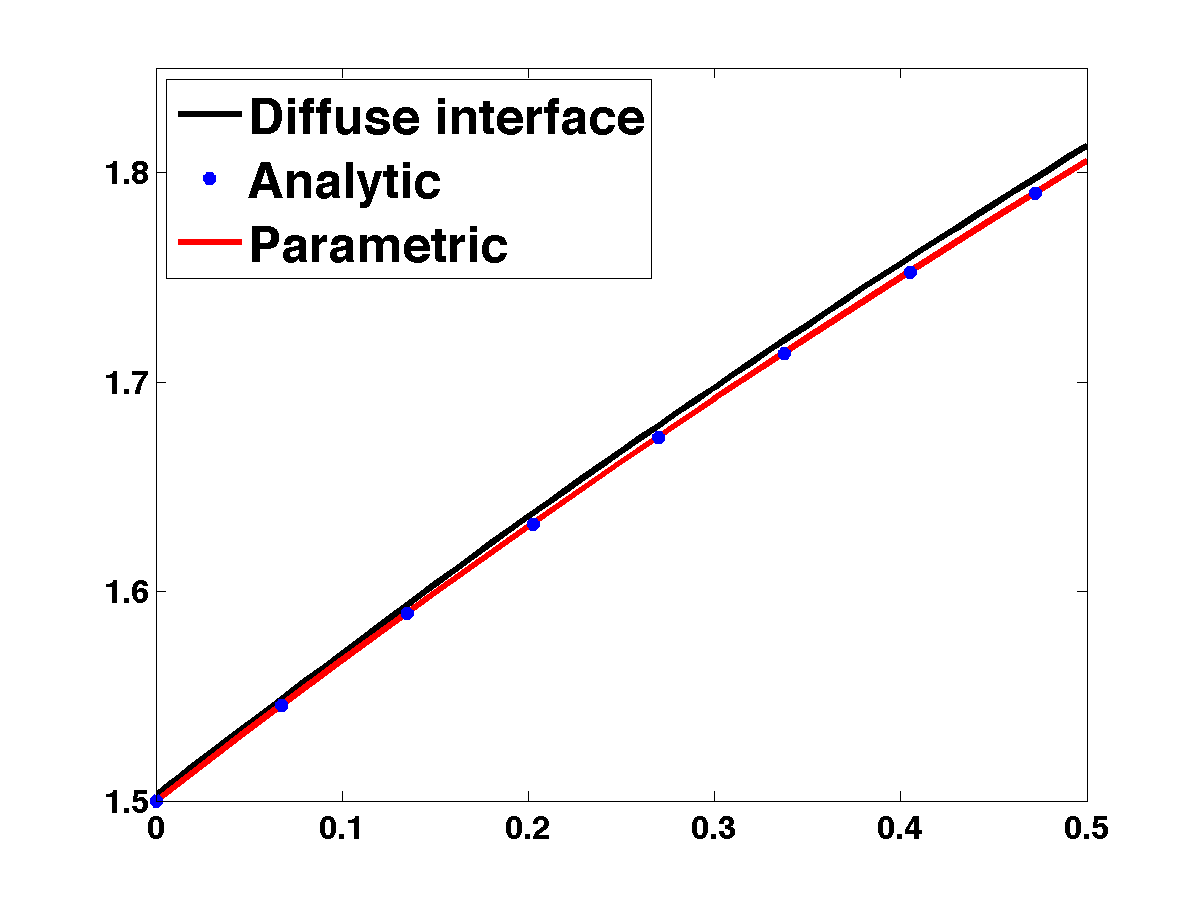}}\hspace{0mm}
\includegraphics[width=.24\textwidth,angle=0]{{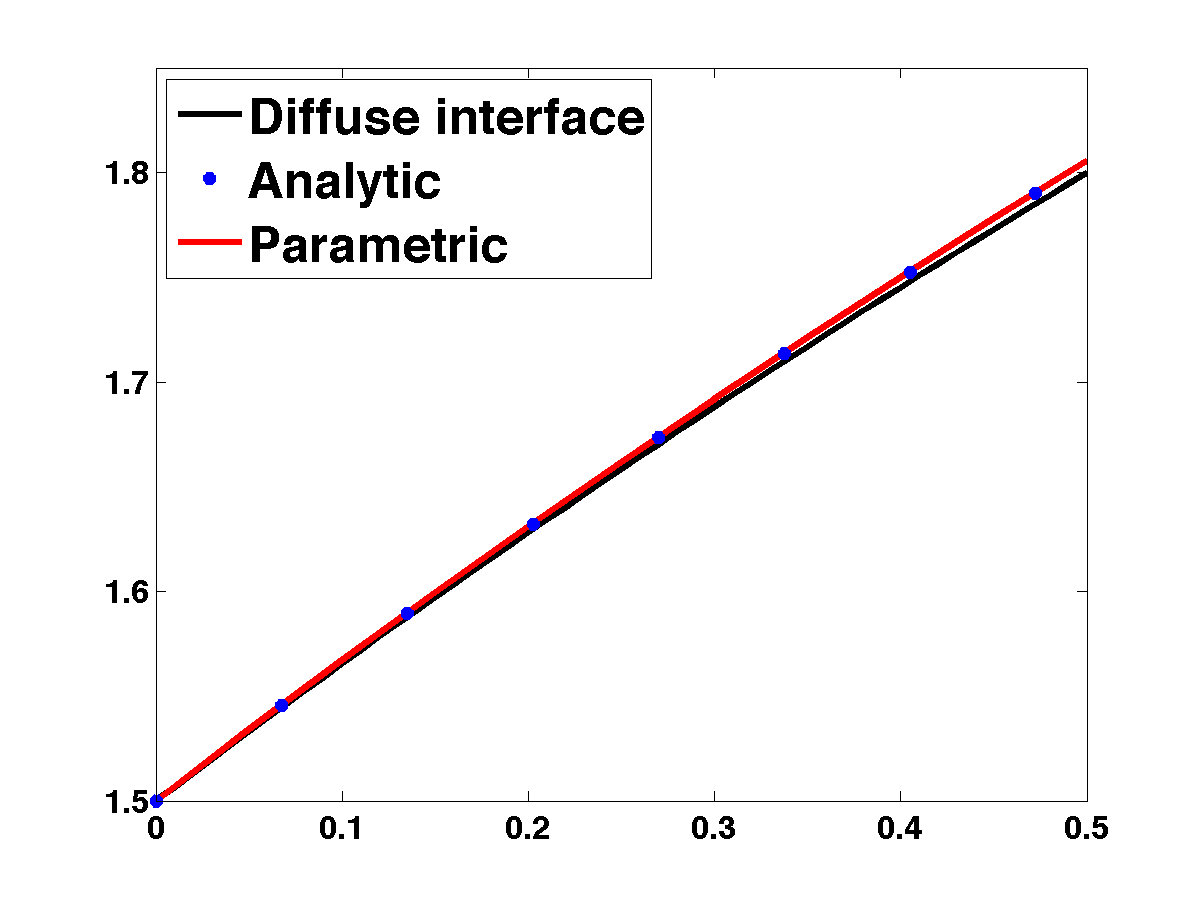}}\hspace{0mm}
\includegraphics[width=.24\textwidth,angle=0]{{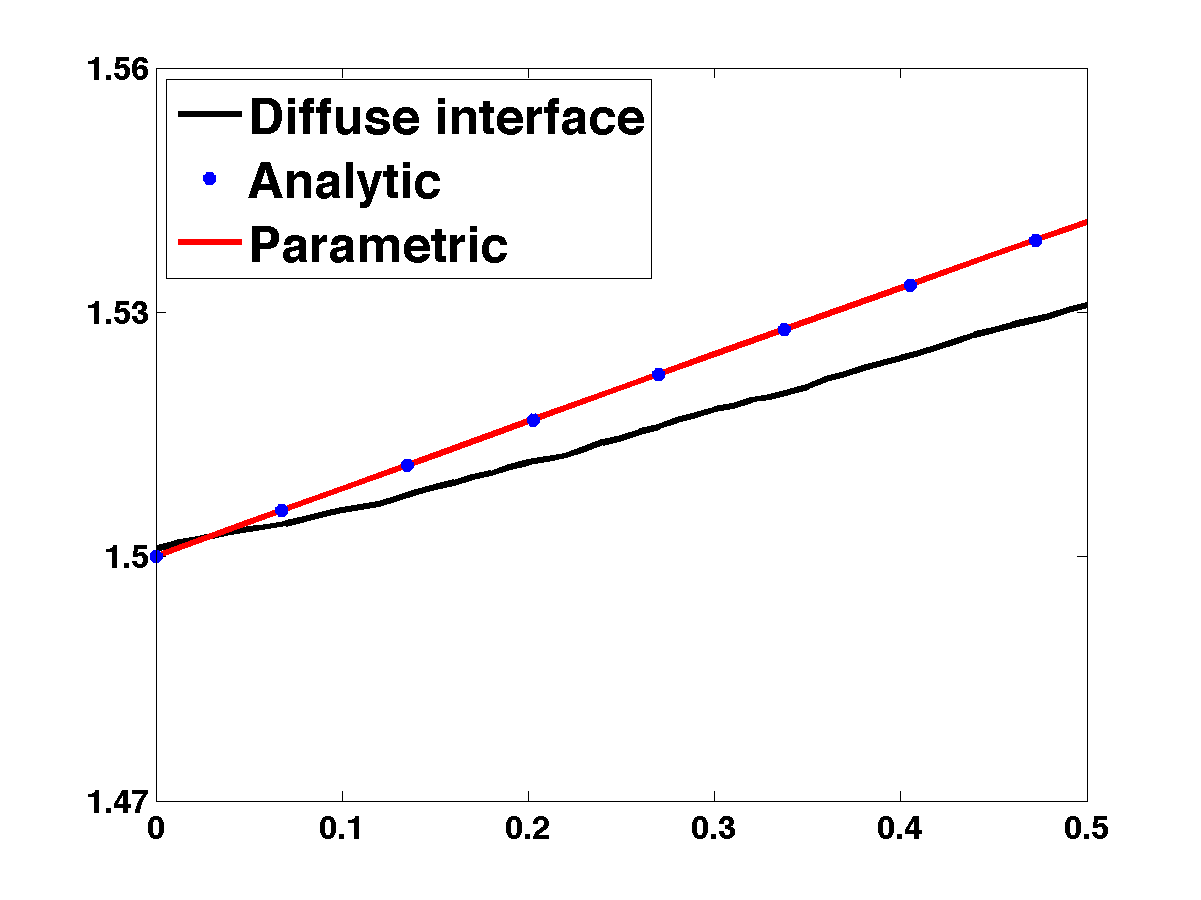}}\hspace{0mm}
\includegraphics[width=.24\textwidth,angle=0]{{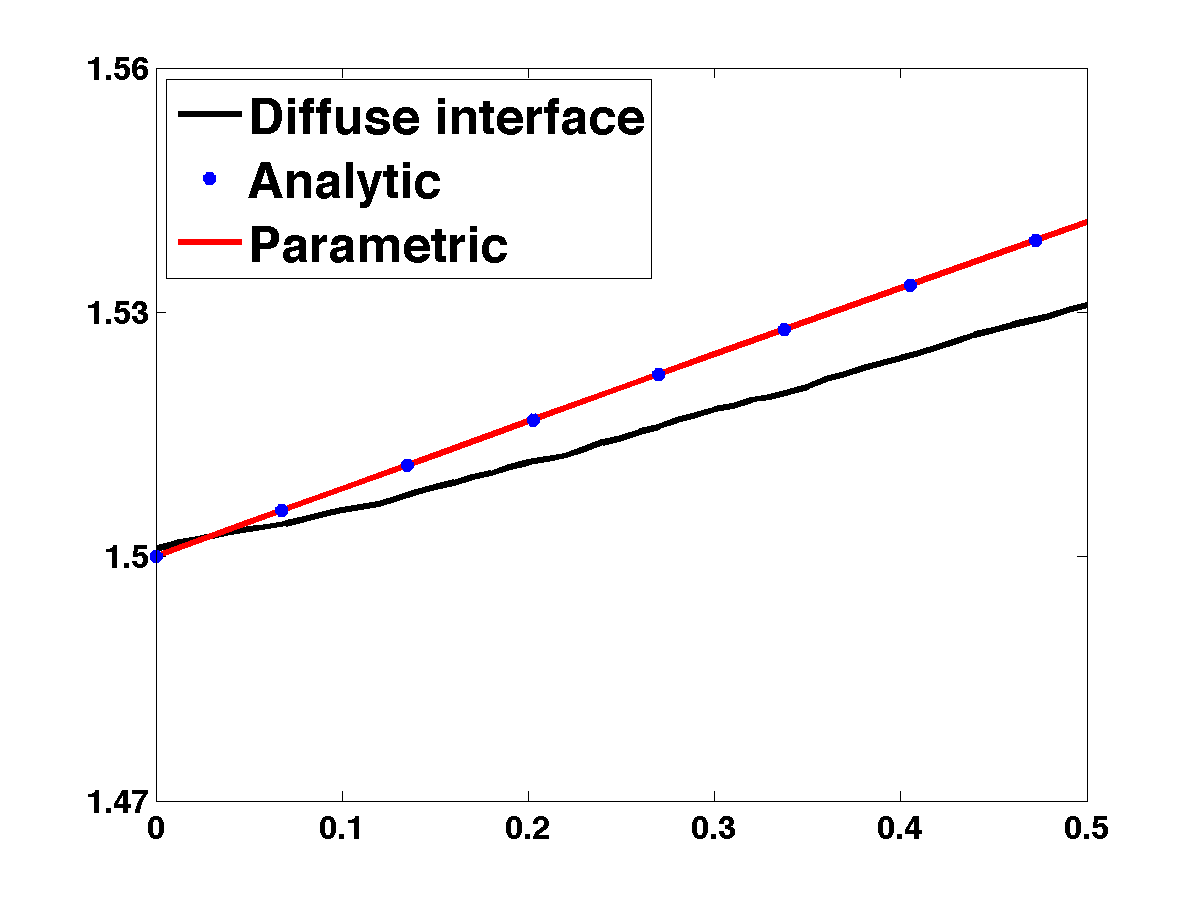}}\hspace{0mm}\\
\includegraphics[width=.24\textwidth,angle=0]{{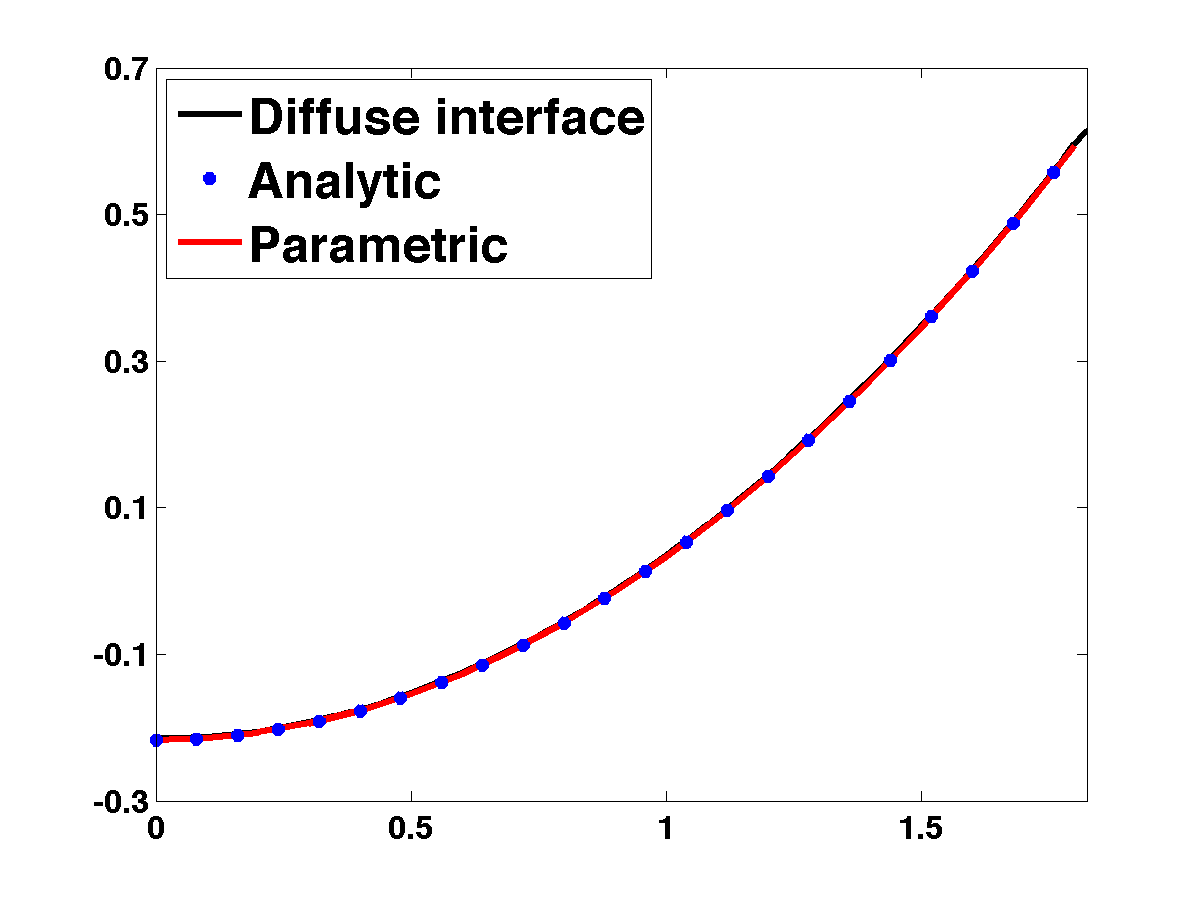}}\hspace{0mm}
\includegraphics[width=.24\textwidth,angle=0]{{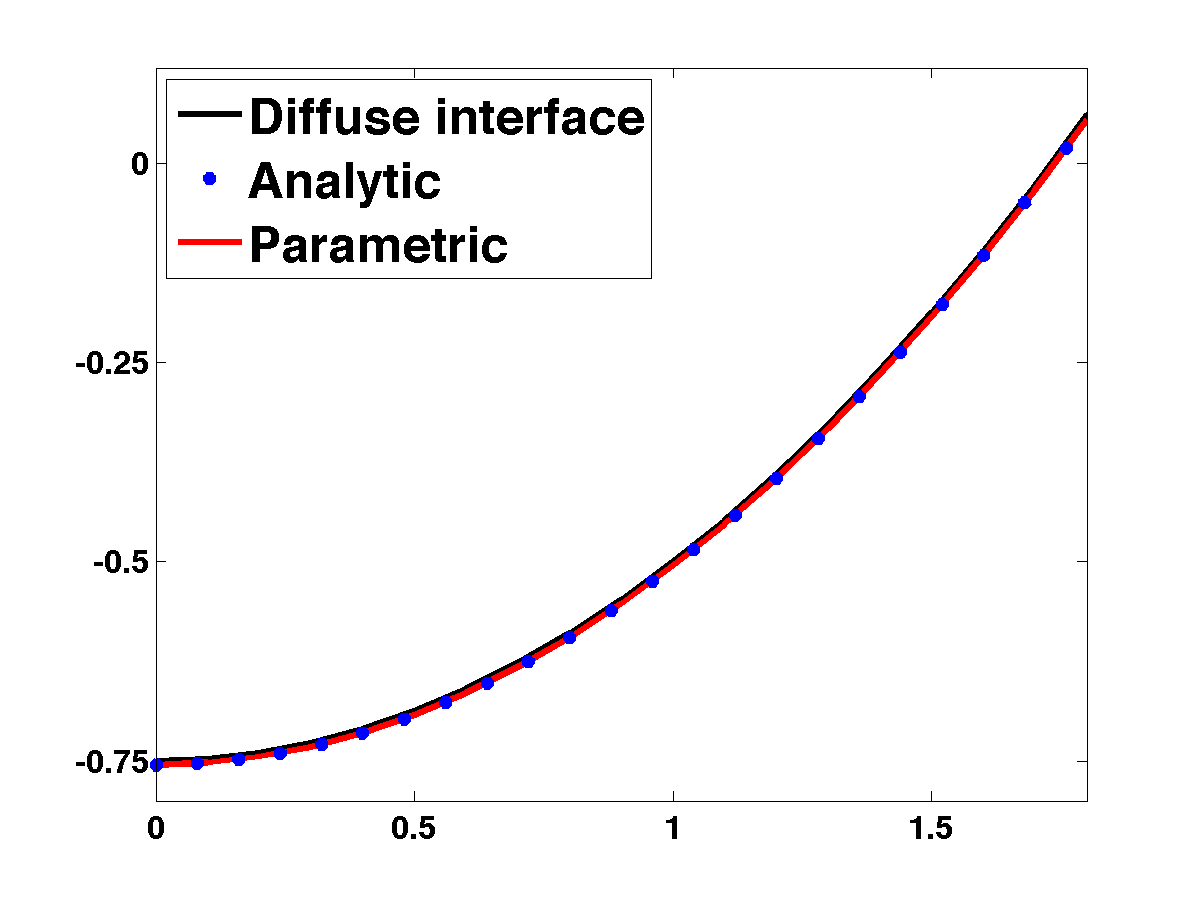}}\hspace{0mm}
\includegraphics[width=.24\textwidth,angle=0]{{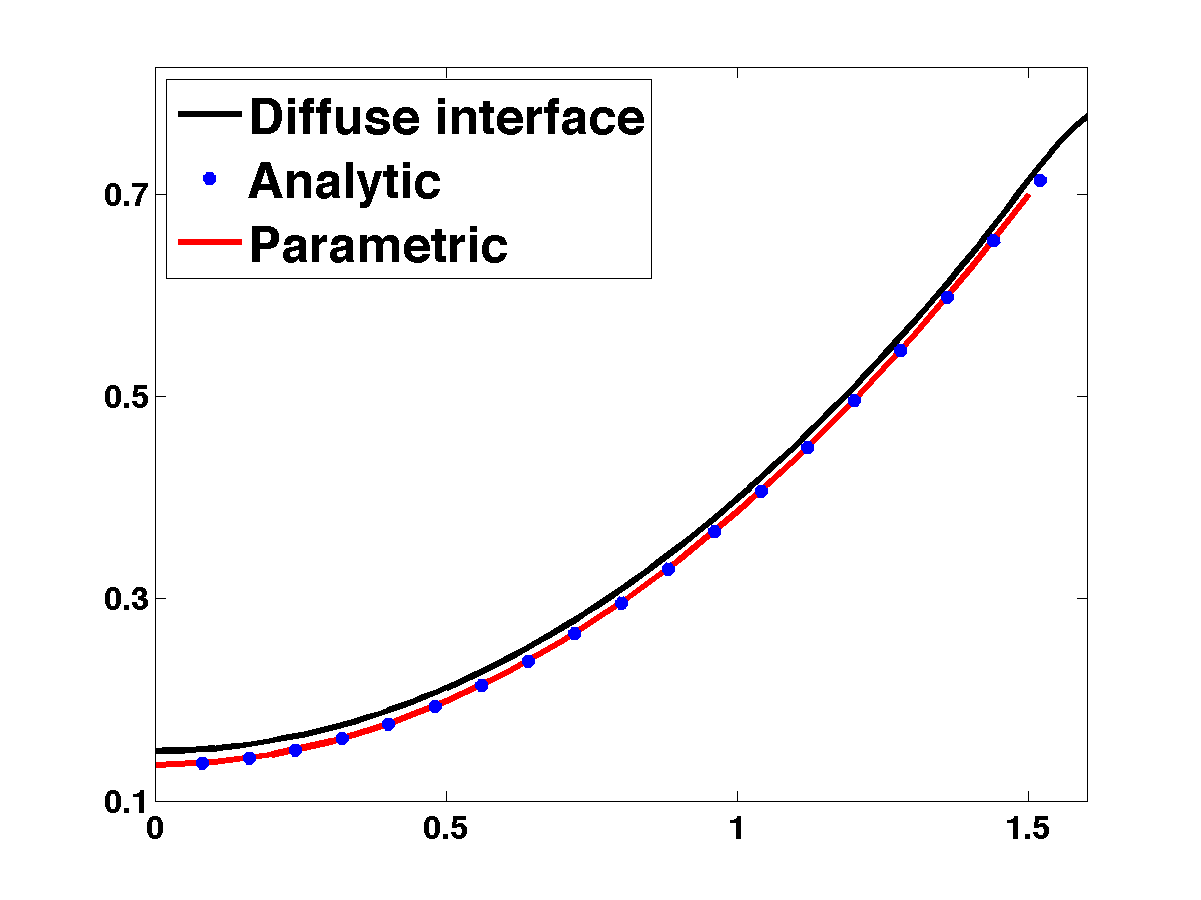}}\hspace{0mm}
\includegraphics[width=.24\textwidth,angle=0]{{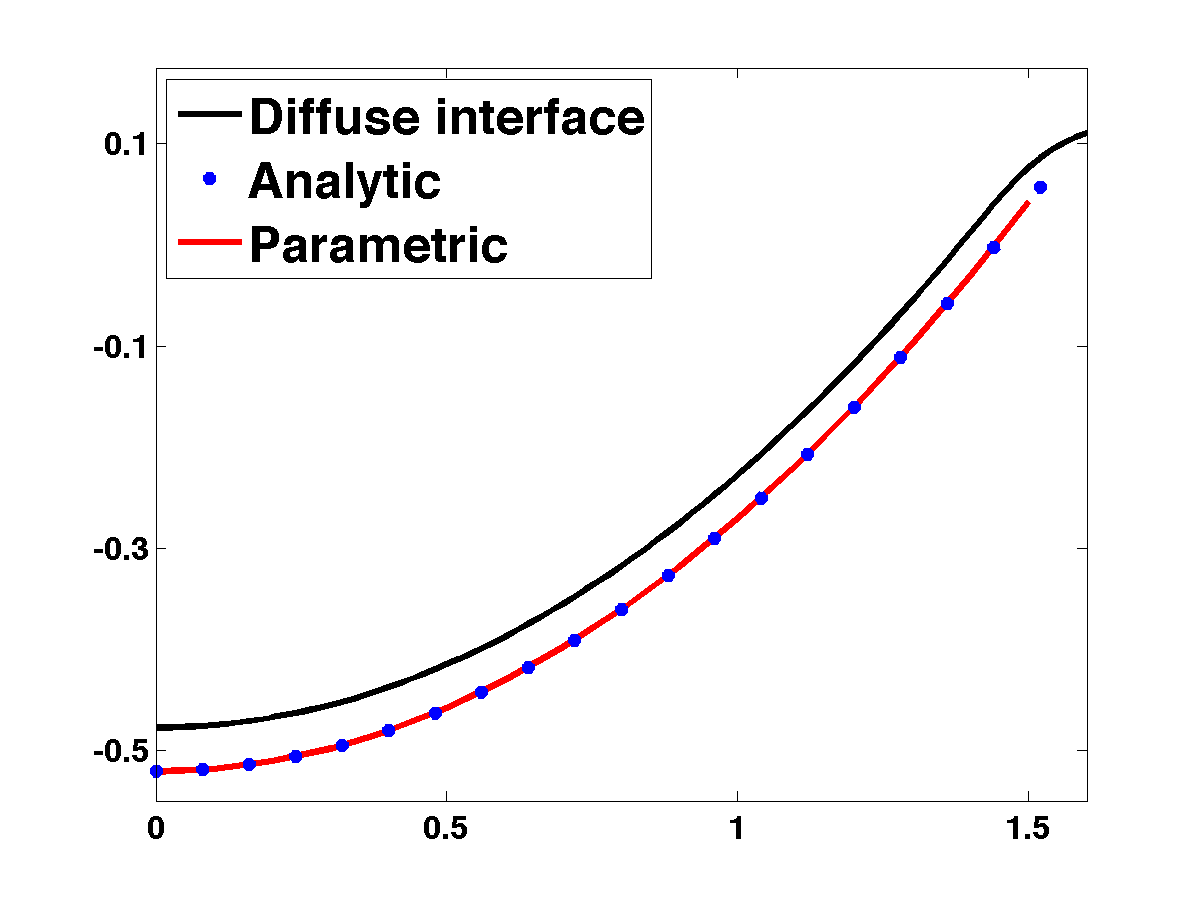}}\hspace{0mm}
\end{center}
\caption{
Comparison of the radius (upper plots) and the pressure $u$ at $t=0.5$ (lower plots) with $\beta=0$, $\alpha=1$, $\gamm=0.1$ (first column), $\beta=0$, $\alpha=\gamm=0.1$, (second column), $\beta=0$, $\alpha=\gamm=1$ (third column) and $\beta=0$, $\alpha=0.1$, $\gamm=1$ (fourth column). The upper plots display $t$ against $R$, on the $x$-- and $y$--axis, respectively. The lower plots display $r$ against $u$, again on the $x$-- and $y$--axis, respectively.
}
\label{f:radial2}
\end{figure}

Figure \ref{f:radial2} presents radially symmetric results with $\beta=0$ and $Q=R_0=1.5$ displayed in the same format as the results in Figure \ref{f:radial}.
In the parametric simulations 
the mesh size at $t=0$ was taken to be  $h\approx 0.1$. 
As in Figure \ref{f:radial} we see that for all four combinations of $\alpha$ and $\gamm$ there is very good agreement between the solutions of the parametric scheme and its analytical solution. In addition we see that for $\beta=0$ , $\alpha=1$, $\beta=0.1$ and $\alpha=\gamm=0.1$ the diffuse--interface solution is much closer to the analytical solution than it was for the original model, we also note that for $\beta=0$ , $\alpha=1$, $\gamm=0.1$ we set $\varepsilon=0.04$ rather than the larger value $\varepsilon=0.075$ that was used for the original model. Similar to the original model, the results of the diffuse--interface model with $\beta=0$, $\alpha=\gamm=1$ and $\alpha=0.1$, $\gamm=1$ are not good and so in the subsequent simulations we will not include any diffuse--interface results using these parameter combinations.

\subsubsection{$\mathbb{R}^2:$ $\beta=0,$ $\alpha=1.0,~\gamm=1.0$}
As in the case for $\gamm=0$, with the initial data chosen, an ellipse with length $0.5$ and height $1.0$, our simulations show that for $\beta=0$, $\alpha = \gamm = 1.0$ the geometry tends to a radially symmetric steady state, with the radius of the resulting circle decreasing as $\beta$ increases.

\subsubsection{$\mathbb{R}^2:$ $\beta=0,$ $\alpha=0.1,~\gamm=1.0$}
The parameters $\gamm=0$, $\alpha=0.1$ and $\beta=1.0$ gave rise to the geometries displayed in Figure \ref{image_a01b10_new}, however for the parameters $\beta=0$, $\alpha=0.1$ and $\gamm=1.0$,  such that the curvature term in the Robin boundary conditions for $u$ is removed, the initial ellipse with length $0.5$ and height $1.0$ shrinks to a point.

\subsubsection{$\mathbb{R}^2:$ $\beta=0,$  $\alpha=1.0,~\gamm=0.1$}
\label{sss:R2_a10b01}

\begin{figure}[htbp]
\begin{center}
\includegraphics[width=.24\textwidth,angle=0]{{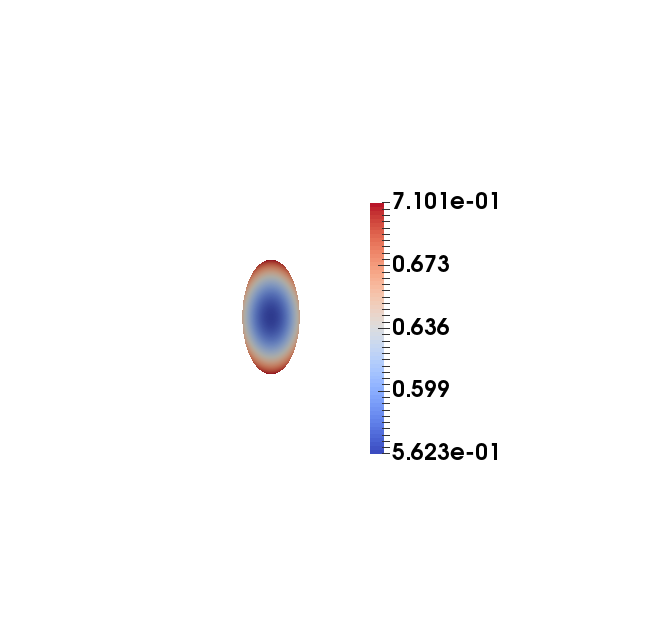}}\hspace{0mm}
\includegraphics[width=.24\textwidth,angle=0]{{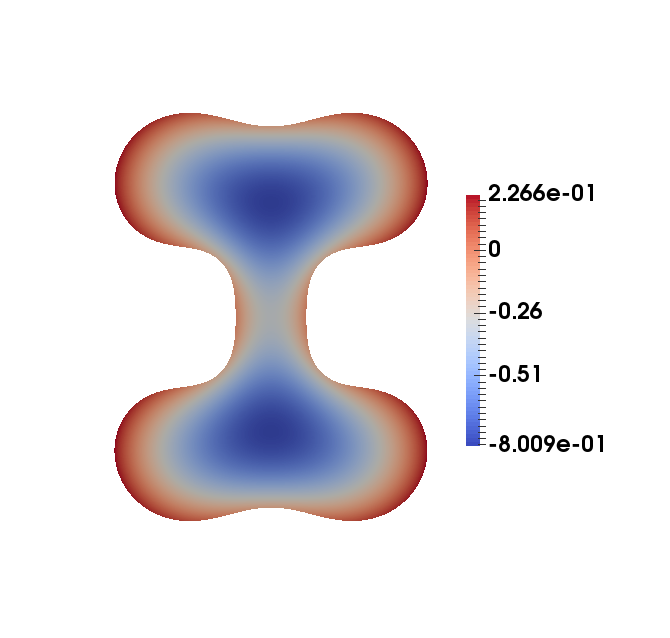}}\hspace{0mm}
\includegraphics[width=.24\textwidth,angle=0]{{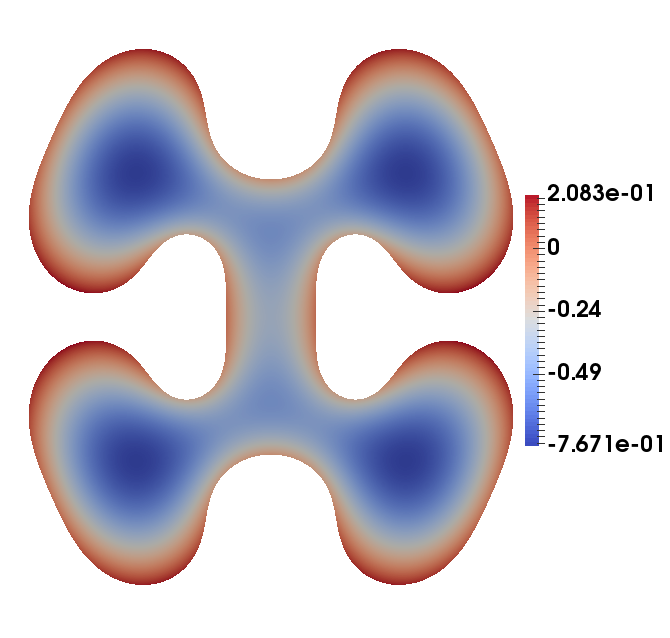}}\hspace{0mm}
\end{center}
\begin{center}
\includegraphics[width=.24\textwidth,angle=0]{{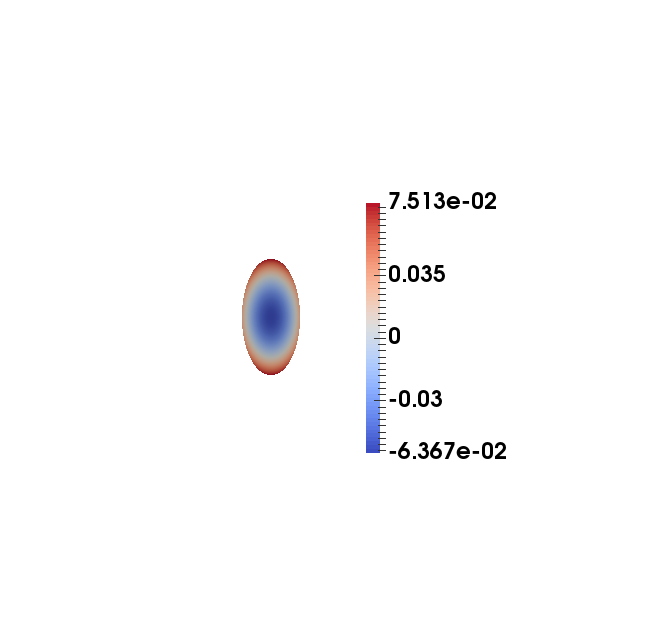}}\hspace{0mm}
\includegraphics[width=.24\textwidth,angle=0]{{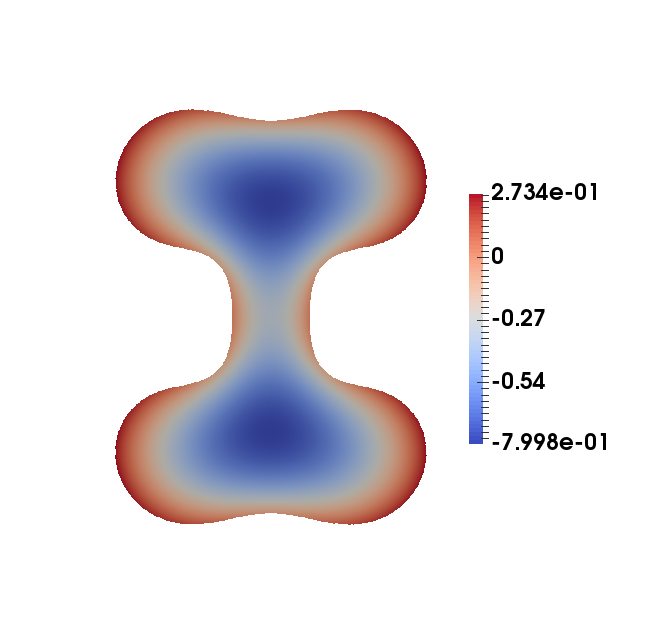}}\hspace{0mm}
\includegraphics[width=.24\textwidth,angle=0]{{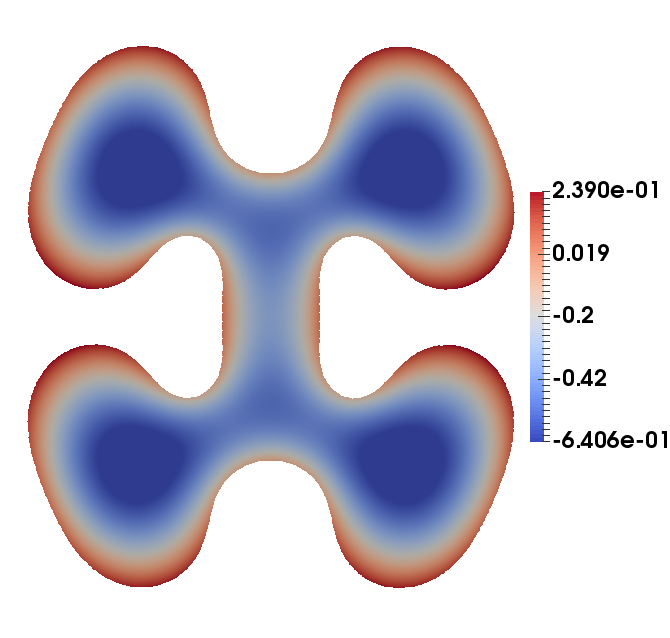}}\hspace{0mm}
\end{center}
\begin{center}
\includegraphics[width=.24\textwidth,angle=0]{{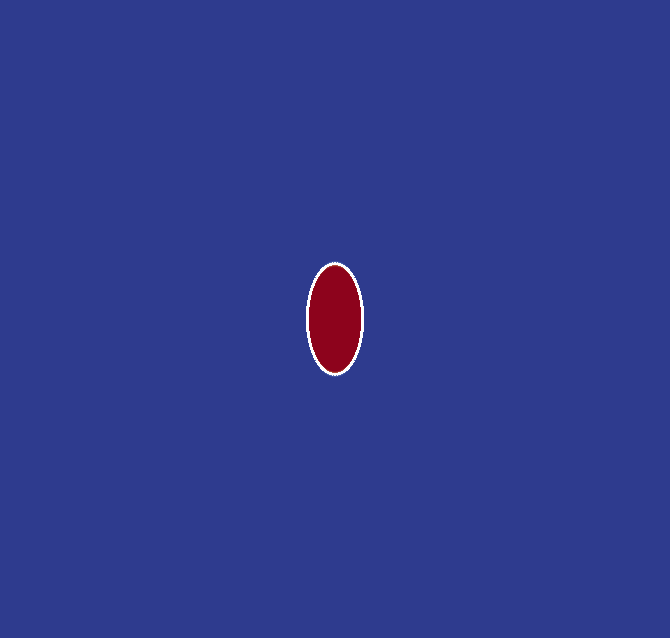}}\hspace{0mm}
\includegraphics[width=.24\textwidth,angle=0]{{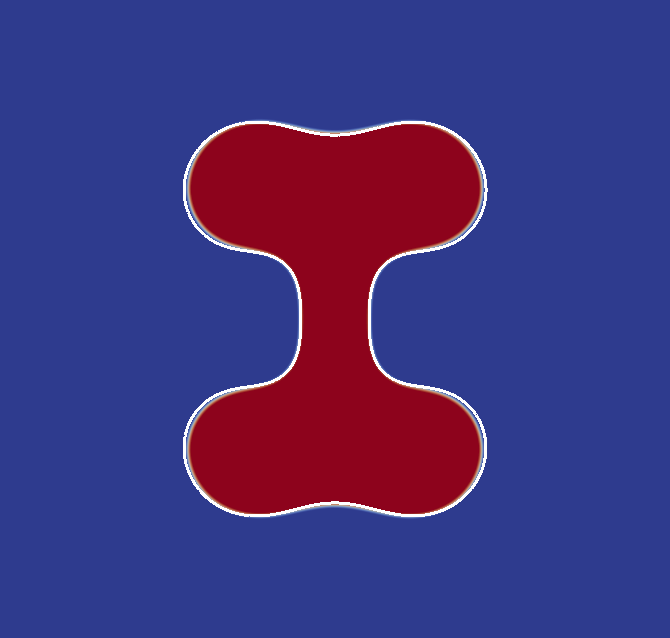}}\hspace{0mm}
\includegraphics[width=.24\textwidth,angle=0]{{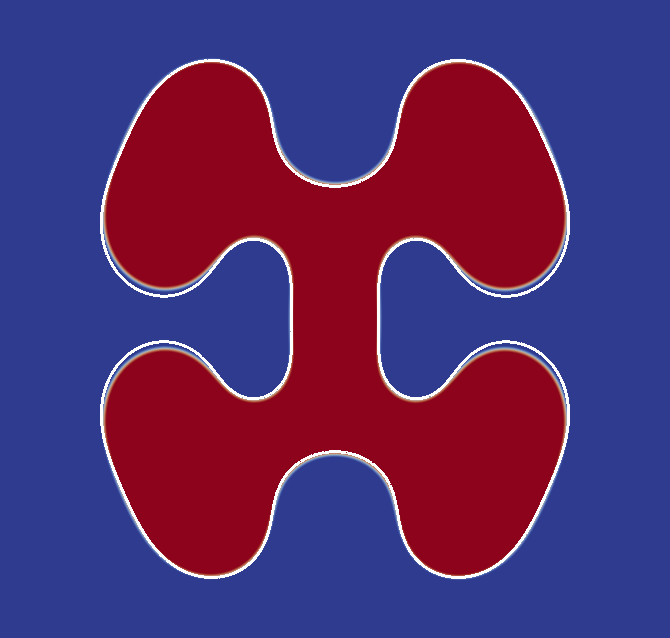}}\hspace{0mm}
\end{center}
\caption{Simulations for $\beta=0$, $\alpha = 1.0$ and $\gamm = 0.1$ at $t=0,30,45$: $u_h$ given by the parametric scheme (top row), $\tilde{u}_h$ given by the diffuse--interface scheme (middle row), $\varphi_h$ (in red and blue) from the diffuse--interface scheme and $\bmcx_h$ (in white) from the parametric scheme (bottom row).} 
\label{image_a10b01_old}
\end{figure}

Figure \ref{image_a10b01_old} presents results with $\beta=0$, $\alpha=1$ and $\gamm=0.1$, displayed in the same format as the results in Figure \ref{image_a10b01_new}. 
The solutions are displayed at times $t = 0, 15,30,45$ (columns one to four respectively). 
As in Figure \ref{image_a10b01_new}, we see good agreement between the parametric and diffuse--interface schemes. When comparing Figure \ref{image_a10b01_old} with Figure \ref{image_a10b01_new} we see that the additional curvature term in the Robin boundary conditions for $u$ gives rise to more a rounded structure in the geometry.

\begin{figure}[htbp]
\begin{center}
\includegraphics[width=.24\textwidth,angle=0]{{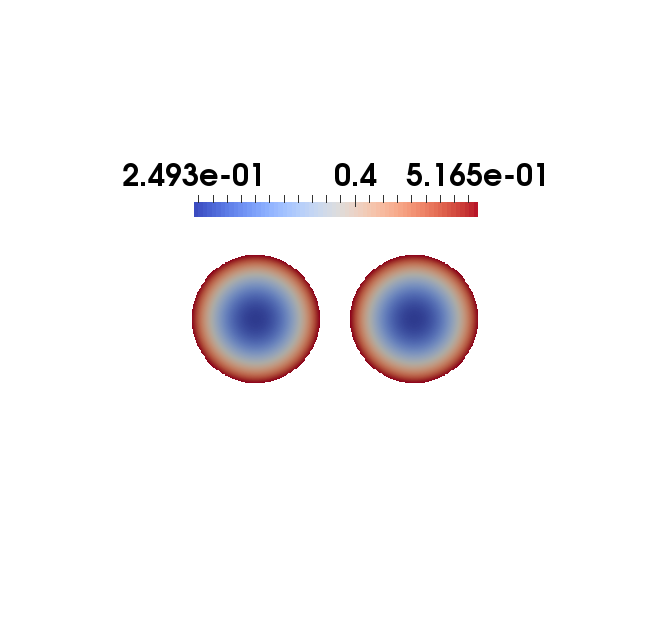}}\hspace{0mm}
\includegraphics[width=.24\textwidth,angle=0]{{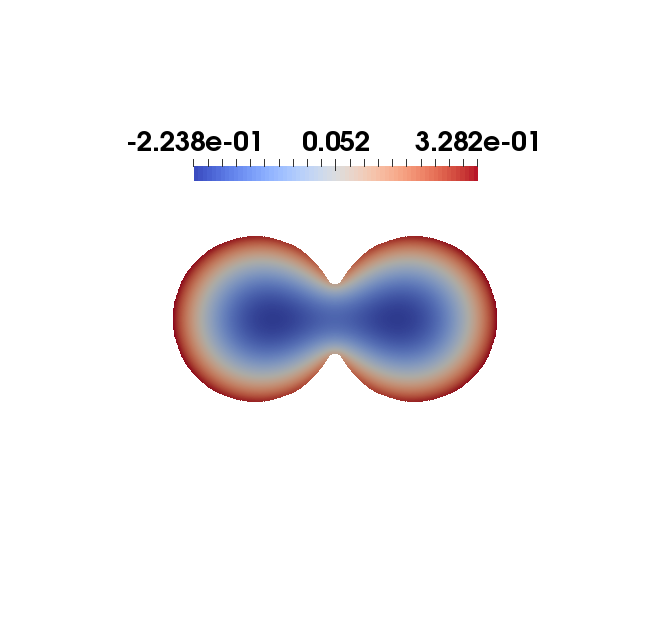}}\hspace{0mm}
\includegraphics[width=.24\textwidth,angle=0]{{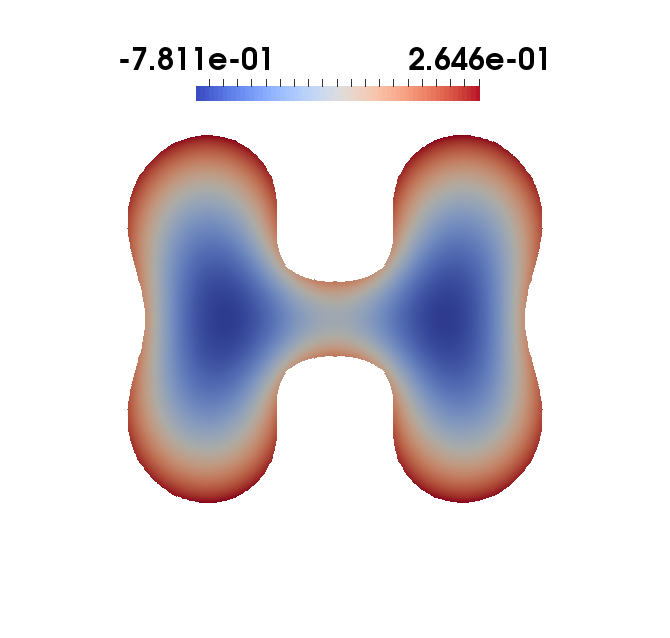}}\hspace{0mm}
\includegraphics[width=.24\textwidth,angle=0]{{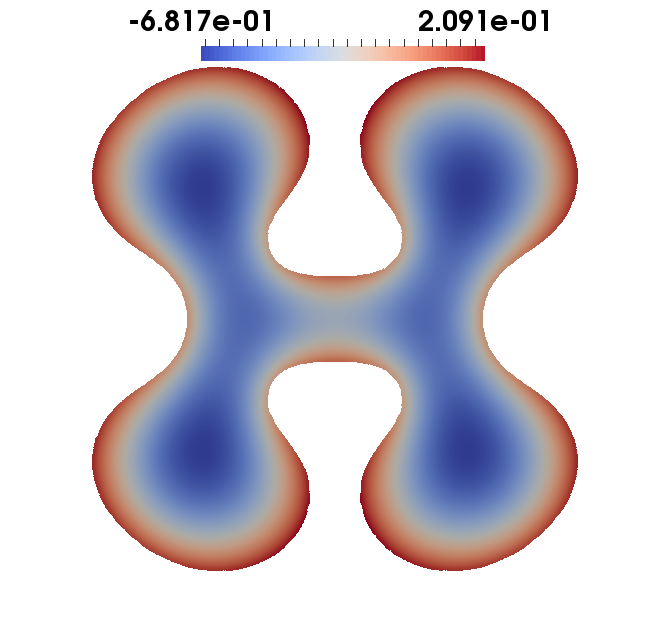}}\hspace{0mm}
\end{center}
\begin{center}
\includegraphics[width=.24\textwidth,angle=0]{{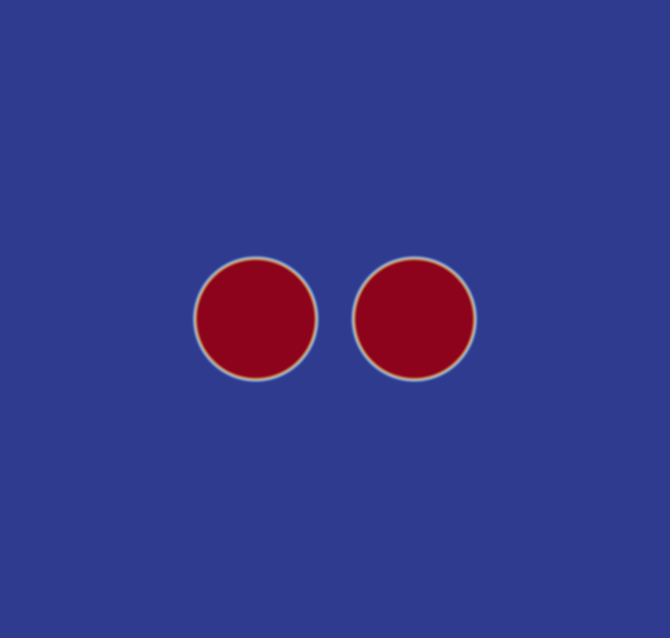}}\hspace{0mm}
\includegraphics[width=.24\textwidth,angle=0]{{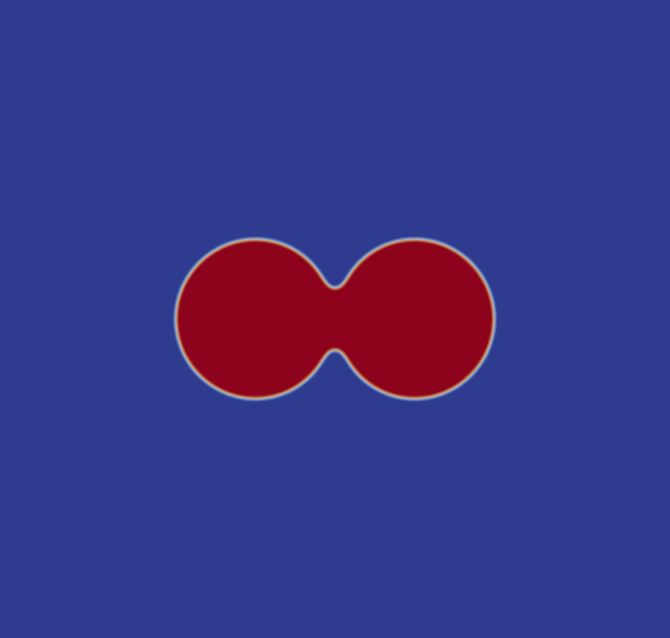}}\hspace{0mm}
\includegraphics[width=.24\textwidth,angle=0]{{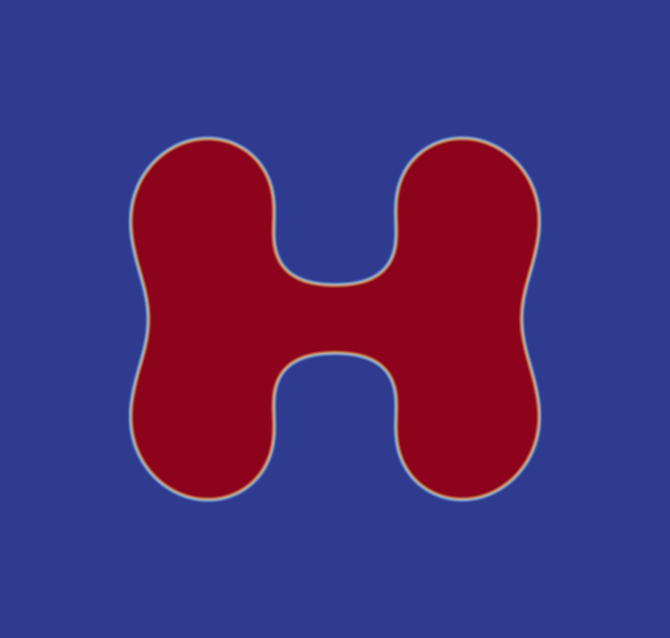}}\hspace{0mm}
\includegraphics[width=.24\textwidth,angle=0]{{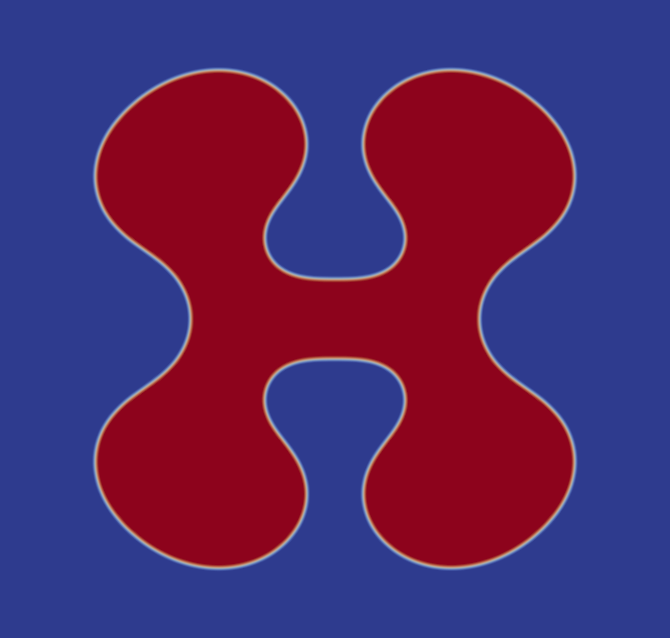}}\hspace{0mm}
\end{center}
\caption{Simulations for $\beta=0$, $\alpha = 1.0$ and $\gamm = 0.1$ at $t = 0, 1, 20,30$ using the diffuse--interface scheme, with 
 $\tilde{u}_h$ displayed in the top row and $\varphi_h$ in the bottom row. Such simulations illustrate the complicated pathologies that can result even for the current simple tissue--growth model.}
\label{image_a10b01_col_old}
\end{figure}

Figure \ref{image_a10b01_col_old} shows the merging of two circular tumours. The results are obtained from the diffuse--interface scheme with $\beta = 0$, $\alpha = 1.0$, $\gamm = 0.1$ and $\varepsilon = 0.04$ and the solutions are displayed at $t = 0, 1, 20, 30$. 
The initial geometry is given by two circles of radius $1.0$, with centres at $(1.3,0)$ and $(-1.3,0)$. This example highlights the power of the diffuse--interface model as the parametric model would not be able to handle directly the topology change that takes place when two tumours merge.

\subsubsection{$\mathbb{R}^2:$ $\beta = 0,$ $\alpha =0.1,~\gamm =0.1$}

\begin{figure}[htbp]
\begin{center}
\hspace{12mm}
\includegraphics[width=.24\textwidth,angle=0]{{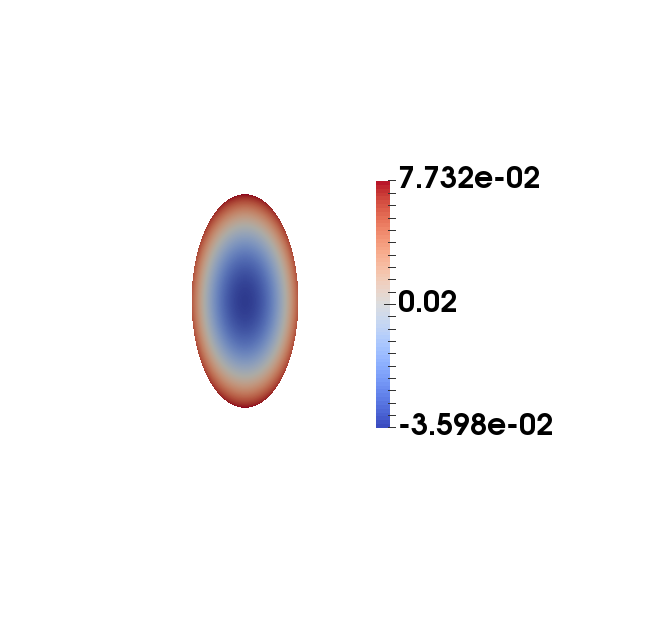}}\hspace{1mm}
\includegraphics[width=.24\textwidth,angle=0]{{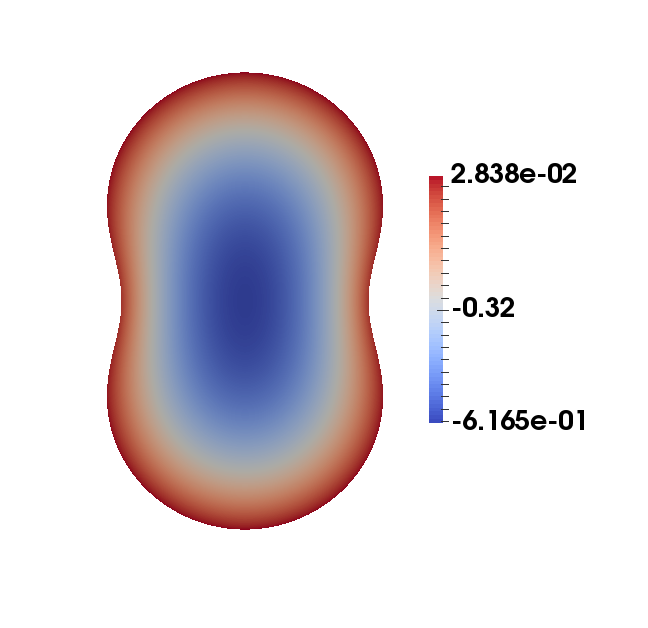}}\hspace{0mm}
\includegraphics[width=.24\textwidth,angle=0]{{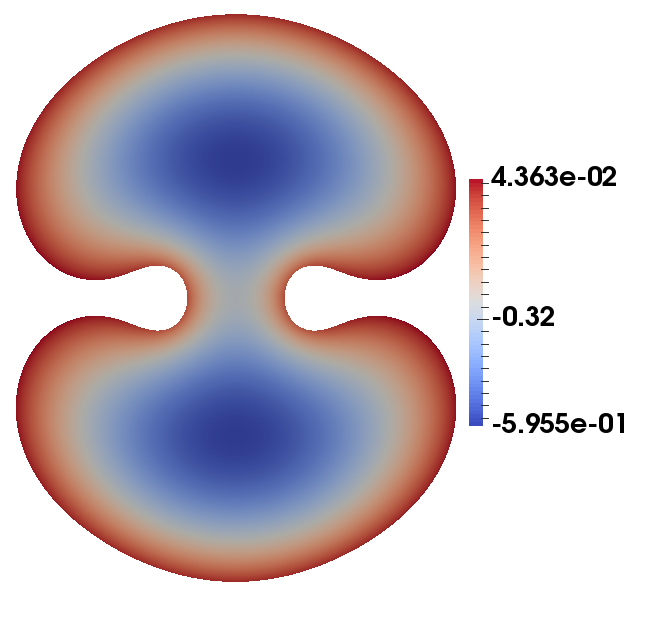}}\hspace{0mm}
\end{center}
\begin{center}
\hspace{12mm}
\includegraphics[width=.24\textwidth,angle=0]{{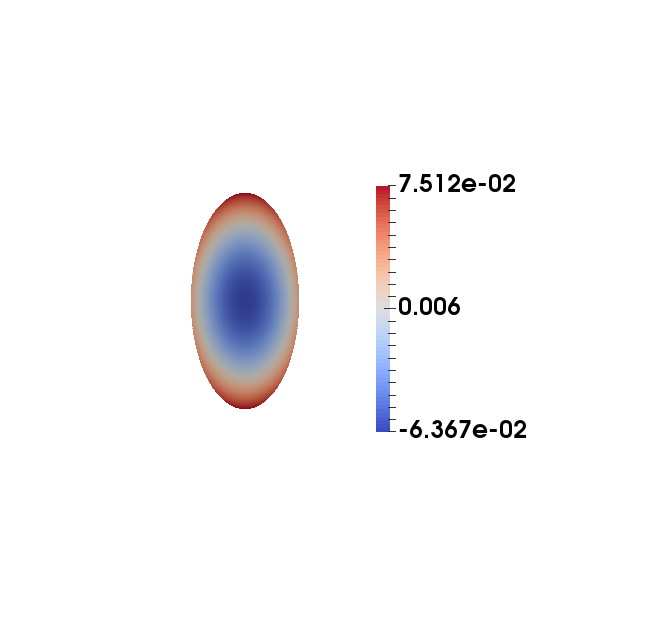}}\hspace{1mm}
\includegraphics[width=.24\textwidth,angle=0]{{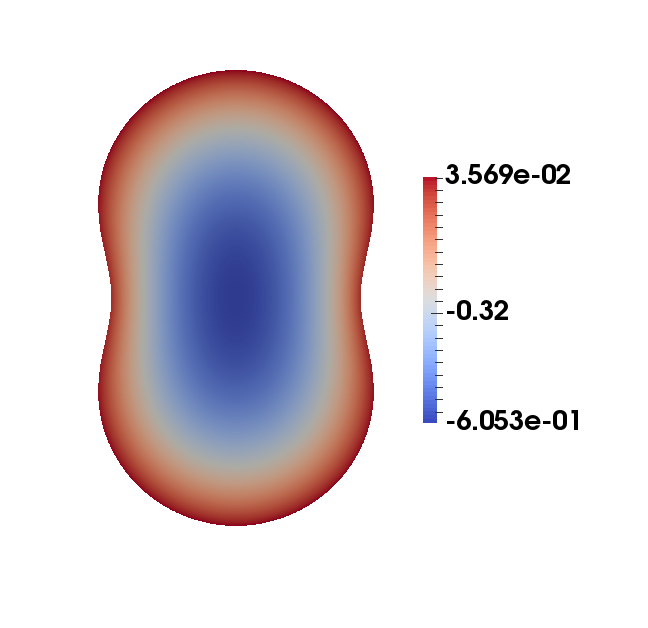}}\hspace{0mm}
\includegraphics[width=.24\textwidth,angle=0]{{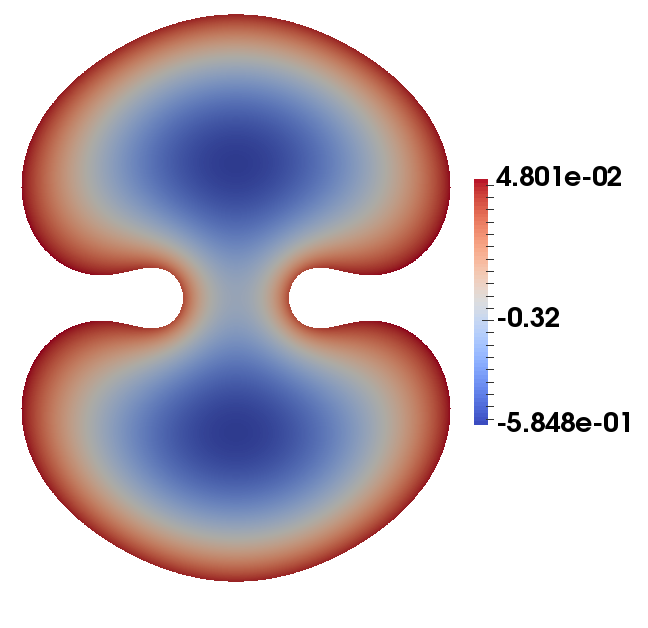}}\hspace{0mm}
\end{center}
\begin{center}
\includegraphics[width=.24\textwidth,angle=0]{{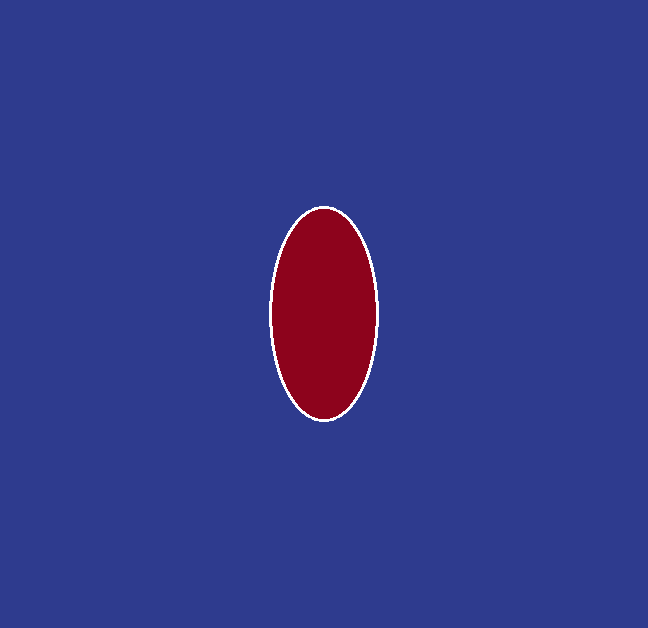}}\hspace{0mm}
\includegraphics[width=.24\textwidth,angle=0]{{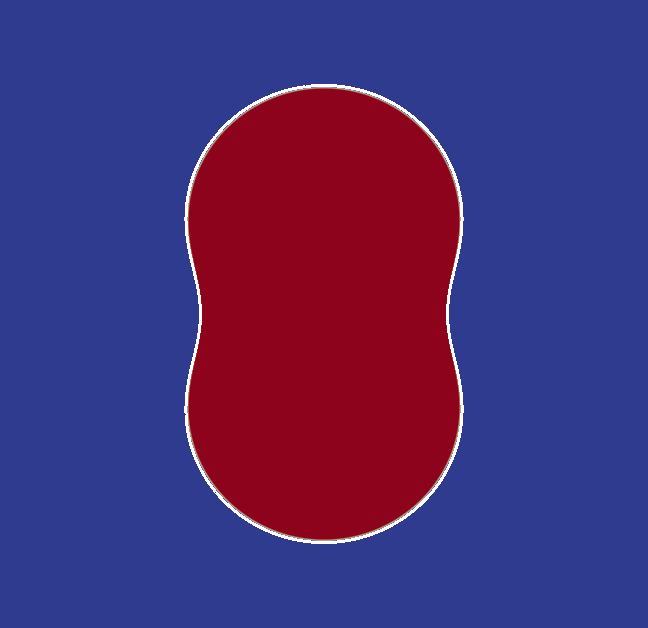}}\hspace{0mm}
\includegraphics[width=.24\textwidth,angle=0]{{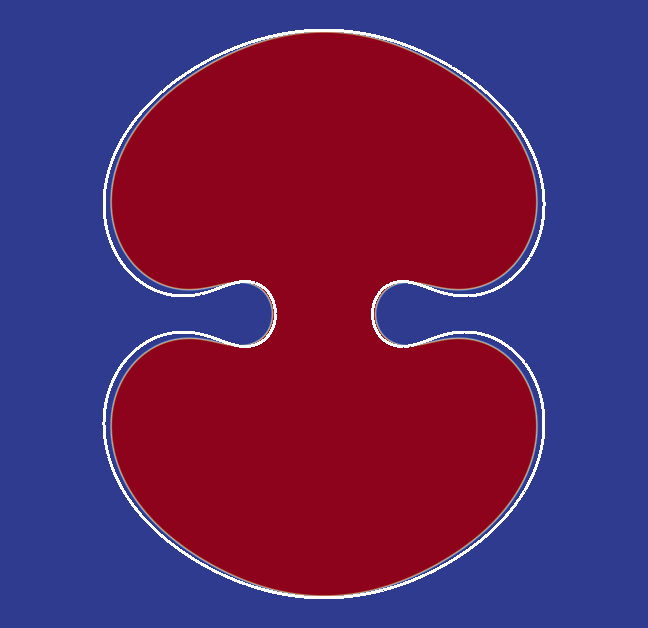}}\hspace{0mm}
\end{center}
\caption{Simulations with $\beta = 0$, $\alpha = 0.1$ and $\gamm = 0.1$ at $t=0,3,7$: $u_h$ given by the parametric scheme (top row), $\tilde{u}_h$ given by the diffuse--interface scheme (middle row), $\varphi_h$ (in red and blue) from the diffuse--interface scheme and $\bmcx_h$ (in white) from the parametric scheme (bottom row).} 
\label{image_a01b01_old}
\end{figure}

Figure \ref{image_a01b01_old} presents results with $\beta = 0$, $\alpha=\gamm=0.1$ displayed at $t = 0,3,7$.
We see good agreement between the parametric and diffuse--interface schemes. This is in contrast to the corresponding simulations with $\gamm=0$,  $\alpha=\beta=0.1$, displayed in Figure \ref{image_a10b01_new}, where the results from the parametric and diffuse--interface schemes are displayed at different times.

\begin{figure}[htbp]
\begin{center}
\includegraphics[width=.275\textwidth,angle=0]{{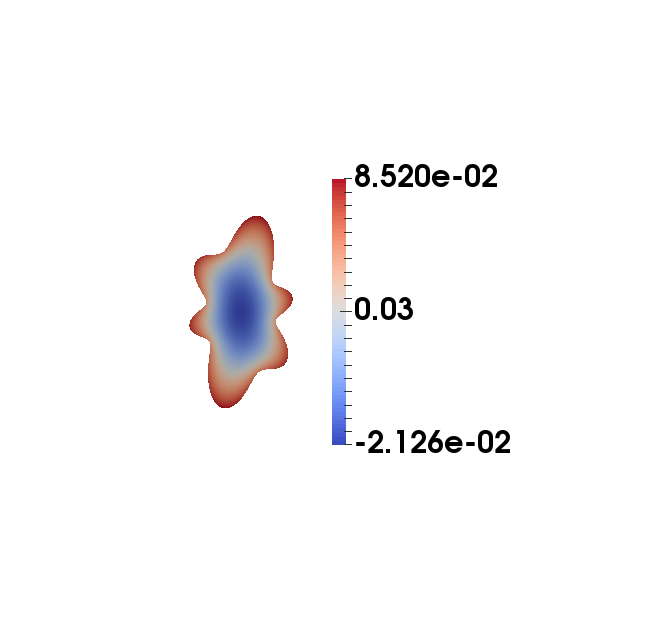}}\hspace{-10mm}
\includegraphics[width=.275\textwidth,angle=0]{{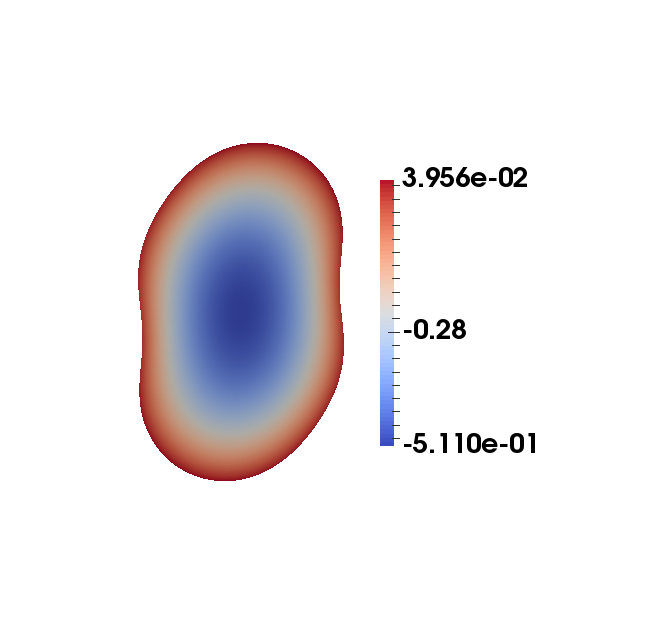}}\hspace{-8mm}
\includegraphics[width=.275\textwidth,angle=0]{{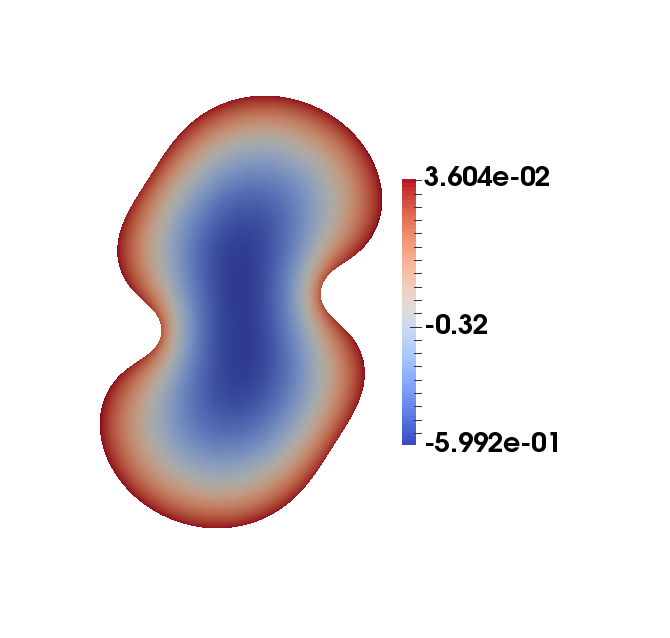}}\hspace{-5mm}
\includegraphics[width=.275\textwidth,angle=0]{{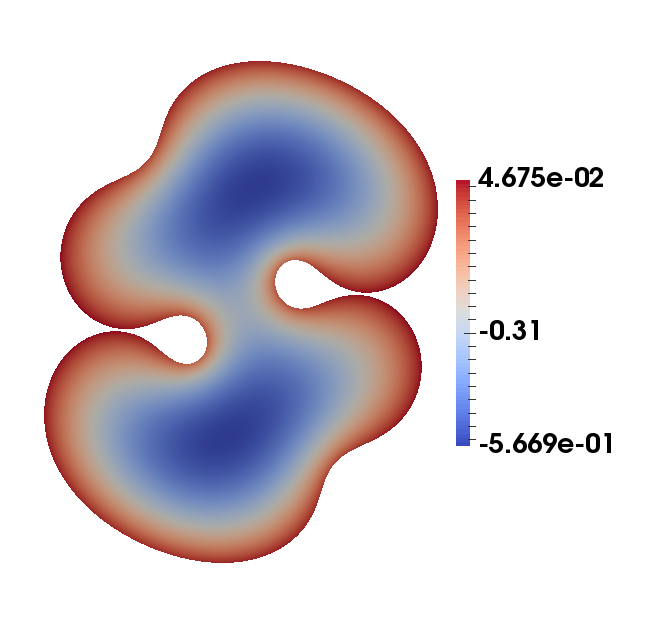}}\hspace{0mm}
\end{center}
\caption{Simulation using the parametric scheme with $\beta = 0$, $\alpha = \gamm = 0.1$ at $t = 0, 2.3, 4.6, 6.9$. }
\label{image_a01b01_nonsym_old}
\end{figure}

In Figure \ref{image_a01b01_nonsym_old} we display results 
using the parametric scheme with $\beta = 0$, $\alpha = \gamm = 0.1$. Here the initial geometry $\Gamma(0)$ is purposefully chosen to reduce symmetry.  To create the initial geometry we perturb an ellipse of length $0.5$ and height $1.0$ by a distance of $0.2 \sin (6 \theta)$ in the normal direction, where $\theta$ is the polar angle. We present the solutions at $t = 0, 2.3, 4.6,6.9$. 

\subsubsection{$\mathbb{R}^2:$ $\beta = 0,$ $\alpha =1.0,~\gamm =0.05,~0.025,~0.01$}
\begin{figure}[htbp]
\begin{center}
\includegraphics[width=.3\textwidth,angle=0]{{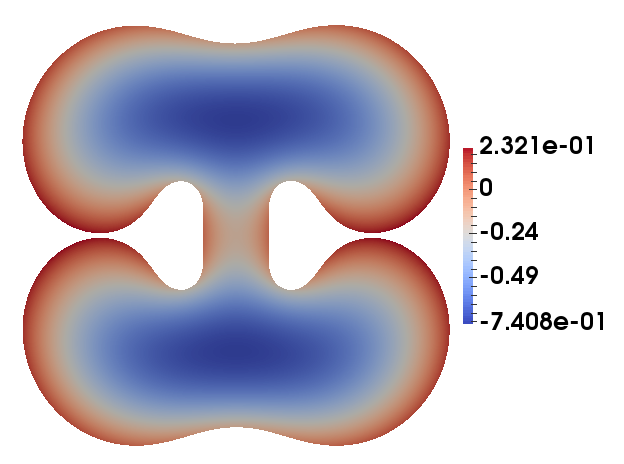}}\hspace{0mm}
\includegraphics[width=.3\textwidth,angle=0]{{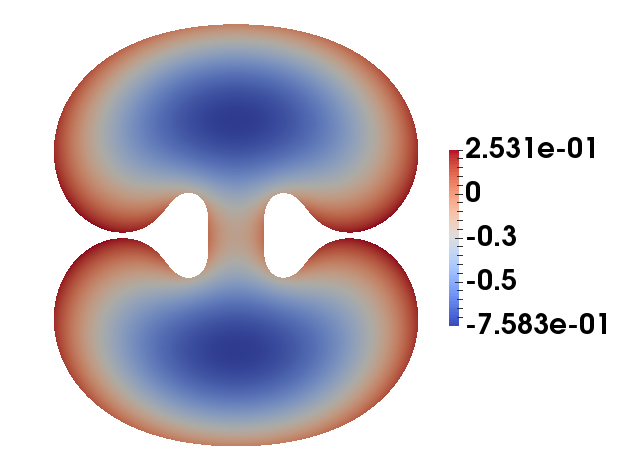}}\hspace{0mm}
\includegraphics[width=.3\textwidth,angle=0]{{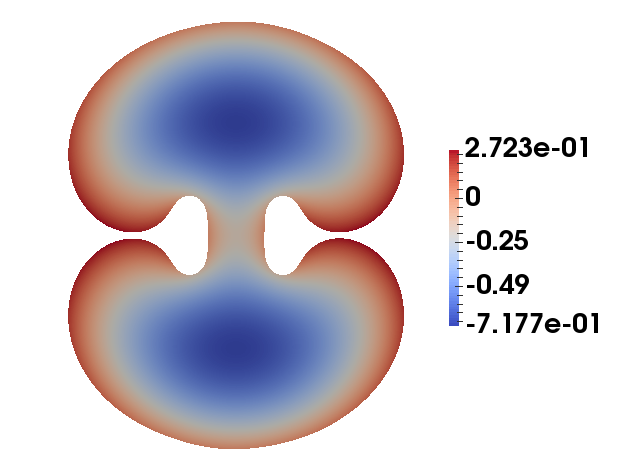}}\hspace{0mm}
\end{center}
\caption{Simulations for $\beta=0$ and $\alpha = 1.0$, with $\gamm=0.05$ at $t=20$ (left plot), $\gamm=0.025$ at $t=14$ (centre plot) and $\gamm=0.01$ at $t=11.6$ (right plot).} 
\label{image_small_gam_old}
\end{figure}

In Figure \ref{image_small_gam_old} we show the effect that reducing $\gamm$ has on the evolution of the tumour. We set $\beta=0$, $\alpha=1.0$ and consider three values of $\gamm$: $\gamm=0.05$ at $t=20$ (left plot), $\gamm=0.025$ at $t=14$ (centre plot) and $\gamm=0.01$ at $t=11.6$ (right plot). From this figure we see that as $\gamma$ is reduced the extent to which the tumour extends in the horizontal direction is also reduced.

\subsubsection{$\mathbb{R}^3:$ $\beta = 0,$ $\alpha = 1.0, ~\gamm = 0.1$}
We set $\beta = 0$, $\alpha = 1.0$, $\gamm = 0.1$ and $Q = 1.25$ and the initial geometry $\Gamma(0)$ is given by the oblate spheroid with equation $ \frac{x^2}{1.0^2}+\frac{y^2}{0.5^2}+\frac{z^2}{1.0^2}=1$ .

In the parametric examples the mesh size was taken to be $h \approx 0.15$ at $t=0$, and we set $\Delta t= 5 \times 10^{-3}$. In the diffuse--interface examples we set $\varepsilon = 0.1$, $h_{max,f} \approx 0.011$, $h_{max,m} \approx 0.045$, $h_{max,c} \approx 8.485$ and we set $\Delta t= 10^{-3}$. 

\begin{figure}[htbp]
\begin{center}
\includegraphics[width=0.21\textwidth,angle=0]{{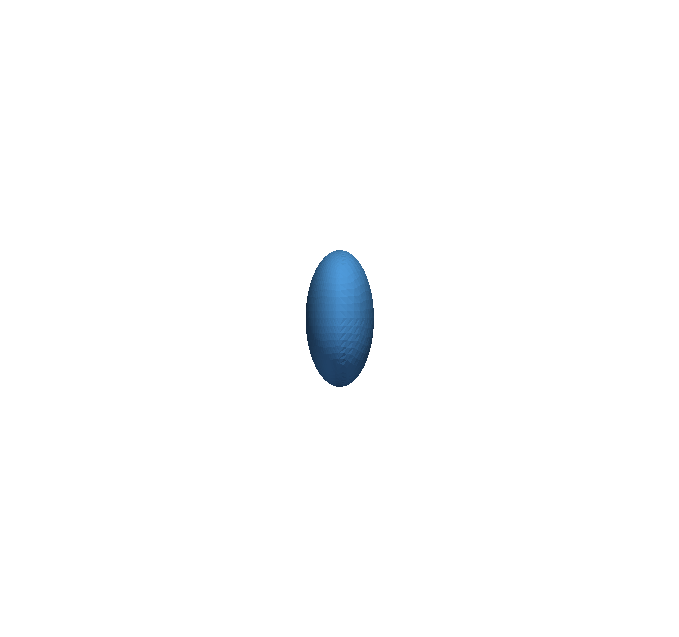}}\hspace{2mm}
\includegraphics[width=0.21\textwidth,angle=0]{{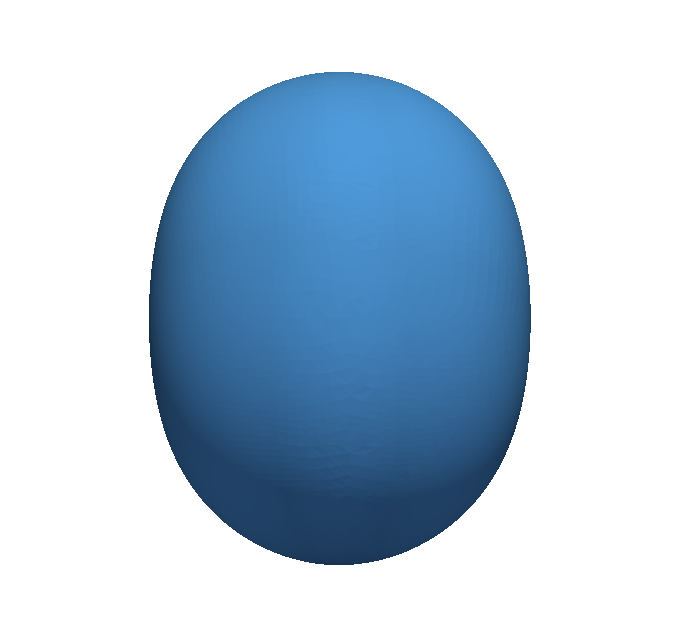}}\hspace{8mm}
\includegraphics[width=0.21\textwidth,angle=0]{{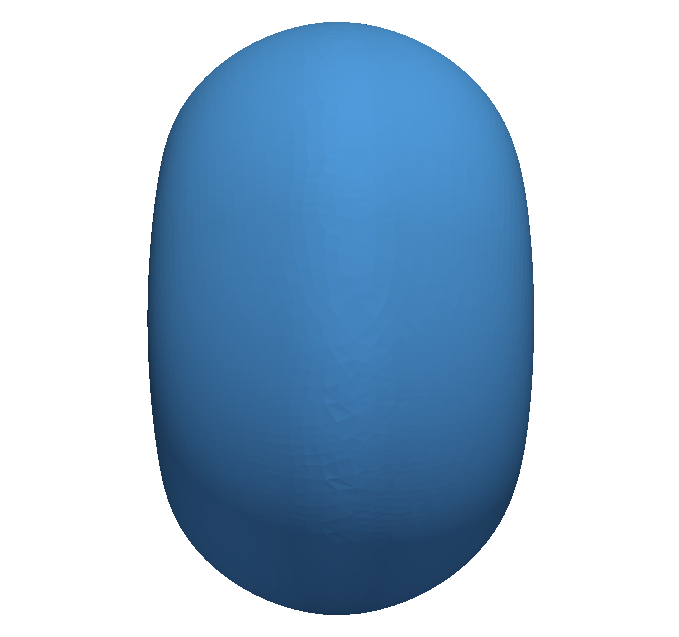}}\hspace{0mm}
\\
\includegraphics[width=0.21\textwidth,angle=0]{{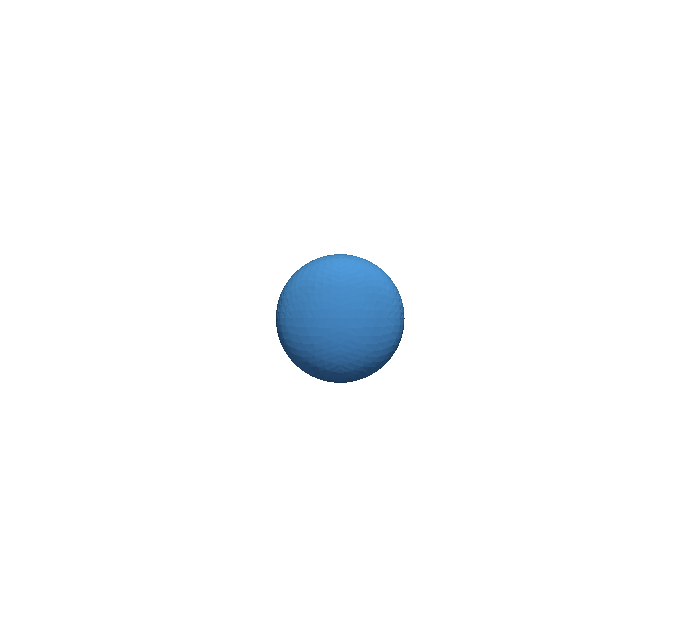}}\hspace{2mm}
\includegraphics[width=0.21\textwidth,angle=0]{{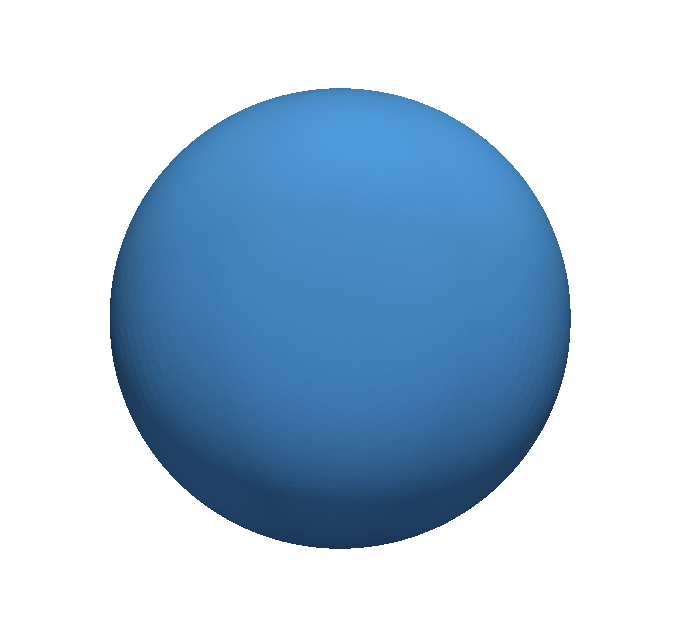}}\hspace{8mm}
\includegraphics[width=0.21\textwidth,angle=0]{{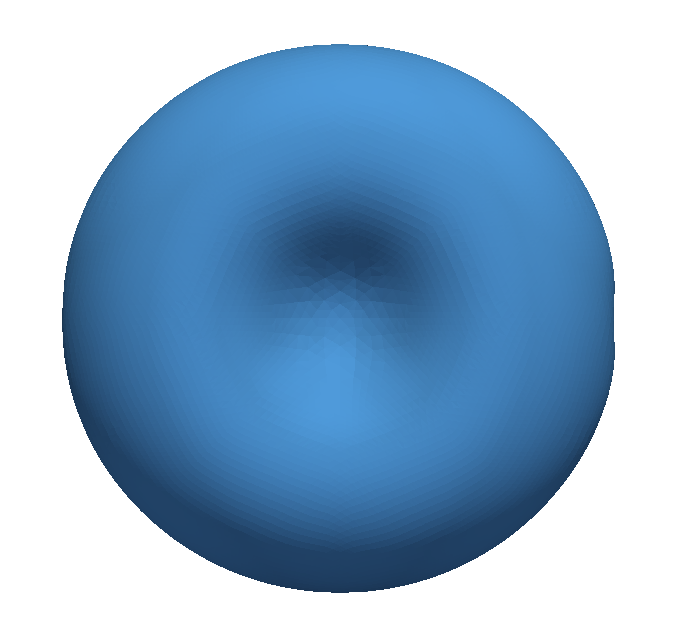}}\hspace{0mm}
\\
\includegraphics[width=0.21\textwidth,angle=0]{{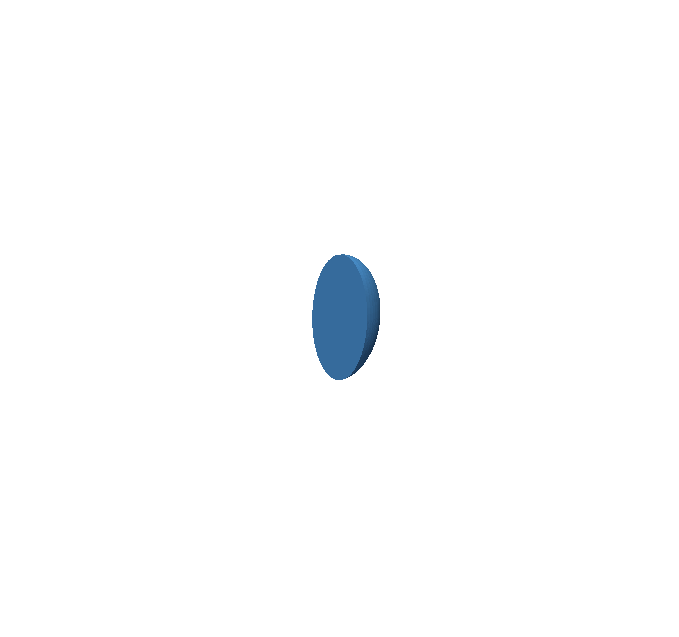}}\hspace{2mm}
\includegraphics[width=0.21\textwidth,angle=0]{{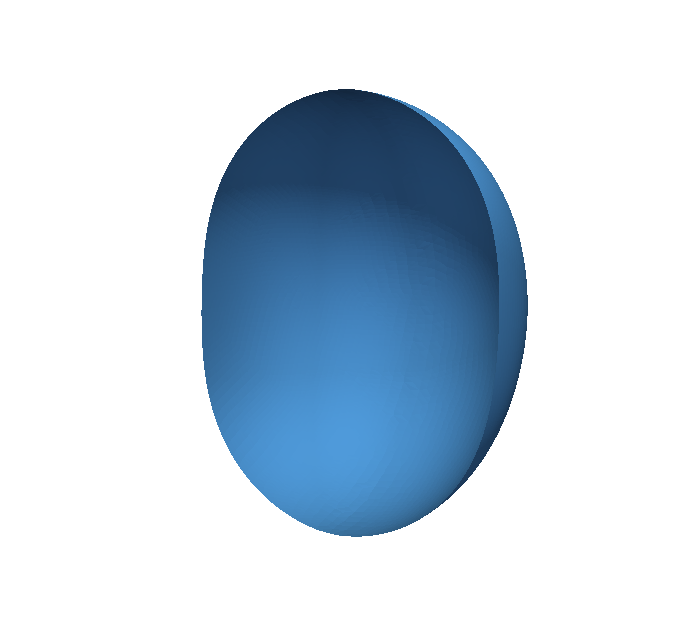}}\hspace{8mm}
\includegraphics[width=0.21\textwidth,angle=0]{{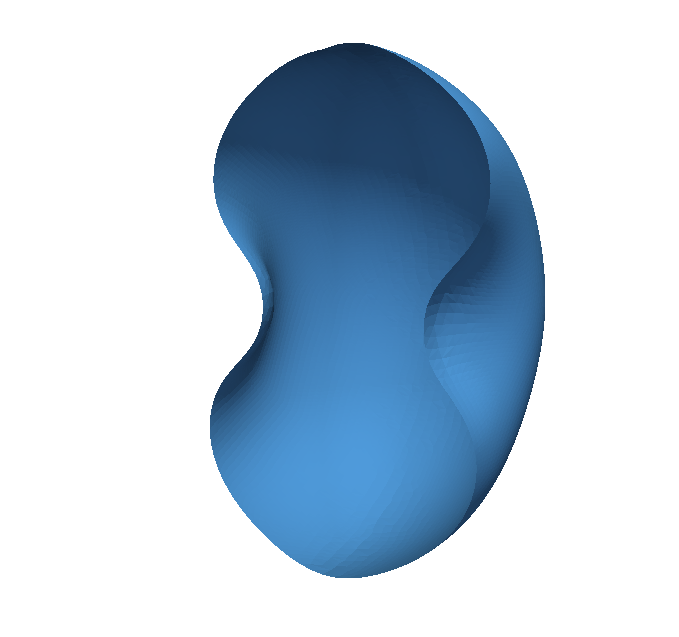}}\hspace{0mm}
\\
\includegraphics[width=0.21\textwidth,angle=0]{{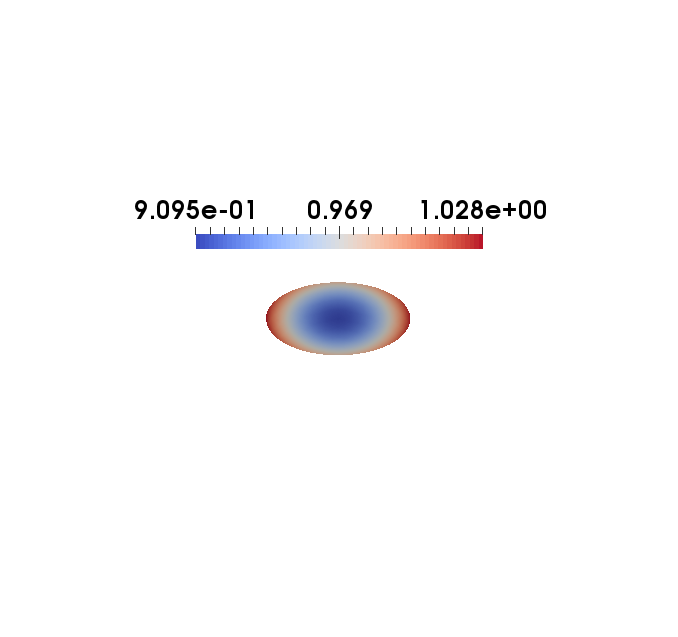}}\hspace{2mm}
\includegraphics[width=0.21\textwidth,angle=0]{{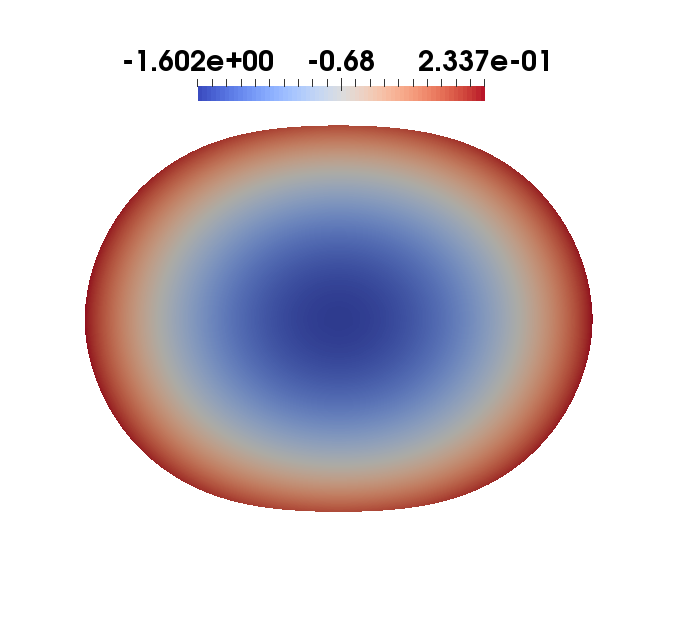}}\hspace{8mm}
\includegraphics[width=0.21\textwidth,angle=0]{{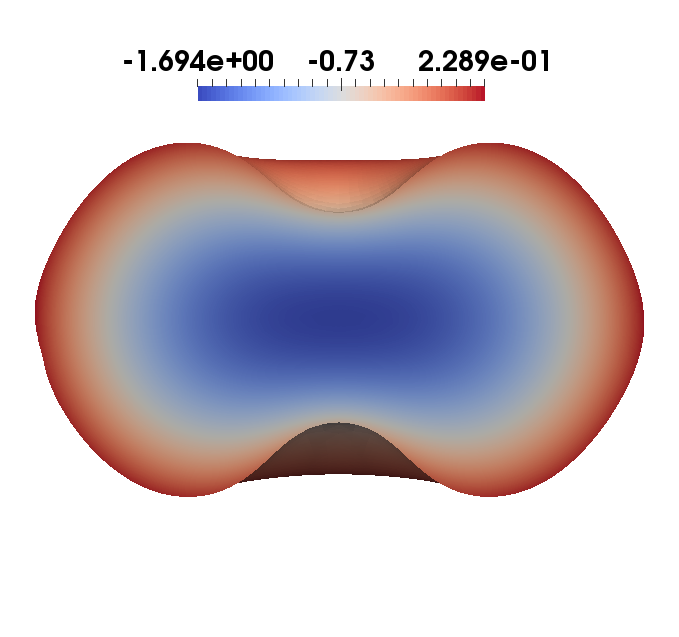}}\hspace{0mm}
\end{center}
\caption{Parametric scheme with $\beta = 0$, $\alpha = 1.0, ~\gamm = 0.1$: 
looking down the $x$ axis (first row), looking down the $y$ axis (second row), cross section in the $z = 0$ plane (third row), and $u_h$ on the plane $z = 0$ (fourth row). Taken at $t = 0, 7, 13$.}
\label{fig_para_3d}
\end{figure}

Figure \ref{fig_para_3d} displays results from the parametric scheme at $t = 0, 7,13$. In this figure we display the surface $\Gamma_h$ in three different orientations: looking down the $x$ axis (first row), down the $y$ axis (second row), and a cross section in the plane $z = 0$ (third row). We also display the solution $u_h$ on a cross section in the plane $z=0$ (fourth row). 

\begin{figure}[htbp]
\begin{center}
\includegraphics[width=0.21\textwidth,angle=0]{{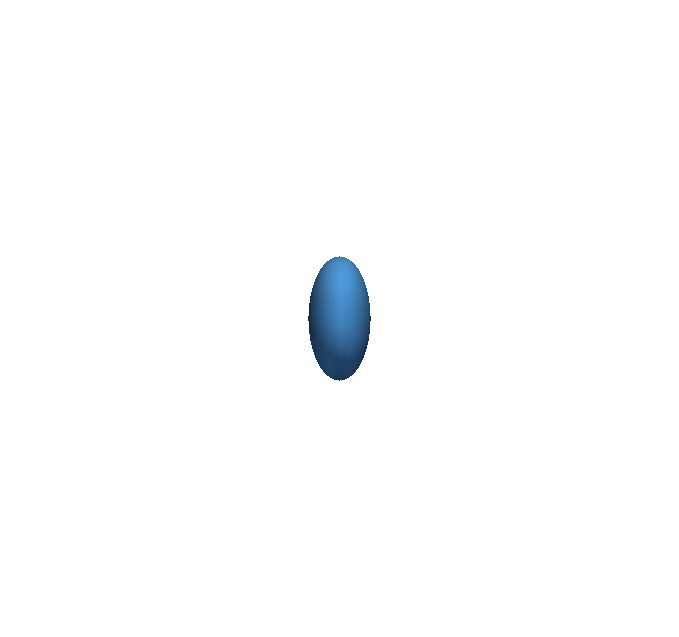}}\hspace{2mm}
\includegraphics[width=0.21\textwidth,angle=0]{{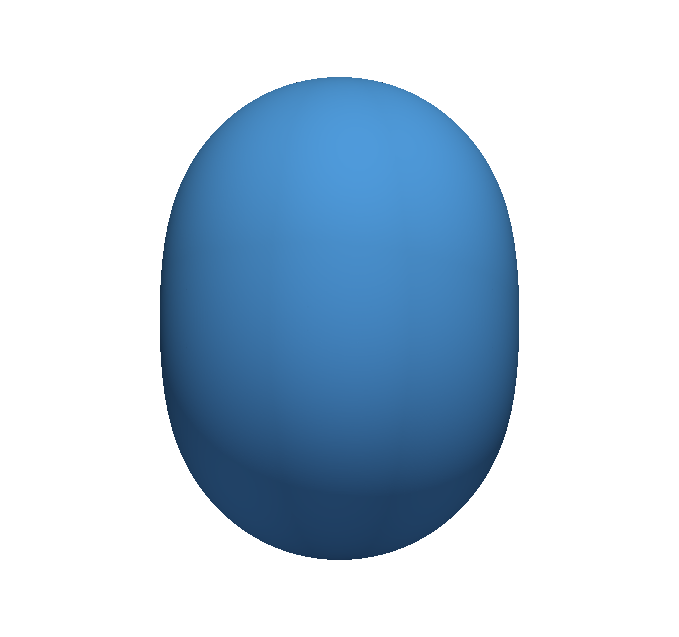}}\hspace{8mm}
\includegraphics[width=0.21\textwidth,angle=0]{{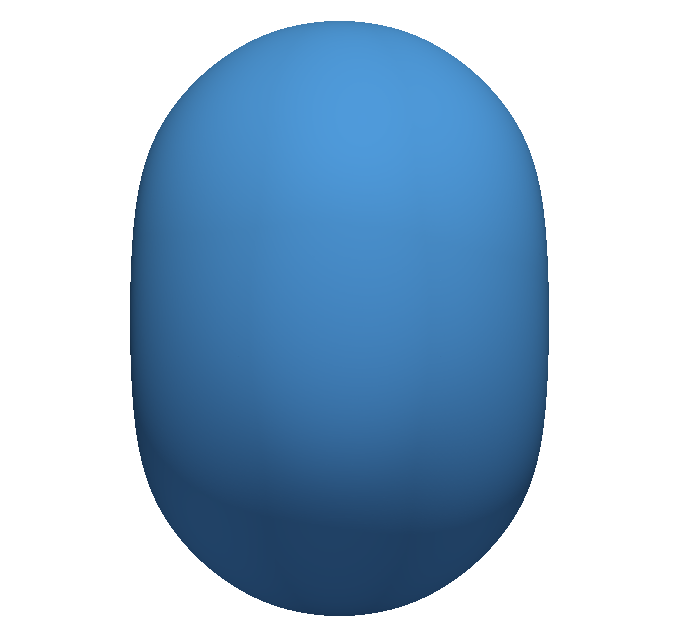}}\hspace{0mm}
\\
\includegraphics[width=0.21\textwidth,angle=0]{{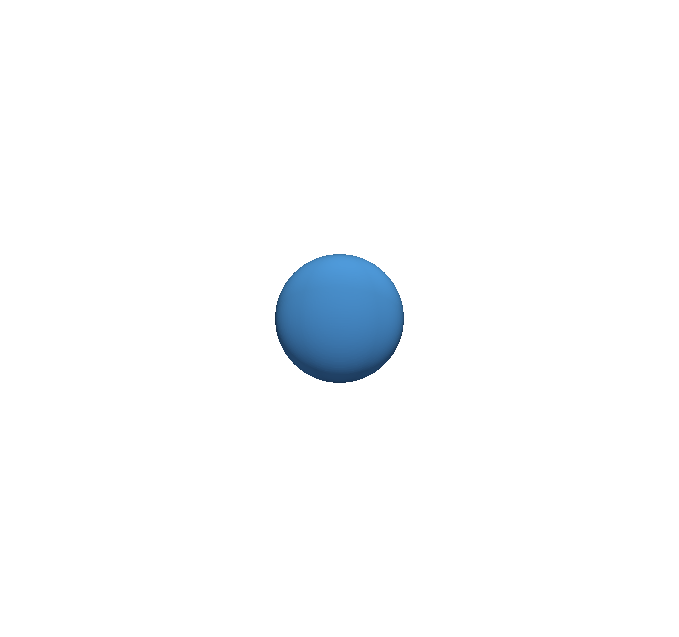}}\hspace{2mm}
\includegraphics[width=0.21\textwidth,angle=0]{{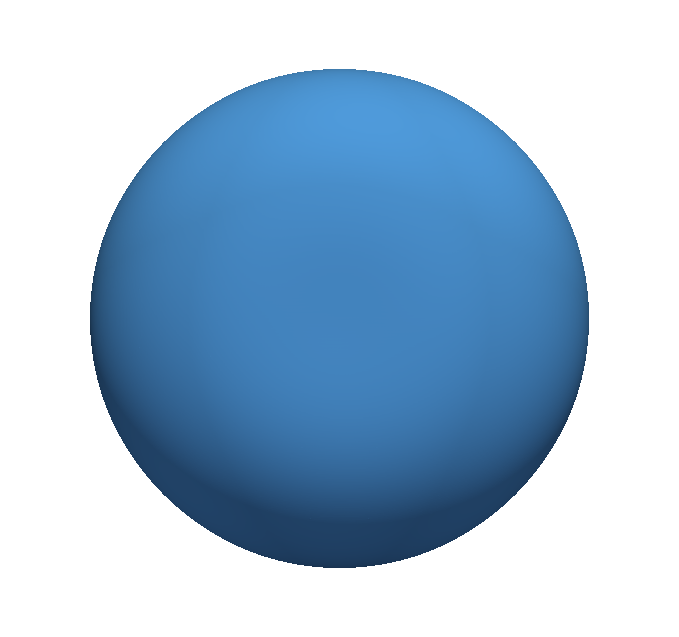}}\hspace{8mm}
\includegraphics[width=0.21\textwidth,angle=0]{{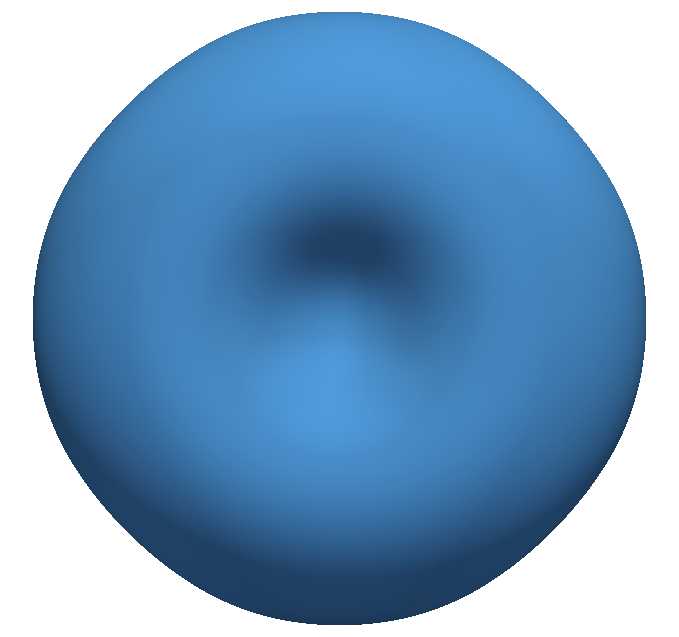}}\hspace{0mm}
\\
\includegraphics[width=0.21\textwidth,angle=0]{{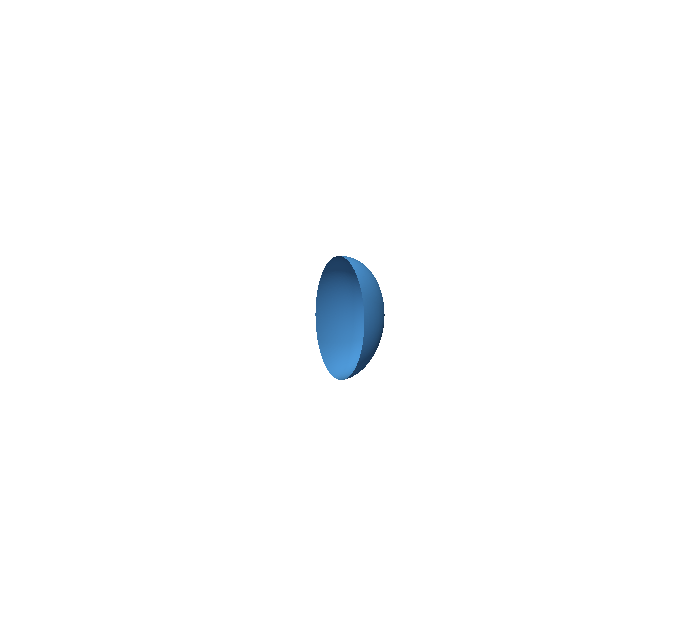}}\hspace{2mm}
\includegraphics[width=0.21\textwidth,angle=0]{{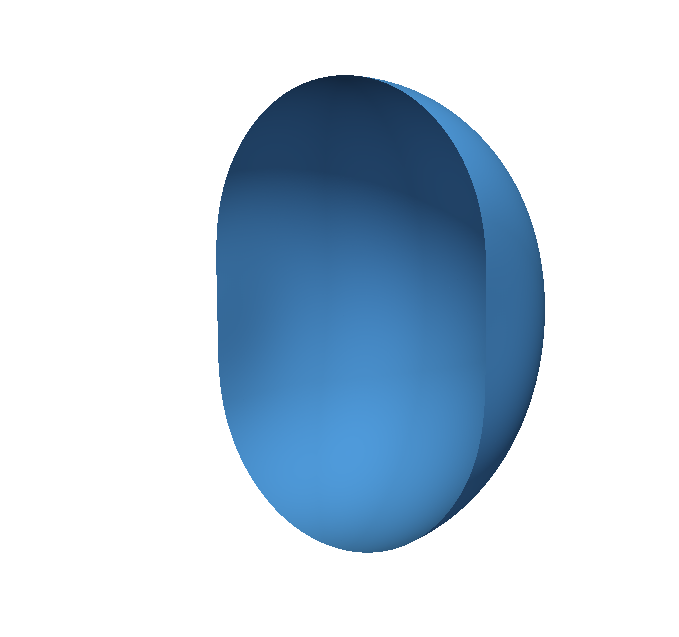}}\hspace{8mm}
\includegraphics[width=0.21\textwidth,angle=0]{{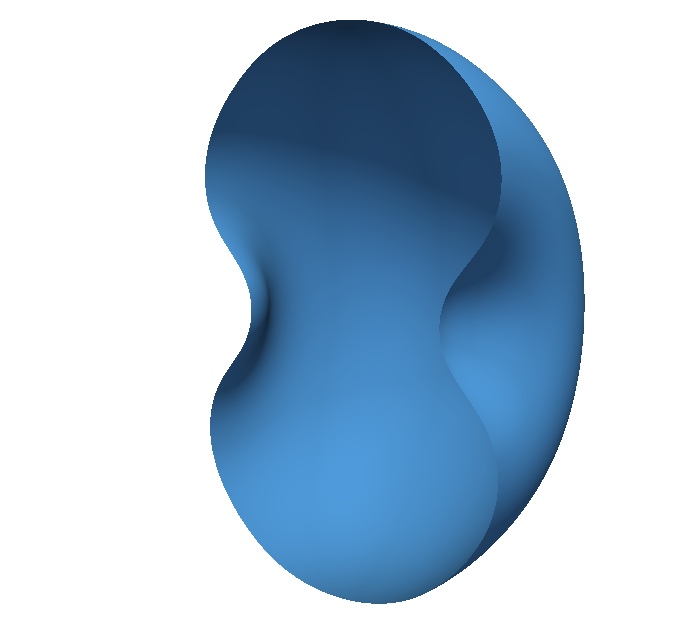}}\hspace{0mm}
\\
\includegraphics[width=0.21\textwidth,angle=0]{{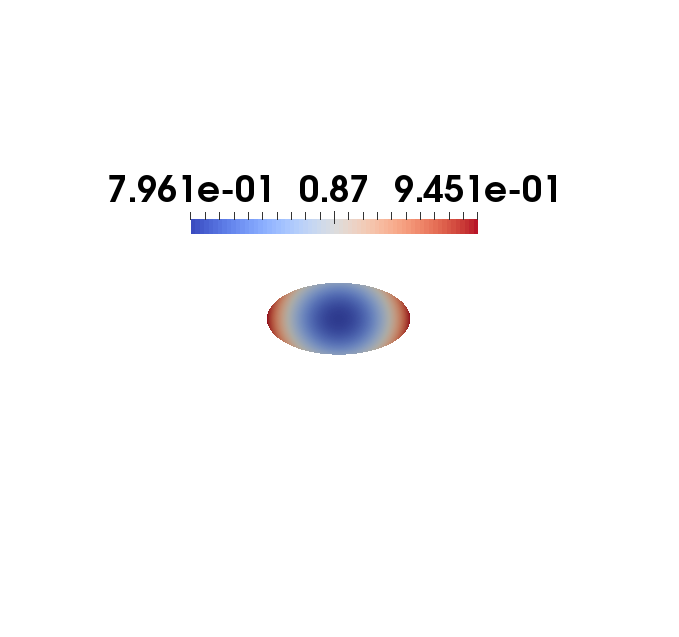}}\hspace{2mm}
\includegraphics[width=0.21\textwidth,angle=0]{{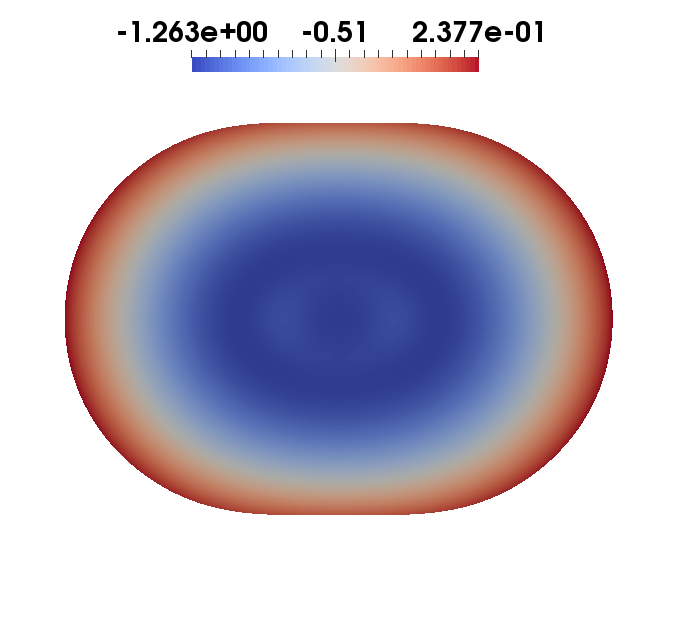}}\hspace{8mm}
\includegraphics[width=0.21\textwidth,angle=0]{{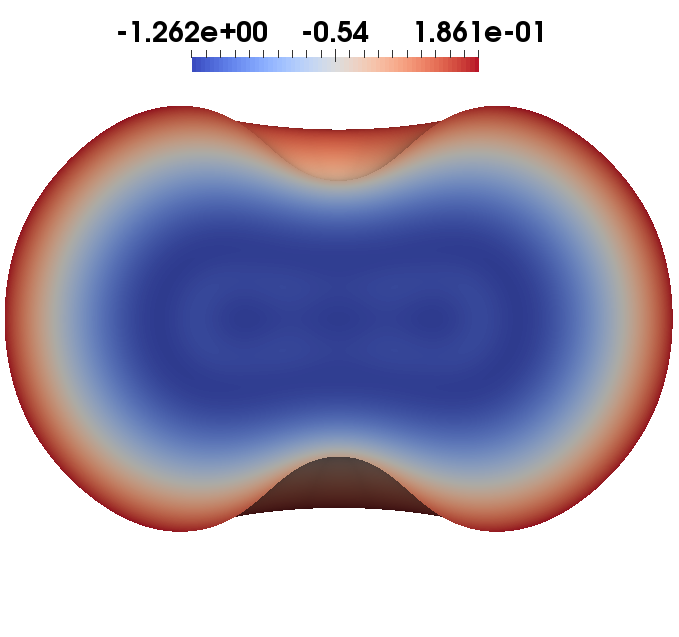}}\hspace{0mm}
\end{center}

\caption{Diffuse--interface scheme with $\beta = 0,$ $\alpha = 1.0, ~\gamm = 0.1$:
looking down the $x$ axis (first row), looking down the $y$ axis (second row), cross section in the $z = 0$ plane (third row), and $\tilde{u}_h$ on the plane $z = 0$ (fourth row). Taken at $t = 0, 7, 13$.}
\label{fig_phase_3d}
\end{figure}

In Figure \ref{fig_phase_3d} we display results from the diffuse--interface scheme. We display the zero level surface of $\varphi_h$ in three orientations: looking down the $x$ axis (first row), down the $y$ axis (second row), and a cross section in the plane $z = 0$ (third row). We also display plots of the solution $\tilde{u}_h$ on the zero level surface of $\varphi_h$ on a cross section in the plane $z=0$ (fourth row). 

\subsection{Summary of the numerical simulations}
From the above results we see that the model gives rise to complicated structures developing from simple elliptical geometries. 
Comparing  Figure \ref{image_a01b01_old} with Figure \ref{image_a01b01_nonsym_old} we see the complexity of the structure increases with the complexity of the initial geometry. 
The additional curvature term in the Robin boundary condition for the pressure that is present when $\beta\neq 0$, results in a more compact rounded geometry, compare Figures \ref{image_a10b01_new} and \ref{image_a01b01_new} with the corresponding Figures \ref{image_a10b01_old} and \ref{image_a01b01_old}. From Figures \ref{image_a10b01_new}, \ref{image_a01b01_new}, \ref{image_a10b01_old}  and \ref{image_a01b01_old} we see that significantly larger values of the pressure, $u$, on the boundary are achieved with $\alpha=1$ compared to $\alpha=0.1$, whereas the difference between the minimum values in the interior is much smaller.

In Figure \ref{image_a10b01_col_old} simulations from the diffuse--interface model show two tumours merging to form a single tumour. The change in topology that occurs in this example is naturally handled by the diffuse--interface model and demonstrates the power of this model, in addition it illustrates the complicated pathologies that can result even for the current simple tissue--growth model 

\section{Discussion}
We conclude with some observations, noting possible extensions to the work and highlighting open problems. 

The formulation (\ref{eq_f2_uOnOmega_intro}) -- (\ref{eq_f2_vOnGamma_intro}) derived and analysed above represents arguably the simplest macroscale spatially structural model for tissue growth (another of comparable complexity replaces Darcy flow with Stokes flow) and hence bucks the current trend in which, for good reason, numerous additional mechanisms are incorporated with the goal of achieving better biological realism. It nevertheless exhibits growth morphologies sharing features of those much more complicated models, allowing the associated mechanisms much more readily to be identified and mitigating against possible pitfalls of assuming responsibility for behaviour that is in fact rather generic to specific biological processes. 

The biological aspects implicit in (\ref{der_L1}) pertain primarily to the constitutive assumptions made on $k_b(c)$, $k_d(c)$, $D(n)$ and $K(c)$, on the interpretation of $\lambda$ and on the choice of Darcy flow. Obvious generalisations would include the inclusion of active cell--cell interactions (necessitating the adoption of genuinely multiphase models), the adaption of more complex and realistic constitutive laws for the mechanics and the proper treatment of the exterior domain - notably, replacing the Dirichlet condition on $c$ in (\ref{der_L2}) by describing exterior nutrient transport would allow the inclusion of Mullins--Sekerka instabilities as well as those of Saffron--Taylor type captured above (whereby the exterior is in effect inviscid and fingers are prone to develop due to a self-reinforcing mechanism whereby for the sink it is easier to pull in tissue in directions for which there is less to pull). We believe the current framework provides a firm and well--characterised basis on which to incorporate such additional effects.

From a mathematical perspective, open problems for rigorous analysis abound in this context, including the derivation of the thin--rim limit, the analysis of the travelling--wave boundary value problem and the derivation and analysis of the thin--film evolution equation from the appendix, the last of these having been successfully applied in many contexts, including in biological ones such as the growth of bacteria biofilms. From both the formal and rigorous point of view more could be done by way of weakly nonlinear analysis on the basis of the linear--stability results and plausibly exploiting bifurcatia--with--symmetry approaches. Changes of topology result from evolution beyond that demonstrated in Figure \ref{image_a01b01_new}, for example, with the parts of the boundary that become isolated from the exterior requiring specific modelling. 
We record here the simplest such scenario, namely a concentric circular annulus $s_1(t)<r<s_2(t)$ containing passive material in $0<r<s_1(t)$ that is devoid of nutrient, so that
$$
\frac1{r}\frac{\partial}{\partial r}\left(r\frac{\partial u}{\partial r}\right)=1
$$
\begin{eqnarray*}
&\mbox{at}~r=s_1(t),&u=\frac{\alpha\beta}{s_1},~\dot{s}_1=-\frac{\partial u}{\partial r}\\
&\mbox{at}~r=s_2(t),&u=\frac{\alpha\beta}{s_2}+\alpha\dot{s}_2,~\dot{s}_2=Q-\frac{\partial u}{\partial r}
\end{eqnarray*}
(since no kinetic--undercooling regularisation is present at $r=s_1(t)$, the limit $\alpha\to0$ with $\alpha\beta=O(1)$ needs to be taken there). Hence 
$$
u=\frac14r^2+A(t)\ln r+B(t)
$$
with $s_1,s_2,A$ and $B$ given by the system
$$
\dot{s}_1=-\frac12s_1-\frac{A}{s_1},~~~\dot{s}_2=Q-\frac12s_2-\frac{A}{s_2},
$$
$$
\frac14s_1^2+A\ln s_1+B=-\frac{\alpha\beta}{s_1},~~~\frac14s_2^2+A\ln s_2+B=-\frac{\alpha\beta}{s_2}+\alpha\dot{s}_2,
$$
the most revealing aspect of which is its large--time behaviour
$$
s_2-s_1\sim Q,~~\left(1+\frac{\alpha}{Q}\right)s_2\sim \frac12 Qt~~\mbox{as }t\to \infty,
$$
which is equivalent to the corresponding one--dimensional solution that provides the dominant balance in the growing but asymptotically constant thickness, annulus. The stability or otherwise of this solution highlights more general, and challenging, open questions regarding the full bifurcation structure of solutions in order to characterise the variety of possible large--time outcomes.
Finally we note that for $\beta=0$ time--reversal ($t\to-t$, $u\to-u$, $V\to -V$) of (\ref{eq_f2_uOnOmega_intro}) -- (\ref{eq_f2_vOnGamma_intro})  leads to a meaningful problem comprising a volumetric source and surface sink of possible reference to drug treatment from the exterior of nutrient--rich tissue, though the assumption on the viscosity ratio would need reversing if the kinetic--undercooling term is to be retained. 

\vspace{3mm}

{\bf Acknowledgements}\\
The authors would like to thank the Isaac Newton Institute for
Mathematical Sciences for its hospitality during the programme 
{\it Coupling Geometric PDEs with Physics for Cell Morphology, Motility and Pattern Formation}
supported by EPSRC Grant Number EP/K032208/1. JE gratefully acknowledges the support of the EPSRC grant 1507261, VS gratefully acknowledges the support of the Leverhulme Trust  Research Project Grant (RPG-2014-149) and JK and VS would like to thank the Mathematisches Forschungsinstitut Oberwolfach (MFO) for its hospitality during the programme  {\it 1704: Emerging Developments in Interfaces and Free Boundaries}.

\section*{Appendix: The thin--film limit}
\label{s:thin_film}

In this appendix we apply thin--film methods to (\ref{eq_f2_uOnOmega_intro}) -- (\ref{eq_f2_vOnGamma_intro}) in a traditional way (indeed, when the radially symmetric solution is unstable, this is perhaps the most natural next regime to study, motivation being provided by numerical results such as those in Figure \ref{f:stab}) to derive the novel evolution equation (\ref{thin_film:NH1}), but supplement this derivation with a warning about its limited applicability. We consider here a tumour that is initially thin and, for brevity, two--dimensional and symmetric about the $x$--axis. Thus we take $\Gamma$ to be given in $y > 0$ by 
\[
	y = \epsilon h(x, t) ,
\]
with $0 < \epsilon \ll 1$, and rescale via $y = \epsilon Y$ to give
\[
	\frac{\partial ^2 u}{\partial Y^2} + \epsilon^2 \frac{\partial^2 u}{\partial x^2} = \epsilon^2 , 
\]
\begin{equation}
 \tag{A.1}
\label{thin_film:NH1}
\frac{\partial u}{\partial Y} = 0 , \quad \mbox{ at } Y = 0,  
\end{equation}
\[
	\epsilon^2 \frac{\partial h}{\partial t}  = \epsilon^2 Q \left( 1 + \epsilon^2 \left( \frac{\partial h}{\partial x}\right) ^2 \right)^{\frac{1}{2}} - \frac{\partial u}{\partial Y} + \epsilon ^2 \frac{\partial h}{\partial x} \frac{\partial u}{\partial x}   \quad \mbox{ at } Y = h(x, t),
\]
\[ 
	\frac{1}{\left( 1 + \epsilon^2 (\frac{\partial h}{\partial x})^2 \right) ^{\frac{1}{2}}} \frac{\partial h}{\partial t} = \frac{u}{\alpha} + \beta \frac{1}{\left( 1 + \epsilon^2 (\frac{\partial h}{\partial x})^2 \right)^{\frac{3}{2}}} \frac{\partial^2 h}{\partial x^2}    \quad \mbox{ at } Y = h(x, t),
\]
where to obtain a distinguished limit we have also rescaled $Q$ and $\alpha$, replacing $Q$ by $\epsilon Q$ and $\alpha $ by $\alpha / \epsilon$.

Integrating with respect to $Y$ from $0$ to $h$ gives 
\[
	\frac{\partial}{\partial x} \left( \int_0^h \frac{\partial u}{\partial x} \diff Y \right)
	= \frac{\partial h}{\partial t} - Q \left( 1+ \epsilon^2 \left( \frac{\partial h}{\partial x} \right) ^2 \right) ^{\frac{1}{2}} + h ,
\]
where the final boundary condition has yet to be used; taking $\epsilon \rightarrow 0$ gives at leading order that 
\[
	u = \alpha \left( \frac{\partial h}{\partial t} - \beta \frac{\partial^2 h }{\partial x^2} \right) ,
\]
\[
	\frac{\partial h}{\partial t} = \frac{\partial }{\partial x} \left( h \frac{\partial u}{\partial x} \right) + Q - h,
\]
so an evolution equation with both a pseudoparabolic and a fourth--order regularisation, namely
\begin{equation}
 \tag{A.2}
\label{thin_film:NH2}
	\frac{\partial h}{\partial t} = \alpha \frac{\partial }{\partial x} \left( h \left( \frac{\partial ^2 h }{\partial x \partial t } - \beta \frac{\partial^3 h }{\partial x^3 } \right) \right) + Q - h, 
\end{equation}
ensues. While interesting in its own right, (\ref{thin_film:NH2}) contains no destabilising terms, contrary to the conclusions above about the prevalence of instabilities. To explain this conundrum, which amounts to a warning against the naive application of thin--film approaches ((\ref{thin_film:NH2}) does not in and of itself provide such a warning, being a composite of well--established types), we now complement the current analysis by characterising the stability properties of the one--dimensional solution (the results being of more general local relevance for slender initial data), retaining for subsequent transparency the scalings in (\ref{thin_film:NH1}); $\epsilon$ amounts to an artificial small parameter that can be removed from (\ref{thin_film:NH3}) by reverting the previous scalings via $\alpha \rightarrow \epsilon \alpha$ and $h_p \rightarrow h_p / \epsilon$.

Thus the planar solution $u = u_p(Y, t)$, $h = h_p(t)$, in which surface tension of course plays no role, is given by
\[
	\dot{h}_p = Q - h_p, \quad u_p = \frac{\epsilon^2}{2}{Y^2} - \frac{\epsilon^2}{2} h_p^2 + \alpha \dot{h}_p .
\]
Perturbing in the form 
\[
	u \sim u_p(Y, t) + \delta U(x, Y, t), \quad h \sim h_p(t) + \delta H(x, t) ,
\]
and retaining terms that are linear in $\delta$ only gives
\[
 	\frac{\partial^2 U }{\partial Y^2 } + \epsilon^2 \frac{\partial^2 U }{\partial x^2} = 0, 
\]
\[
	\frac{\partial U}{\partial Y} = 0 \quad \mbox{ at } Y = 0, 
\]
\[
	\epsilon^2 \frac{\partial H}{\partial t}  = - \epsilon^2 H - \frac{\partial U}{\partial Y} 
	, 
	\quad \frac{\partial H}{\partial t} = \frac{1}{\alpha} (\epsilon^2 h_p H + U) + \beta \frac{\partial^2 H}{\partial x^2} 
	\quad \mbox{ at } Y = h_p(t)
	.
\]
Hence for 
\[
	U = A_n (t) \cosh(\epsilon \lambda Y) 
	\begin{cases}
		\cos(\lambda x) \\
		\sin(\lambda x)
	\end{cases}	
	,
	\quad
	H = B_n (t) 
	\begin{cases}
		\cos(\lambda x) \\
		\sin(\lambda x)
	\end{cases}
	,
\]
we obtain
\[
	\epsilon \frac{\diff B_n}{\diff t} = - \epsilon B_n - \lambda A_n \sinh(\epsilon \lambda h_p) ,
	\quad
	\frac{\diff B_n}{\diff t} = \frac{1}{\alpha} (\epsilon^2 h_p B_n + A_n \cosh(\epsilon \lambda h_p)) - \beta \lambda ^2 B_n 
	, 
\]
so that
\begin{equation}
 \tag{A.3}
\label{thin_film:NH3}
	(\epsilon + a \lambda \tanh(\epsilon \lambda h_p) \dot{B}_n )
	 = 
	 - \alpha \beta \lambda^3 \tanh(\epsilon \lambda h_p)B_n
	 - \epsilon B_n 
	 + \epsilon^2 \lambda h_p \tanh(\epsilon \lambda h_p) B_n
	 .
\end{equation}
In the limit $\epsilon \rightarrow 0$, $\lambda = O(1)$, this recovers the linear stability result for (\ref{thin_film:NH2}), namely 
\[
	(1 + a \lambda^2 h_p) \dot{B}_n \sim - a \beta \lambda ^4 h_p B_n - B_n ,
\]
but for short wavelengths, $\lambda = \mu / \epsilon$, and small $\beta$, $\beta = \epsilon^4 b$, we obtain a different distinguished limit which is in no way foreshadowed by the thin--film analysis, namely
\[
	\alpha \mu \frac{\diff B_n}{\diff t} = \mu h_p B_n - a b \mu^3 B_n - \coth(\mu h_p) B_n ,
\] 
wherein $T = \epsilon t$, in which the first term on the right--hand side will lead to instability for sufficiently large $h_p$, limiting the range of applicability of the thin--film approach and reflecting a more general property of (\ref{thin_film:NH3}) that can also be exhibited by other distinguished limits. Thus the linear--stability analysis represents an essential adjunct to the thin--film description in characterising the evolution from slender initial data. With regard to the interpretation of the numerical results above, it is also important to note that when $Q$ in (\ref{thin_film:NH2}) is large the domain may in any case cease to be slender simply due to the associated growth in the $y$ direction; moreover, a thin--film analysis is not of course applicable close to the tips.

\bibliographystyle{siam}

\begin{thebibliography}{10}

\bibitem{alam}
\textsc{Abels, H. Lam, K.F., \& Stinner, B.}
\newblock {Analysis of the diffuse domain approach for a bulk--surface coupled
  PDE system}.
\newblock {\em SIAM J. Numer. Anal.}, 47(5):3687--3725, 2015.

\bibitem{Barrett2005}
\textsc{Barrett, J.W., N{\"{u}}rnberg, R., \& Styles, V}
\newblock {Finite element approximation of a phase--field model for void electromigration}.
\newblock {\em SIAM J. Numer. Anal.}, 42(2):738--772, 2005.

\bibitem{Blowey_Elliott_double_obs}
\textsc{Blowey J.F., \& Elliott, C.M.}
\newblock {A phase--field model with a double obstacle potential}.
\newblock 1994.

\bibitem{Burger2014}
\textsc{Burger, M., Schlottbom, M., \& Elvetun, O.L.}
\newblock {Analysis of the diffuse domain method for second order elliptic
  boundary value problems}.
\newblock {\em Found. Comput. Math.}, 17(3):627--674, 2017.

\bibitem{byrne_chaplain_SI}
\textsc{Chaplain, M.A.J., \& Byrne, H.M.}
\newblock {Free boundary value problems associated with the growth and
  development of multicellular spheroids}.
\newblock {\em Eur. J. Applied Math.}, 8(6):639--658, 1997.

\bibitem{cristini}
\textsc{Cristini, V., Li, X., Lowengrub, J.S., \& Wise, S.M.}
\newblock {Nonlinear simulations of solid tumor growth using a mixture model:
  invasion and branching}.
\newblock {\em J. Math. Biol.}, 58(4--5):723--763, 2009.

\bibitem{Deckelnick2005}
\textsc{Deckelnick, K., Elliott, C.M., \& Dziuk, G.}
\newblock {Computation of geometric partial differential equations and mean
  curvature flow}.
\newblock {\em Acta Numerica}, 14:139--232, 2005.

\bibitem{digm}
\textsc{Deckelnick, K., \& Elliott, C.M.}
\newblock {An existence and uniqueness result for a phase-field model of
  diffusion-induced grain-boundary motion}.
\newblock {\em Proceedings of the Royal Society of Edinburgh}, pages
  1323--1344, 2001.

\bibitem{Dziuk1990}
\textsc{Dziuk, G.}
\newblock {An algorithm for evolutionary surfaces}.
\newblock {\em Numerisch Mathematik}, 58:603--611, 1991.

\bibitem{Elliotta}
\textsc{Elliott, C.M., \& Fritz, H.}
\newblock {On algorithms with good mesh properties for problems with moving
  boundaries based on the Harmonic Map Heat Flow and the DeTurck trick}.
\newblock {\em SMAI J. Comput. Math.}, 2:141--176, 2016.

\bibitem{Eyles}
\textsc{Eyles, J.}
\newblock{Numerical analysis and simulations of a tractable model for tumour growth}.
\newblock{PhD thesis}, University of Sussex, 2019.

\bibitem{chdTumourModel}
\textsc{Garcke, H.,  Lam, K.F, Sitka, E., \& Styles, V.}
\newblock {A Cahn--Hilliard--Darcy model for tumour growth with chemotaxis
  and active transport}.
\newblock {\em Math. Models Meth. Appl. Sci.}, 26(6),
  2016.

\bibitem{GMSH_paper}
\textsc{Geuzaine, C., \&  Remacle, J.F.}
\newblock {Gmsh: a three--dimensional finite element mesh generator with
  built--in pre-- and post--processing facilities}.
\newblock {\em Int. J. Numer. Meth. Eng.},
  79(11):1309--1331, 2009.

\bibitem{greenspan1}
\textsc{Greenspan, H.P.}
\newblock {Models for the growth of a solid tumour by diffusion}.
\newblock {\em Stud. Appl. Math.}, 52:317--340, 1972.

\bibitem{greenspan2}
\textsc{Greenspan, H.P.}
\newblock {On the growth and stability of cell cultures and solid tumours}.
\newblock {\em J. Theor. Biol.}, 56:229--242, 1976.

\bibitem{paraview_book}
\textsc{Hansen, C.D., \& Johnson, C.R.}
\newblock {\em {Visualization Handbook}}.
\newblock Elsevier--Butterworth Heinemann, 2005.

\bibitem{jkow}
\textsc{King, J.R.}
\newblock {Interface dynamics in biological tissue growth}.
{\sc C.M. Elliott, Y. Giga, M. Hinze and V. Styles (Eds.)}, (2017), 
{\it Emerging Developments in Interfaces and Free Boundaries}, 
Oberwolfach Reports, {\bf 14}, 267--338.


\bibitem{king_franks_04}
\textsc{King, J.R., \& Franks, S.J.}
\newblock {Mathematical analysis of some multi--dimensional tissue growth models}.
\newblock {\em Eur. J. Appl. Math.}, 15, 273--295, 2004.
  

\bibitem{king_franks_paper}
\textsc{King, J.R., \& Franks, S.J.}
\newblock {Mathematical modelling of nutrient--limited tissue growth}.
\newblock {\em Free Boundary Problems. International Series of Numerical
  Mathematics}, 154, 273--282, 2006.
  
  
\bibitem{king_franks_book}
\textsc{King, J.R., \& Franks, S.J.}
\newblock {Stability properties of some tissue growth models}.
\newblock 175--182, in Mathematical Modelling of Biological Systems, Volume I, Eds. A Deutsch, L Brusch, H Byrne, G Vries, H Herzel, Springer 2007.  

\bibitem{ALBERTABook}
\textsc{Schmidt, A., \&  Siebert, K.G.}
\newblock {\em {Design of adaptive finite element software}}.
\newblock Springer Berlin Heidelberg, 2006.

\bibitem{ok_ho_la_03}
\textsc{Ockendon, J.R., Howison, S.D., \& Lacey, A.A.}
\newblock {Mushy regions in negative squeeze films}.
\newblock {\em Quart. J. Mech. Appl. Math.}, 56(3), 361--379, 2003.



\bibitem{ho_92}
\textsc{Howison, S.D.}
\newblock {Complex variable methods in Hele--Shaw moving boundary problems.}.
\newblock {\em Eur. J. Appl. Math.}, 3, 209--224, 1992.



\bibitem{Norris_king_byrne_06}
\textsc{Norris, E.S., King, J.R., \& Byrne, H.M.}
\newblock {Modelling the response of spatially structured tumours to chemotherapy: Drug kinetics}.
\newblock {\em Math. Comp. Model.}, 43, 820--837, 2006.
\end{thebibliography}

\end{document}